\documentclass[aps,twocolumn]{revtex4}

\RequirePackage[T2A]{fontenc}

\usepackage{amssymb}
\usepackage{amsmath}
\usepackage{bm}
\usepackage[dvips]{epsfig}
\usepackage{graphicx}

\begin{document}

\title{On the Novikov problem for superposition of periodic potentials}

\author{A.Ya. Maltsev}

\affiliation{
\centerline{\it L.D. Landau Institute for Theoretical Physics}
\centerline{\it 142432 Chernogolovka, pr. Ak. Semenova 1A,
maltsev@itp.ac.ru}}

\begin{abstract}
 We consider the Novikov problem, namely, the problem of describing 
the level lines of quasiperiodic functions on the plane, for a special 
class of potentials that have important applications in the physics 
of two-dimensional systems. Potentials of this type are given 
by a superposition of periodic potentials and represent quasiperiodic 
functions on a plane with four quasiperiods. Here we study an 
important special case when the periodic potentials have the same 
rotational symmetry. In the generic case, their superpositions 
have ``chaotic'' open level lines, which brings them close to random 
potentials. At the same time, the Novikov problem has interesting 
features also for ``magic'' rotation angles, which lead to the 
emergence of periodic superpositions.
\end{abstract}

\maketitle

\section{Introduction}

 We consider the problem of S.P. Novikov, namely, the description 
of level lines of quasiperiodic potentials on a plane, 
for an important special class of such potentials.

 In general, the Novikov problem consists in describing the 
geometry of the level lines
\begin{equation}
\label{LevelLines}
f (x, y) \,\,\, = \,\,\, \text{\rm const} 
\end{equation}
of a quasiperiodic function $\, f (x, y) \, $ in $\, \mathbb{R}^{2} \, $ 
with a given number of quasiperiods. By a quasiperiodic function in 
$\, \mathbb{R}^{2} \, $ with $N$ quasiperiods we mean the restriction 
of an $N$ - periodic function $\, F (z^{1}, \dots , z^{N}) \, $ in 
$\, \mathbb{R}^{N} \, $ to $\, \mathbb{R}^{2} \, $ for any affine 
embedding $\, \mathbb{R}^{2} \subset \mathbb{R}^{N} \, $. The function 
$\, F (z^{1}, \dots , z^{N}) \, $ is assumed to be sufficiently smooth.

 The Novikov problem was first formulated for the case of 
3 quasi-periods (\cite{MultValAnMorseTheory}), where it is equivalent 
to describing the intersection lines of an arbitrary 3-periodic surface 
in $\, \mathbb{R}^{3} \, $ by a family of parallel planes of a given
direction. This case was studied in great detail in the Novikov 
topological school 
(\cite{zorich1,dynn1992,Tsarev,dynn1,zorich2,DynnBuDA,dynn2,dynn3}). 
At present, a complete classification of various types of level lines 
of functions with three quasiperiods on the plane has been obtained. 
The Novikov problem with three quasiperiods is, in fact, directly 
related to the theory of galvanomagnetic phenomena in metals, and 
the corresponding mathematical theorems have observable consequences 
in the magnetic conductivity of metals with complex Fermi surfaces 
(see \cite{PismaZhETF,UFN,BullBrazMathSoc,JournStatPhys}).

 The most profound results in the Novikov problem with 4 quasiperiods 
were obtained in the works \cite{NovKvazFunc,DynNov}. In general, 
however, the problem with 4 quasiperiods looks much more complicated 
and has not been studied as deeply as the problem with 3 quasiperiods. 
Here we will consider the situation related to the Novikov problem 
with 4 quasiperiods.

\vspace{1mm}

 As can be seen, the most non-trivial part of Novikov's problem 
is the description of open (non-closed) level lines of the function
$\, f (x, y) \, $. A special role in the description of open level 
lines of quasiperiodic functions is played by a certain type of open 
lines (\ref{LevelLines}), which can be called ``topologically regular''.

\begin{figure}[t]
\begin{center}
\includegraphics[width=\linewidth]{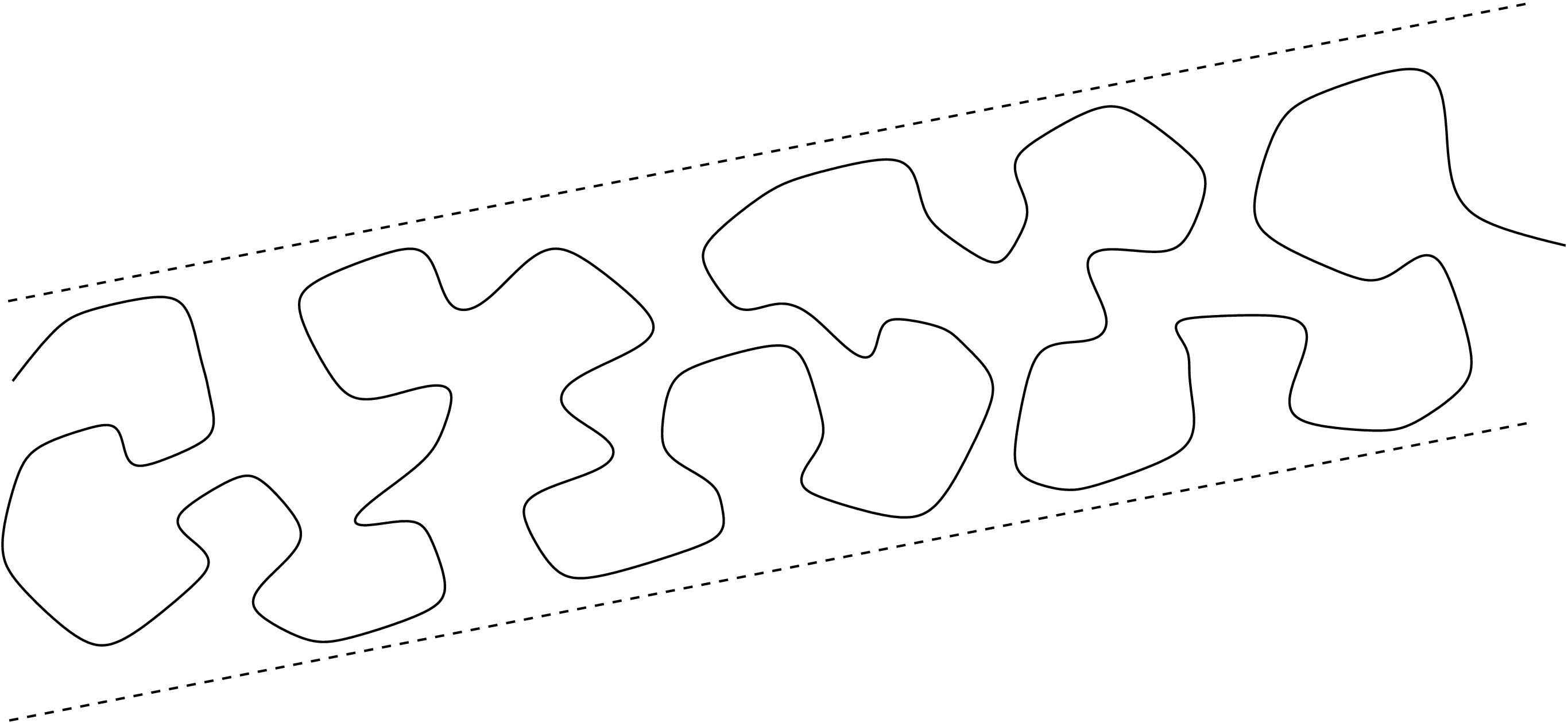}
\end{center}
\caption{The shape of a topologically regular open level line 
of a quasi-periodic function (schematically)}
\label{TopRegLine}
\end{figure}

 The main feature of topologically regular level lines is their 
remarkable geometric properties. Namely, each open topologically 
regular level line (\ref{LevelLines}) in $\, \mathbb{R}^{2} \, $ 
lies in a straight strip of finite width and passes through it 
(Fig. \ref{TopRegLine}). Note that topologically regular level lines, 
generally speaking, are not periodic.

 The second important feature of topologically regular level 
lines is their stability with respect to small variations 
in the parameters of the problem, as well as their connection with 
certain topological invariants corresponding to each stable family 
of such level lines.

 The parameters of the problem
\begin{equation}
\label{LevelLinesRest}
f (x, y) \,\,\, = \,\,\, F |_{\mathbb{R}^{2}} \,\,\, = \,\,\, c 
\end{equation}
may be, for example, the direction of embedding of $\, \mathbb{R}^{2} \, $ 
into the space $\, \mathbb{R}^{N} \, $, as well as the value $c$ of 
the function $\, f (x, y) \, $. In addition, in many problems there 
may also be variations of the function $\, F (z^{1}, \dots, z^{N}) \, $ 
(depending, possibly, on some parameters). A separate set of the 
parameters of the problem (\ref{LevelLinesRest}) is specified by 
parallel shifts of the space $\, \mathbb{R}^{2} \, $ in 
$\, \mathbb{R}^{N} \, $ for a fixed embedding direction 
$\, \xi \in G_{N,2} \, $. As we will see below, it is most convenient 
to study the problem (\ref{LevelLinesRest}) for the entire family of 
planes of a given direction in the space $\, \mathbb{R}^{N} \, $.

 As we have already said, topologically regular open level lines 
are stable with respect to small variations of all the parameters 
of the problem (\ref{LevelLinesRest}). As a consequence, they arise 
in stable ``families'', each of which corresponds to a certain 
region (Stability Zone) $\, \Omega_{\alpha} \, $ in the parameter 
space. In fact, the specificity of topologically regular level lines 
is such that they arise simultaneously in all planes 
$\, \mathbb{R}^{2} \subset \mathbb{R}^{N} \, $ of a given direction, 
so that the region $\, \Omega_{\alpha} \, $ always includes all the 
parallel shifts of $\, \mathbb{R}^{2} \, $ in $\, \mathbb{R}^{N} \, $. 
Moreover, when topologically regular level lines appear at one of 
the levels $\, f (x, y) = c \, $, the open level lines at other 
levels (wherever they appear) are also topologically regular 
and have the same mean direction. It can be seen, therefore, 
that the Stability Zones $\, \Omega_{\alpha} \, $ can, in fact, 
be defined as regions in $\, G_{N,2} \, $ 
(for a given $F (z^{1}, \dots, z^{N})$), implying that they include 
all other parameters of the problem as well.

 Topological invariants of regular open level lines of 
$\, f (x, y) \, $ are related to their mean direction and represent 
(irreducible) sets of $N$ integers $\, (m^{1}, \dots, m^{N}) \, $ for 
functions with $N$ quasiperiods. Namely, the mean direction of all 
topologically regular open level lines of $\, f (x, y) \, $ in 
$\, \mathbb{R}^{2} \, $ is always given by the intersection 
of $\, \mathbb{R}^{2} \subset \mathbb{R}^{N} \, $ with some 
(a priori unknown) integer hyperspace in $\, \mathbb{R}^{N} \, $
$$m^{1} z^{1} \, + \, m^{2} z^{2} \, + \, \dots \, + \, m^{N} z^{N}
\,\,\, = \,\,\, 0 \,\,\, , $$
$(m^{1}, \dots , m^{N}) \, \in \, \mathbb{Z}^{N} \, $. 

 The numbers $\, (m^{1}, \dots, m^{N}) \, $ do not change under 
small variations of parameters and are the same for the entire 
domain $\, \Omega_{\alpha} \, $ associated with a stable family of 
topologically regular level lines. In general, the set of domains 
$\, \Omega_{\alpha} \, $ and the corresponding numbers 
$\, (m^{1}_{\alpha}, \dots, m^{N}_{\alpha}) \, $ form a rather complex 
structure in the manifold $\, G_{N,2} \, $ 
(for a fixed function $F (z^{1}, \dots, z^{N})$).

 The union of Stability Zones (with boundaries) in $\, G_{N,2} \, $
$${\cal U}_{F} \,\,\, = \,\,\, \bigcup \, \overline{\Omega}_{\alpha}
\,\,\, \subset \,\,\, G_{N,2} $$
corresponds to the emergence of topologically regular open level 
lines (\ref{LevelLinesRest}) for a function $F (z^{1}, \dots, z^{N})$.

 All open level lines (\ref{LevelLinesRest}) that are not 
topologically regular or periodic will be called chaotic level
lines of the function $\, f (x, y) \, $. As follows from above,
chaotic level lines (\ref{LevelLinesRest}) cannot arise for the 
same parameter values as topologically regular level lines.

 The complement of the set $\, {\cal U}_{F} \, $ in $\, G_{N,2} \, $
$${\cal M}_{F} \,\,\, = \,\,\, 
\left. G_{N,2} \right\backslash {\cal U}_{F} $$
has usually a very complex structure in all 
dimensions $\, N \geq 3 \, $.

\vspace{1mm}

 As an example, we present here a typical picture arising in 
the case of 3 quasiperiods. Instead of the manifold 
$\, G_{3,2} = \mathbb{RP}^{2} \, $ it is more convenient to 
represent here the sphere $\, \mathbb{S}^{2} \, $ (considering 
the planes $\, \mathbb{R}^{2} \subset \mathbb{R}^{3} \, $ oriented) 
and to depict the Zones $\, \Omega_{\alpha} \, $ 
on the ``angular diagram'', indicating the direction of the 
normal to $\, \mathbb{R}^{2} \, $ (on the sphere $\mathbb{S}^{2}$).

 Let us fix a function $\, F (z^{1}, z^{2}, z^{3}) \, $ and assume 
that the plane $\, \mathbb{R}^{2} \subset \mathbb{R}^{3} \, $ does 
not contain integer vectors from $\, \mathbb{R}^{3} \, $. 
Following \cite{dynn3}, the general situation can then be described 
as follows.

\vspace{1mm}

\noindent
1) For any fixed direction $\, \xi \in G_{3,2} \, $ open level 
lines (\ref{LevelLinesRest}) arise in a connected energy interval 
$\, c \in [ c_{1} (\xi) , c_{2} (\xi) ] \, $ which can be contracted 
to a single point $\, c_{1} (\xi) = c_{2} (\xi) = c_{0} (\xi) \, $.

\vspace{1mm}

\noindent
2) Whenever $\, c_{2} (\xi) > c_{1} (\xi) \, $, the open level 
lines (\ref{LevelLinesRest}) are topologically regular and have 
the same mean direction in all planes of direction $\xi$ and at all 
levels $c$ containing open level lines. 

\vspace{1mm}

\noindent
3) For directions $\xi$ with the above condition, the functions 
$\, c_{1} (\xi) \, $ and $\, c_{2} (\xi) \, $ are continuous 
functions on the sphere $\, \mathbb{S}^{2} \, $. Connected 
regions of the set $\, c_{2} (\xi) > c_{1} (\xi) \, $ determine 
the Stability Zones $\, \Omega_{\alpha} \, $ with the corresponding 
topological numbers 
$\, (m^{1}_{\alpha}, m^{2}_{\alpha}, m^{3}_{\alpha}) \, $. 
The condition $\, c_{1} (\xi) = c_{2} (\xi) \, $ holds on the 
boundaries of the Zones $\, \Omega_{\alpha} \, $, as well as on the 
set $\, {\cal M}_{F} \, $ defined above.

\vspace{1mm}

\begin{figure}[t]
\begin{center}
\includegraphics[width=\linewidth]{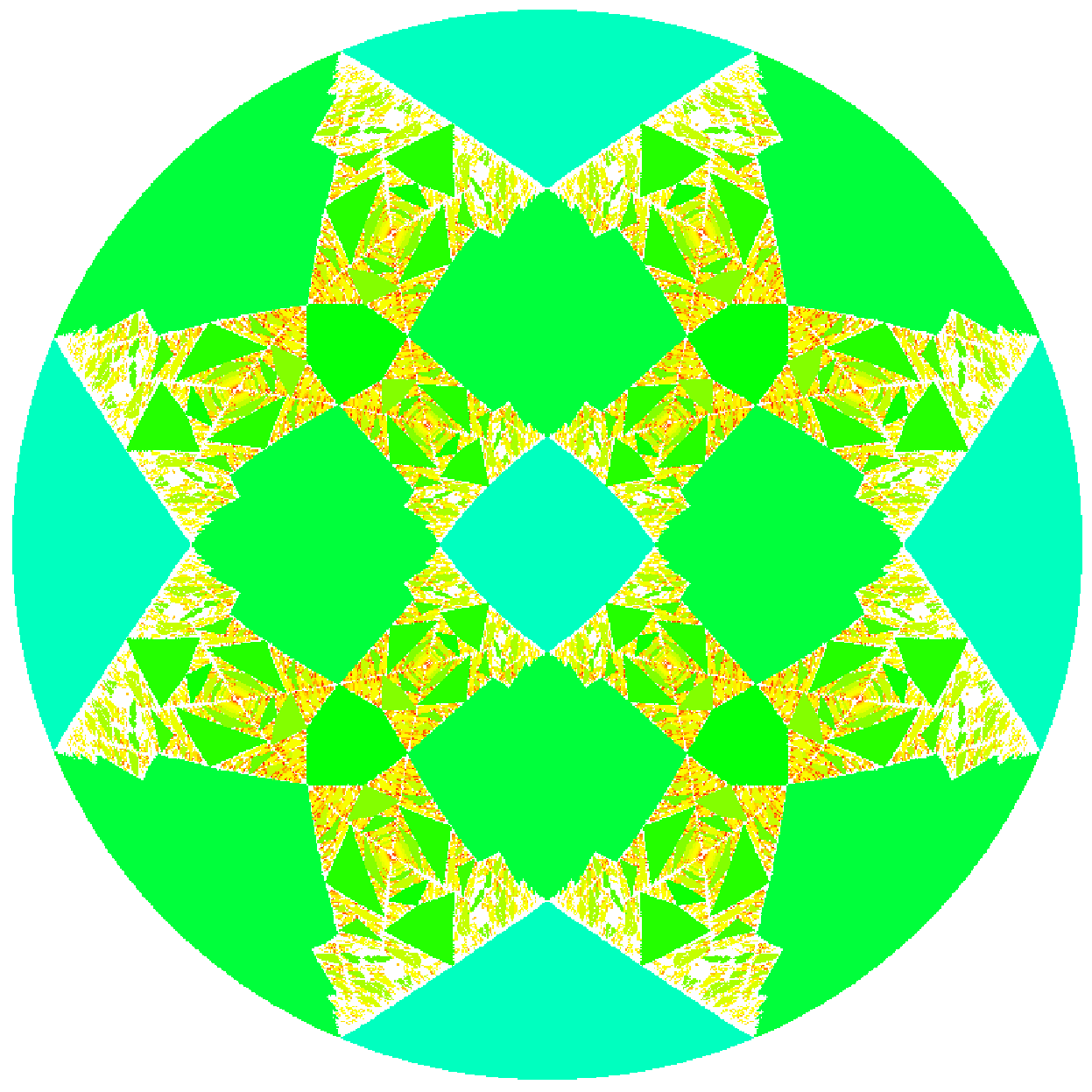}
\end{center}
\caption{The Stability Zones for the function
$F = \cos z^{1} \cos z^{2} + \cos z^{2} \cos z^{3} + 
\cos z^{3} \cos z^{1}$ (\cite{DeLeoObzor}).}
\label{SumPrCos}
\end{figure}

 In the general (non-trivial) case, the number of Zones 
$\, \Omega_{\alpha} \, $ is infinite, and the set 
$\, {\cal M}_{F} \, $ has a complex (fractal) structure 
(see, e.g., Fig. \ref{SumPrCos}). According to the conjecture 
of S.P. Novikov (\cite{DynSyst}), the measure of the 
set $\, {\cal M}_{F} \, $ on $\, \mathbb{S}^{2} \, $ is 
equal to zero, and its Hausdorff dimension is strictly less 
than 2. The first part of this conjecture has been proven 
recently for the most important case in physics 
$\, F ({\bf z}) = F (-{\bf z}) \, $ (I.A. Dynnikov, 
P. Hubert, P. Mercat, A.S. Skripchenko, in preparation). 
The second part of Novikov's conjecture is confirmed by 
serious numerical studies, but has not yet been strictly proven.

\begin{figure}[t]
\begin{center}
\includegraphics[width=\linewidth]{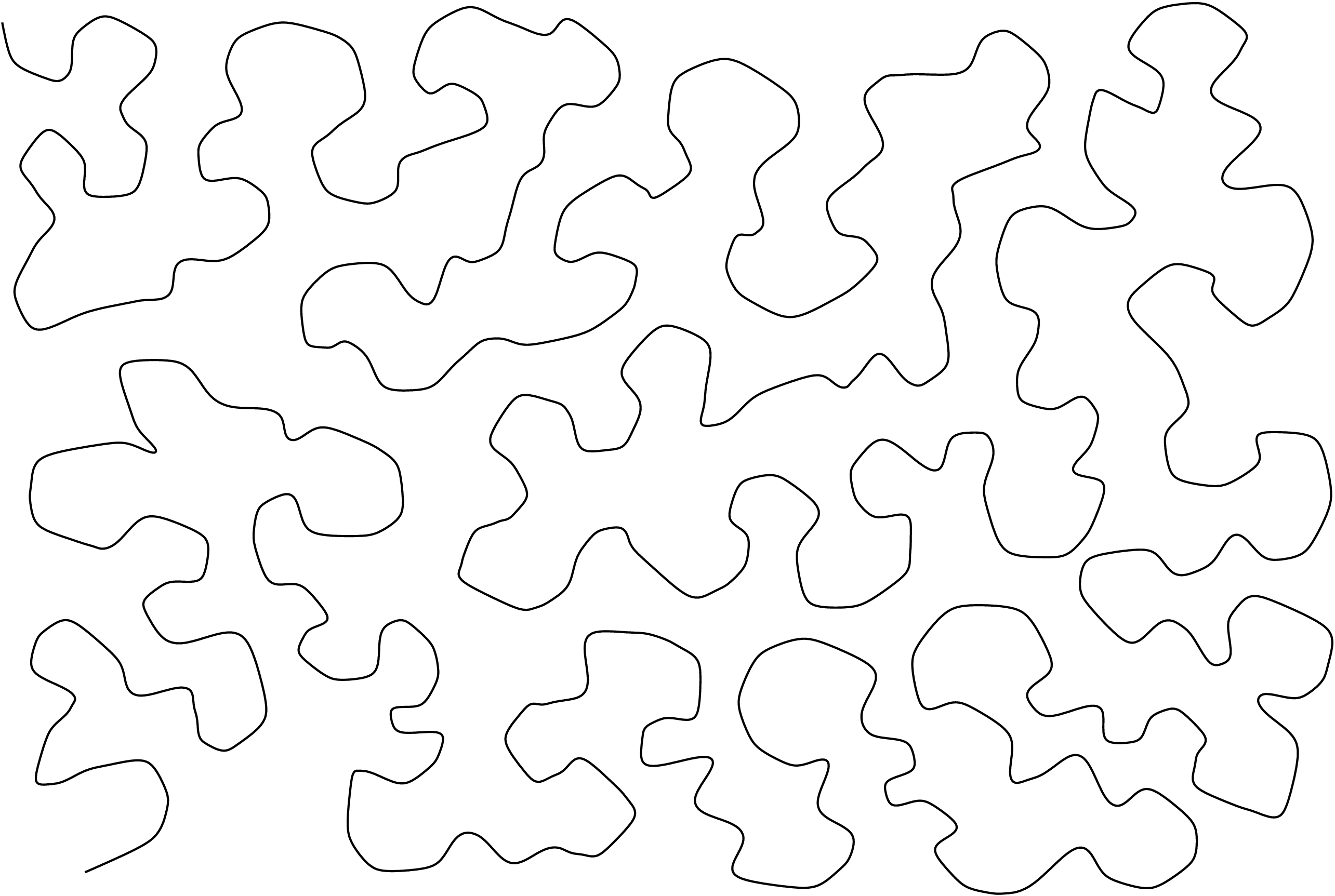}
\end{center}
\caption{Open level line for a purely irrational direction 
$\, \xi \in {\cal M}_{F} \, $ (schematically)}
\label{ChaoticLevelLine}
\end{figure}

 Open level lines, arising at generic directions 
$\, \xi \in {\cal M}_{F} \, $, have truly ``chaotic'' behavior, 
``sweeping'' the entire plane $\, \mathbb{R}^{2} \, $ 
(Fig. \ref{ChaoticLevelLine}). As we have already noted, 
such level lines (of the Dynnikov type) arise only at a 
single level $\, c = c_{0} (\xi) \, $.

 The properties of chaotic level lines are in many ways analogous 
to the properties of level lines of random potentials on the plane 
(see, e.g., \cite{Stauffer, Essam, Riedel, Trugman}). We can say, 
therefore, that the potentials $\, f (x, y) = F |_{\mathbb{R}^{2}} \, $ 
arising for generic $\, \xi \in {\cal M}_{F} \, $ can be considered 
as a model of random potentials on the plane. In contrast, the behavior 
of level lines for $\, \xi \in {\cal U}_{F} \, $ is more consistent 
with the behavior of level lines of periodic potentials.

 In the case of 3 quasiperiods, chaotic level lines arise 
simultaneously in all planes 
$\, \mathbb{R}^{2} \subset \mathbb{R}^{3} \, $ of a direction 
$\, \xi \in {\cal M}_{F} \, $ and, moreover, have the 
same ``scaling'' behavior. In fact, they have many very interesting 
properties, which continue to be studied at the present time (see
\cite{Zorich1996,ZorichAMS1997,Zorich1997,zorich3,DeLeo1,DeLeo2,
DeLeo3,ZorichLesHouches,DeLeoDynnikov1,dynn4,DeLeoDynnikov2,
Skripchenko1,Skripchenko2,DynnSkrip1,DynnSkrip2,
AvilaHubSkrip1,AvilaHubSkrip2,TrMian,DynHubSkrip}).

\vspace{1mm}

 The case of 4 quasiperiods is noticeably more complicated than 
the case of 3 quasiperiods. According to the results 
of \cite{NovKvazFunc,DynNov}, the set of Zones $\, \Omega_{\alpha} \, $ 
is everywhere dense in the space of embedding directions 
$\, \mathbb{R}^{2} \subset \mathbb{R}^{4} \, $.
However, at present nothing is known about the measure of the 
set $\, {\cal U}_{F} \, $ in $\, G_{4,2} \, $ (or about the measure 
of its complement). The set $\, {\cal M}_{F} \, $ apparently 
has a much richer structure here, and the corresponding level 
lines of the functions $\, f (x, y) \, $ have an even more complex 
geometry. Note that chaotic level lines do not necessarily arise 
here at once in all planes $\, \mathbb{R}^{2} \, $ of a given direction, 
which makes this case different from the case $\, N = 3 \, $ 
(as well as from topologically regular level lines). In addition, 
it is unknown whether they arise here only at a single level 
$\, c = c_{0} (\xi) \, $, or whether they can also arise in 
a finite interval $\, c \in [c_{1} (\xi) , c_{2} (\xi)] \, $. 
At the same time, it is obvious that the case $\, N = 4 \, $ 
contains more possibilities for modeling random potentials 
compared to $\, N = 3 \, $.

\vspace{1mm}

 In this paper we study a special class of potentials related 
to the case $\, N = 4\, $. Namely, we consider potentials arising 
from a superposition of two periodic potentials on the plane. 
A general discussion of the Novikov problem for such potentials 
was given in \cite{AnnPhys}, where a picture of the emergence of 
topologically regular and chaotic level lines in this situation 
was presented. In the present paper, however, we consider more 
special potentials, namely, potentials possessing rotational 
symmetry of the same type. Superpositions of potentials of this 
type play a very important role in the theory of ``two-layer'' 
systems and are widely considered in the modern physics literature 
(we cite here only a few of the huge number of papers on this topic 
\cite{Shallcross1,Shallcross2,GeimGrigorieva,TitovKatsnelson,
RShRN,CMFCLK,DindorkarKuradeShaikh,PaulCrowleyFu,BernevigEfetov}).

 From the general point of view, such potentials are not generic,
in particular, their superpositions cannot have topologically regular 
open level lines. Thus, from the point of view of the Novikov problem, 
a superposition of such potentials always (except for rotations 
by ``magic'' angles) corresponds to an embedding direction
$\, \boldmath{\xi} \in {\cal M}_{F} \, $. 
We will show also that open level lines can arise 
here only at a single level $\, V (x, y) = c_{0} \, $, which unites 
such potentials with random potentials, as well as with ``chaotic'' 
potentials with 3 quasiperiods.

 By superposition of potentials in $\, \mathbb{R}^{2} \, $ we shall 
mean here the formation of a complex potential $\, V (x, y) \, $ when 
two periodic potentials $\, V_{1} (x, y) \, $ and $\, V_{2} (x, y) \, $ 
are superimposed in a plane. In this case we shall allow a fairly 
general dependence of the potential $\, V \, $ on $\, V_{1} \, $ 
and $\, V_{2} \, $. In various physical problems the following 
three typical cases naturally arise.
$${\rm (1)} \quad V (x, y) \,\,\, = \,\,\, V_{1} (x, y) 
\,\, + \,\, V_{2} (x, y) $$
(the simplest case of linear superposition)
$${\rm (2)} \quad V (x, y) \,\,\, = \,\,\, Q \, \Big( 
V_{1} (x, y), V_{2} (x, y) \Big) $$
for some smooth function $\, Q \left( V_{1}, V_{2} \right) \, $ 
(the case of nonlinear superposition caused by a local influence 
of potentials on each other);
$${\rm (3)} \quad V (x, y) \,\,\, = \,\,\, Q \left[ 
V_{1}, V_{2} \right] (x, y) $$
for some (smooth) functional $\, Q \left[ V_{1}, V_{2} \right] \, $ 
(the case of possible deformation of lattices due to their 
interaction at finite distances).

\vspace{1mm}

 In all three cases, the potential $\, V (x, y) \, $ is 
a quasiperiodic function on the plane with 4 quasiperiodes, 
i.e. it is given by the restriction of a 4-periodic function 
$\, F (z^{1}, z^{2}, z^{3}, z^{4}) \, $ onto $\, \mathbb{R}^{2} \, $ 
under some embedding $\, \mathbb{R}^{2} \subset \mathbb{R}^{4} \, $. 
Here we will consider embeddings of the form
\begin{equation}
\label{Emb}
(x, y) \quad \rightarrow \quad 
\left( x, y, \, A (x,y) \right) \,\,\, , 
\end{equation}
where
\begin{equation}
\label{Axy}
A (x, y) \,\,\, = \,\,\, \left(
\begin{array}{cc}
\cos \alpha  &  \sin \alpha  \\
- \sin \alpha  &  \cos \alpha 
\end{array}  \right)  
\left( 
\begin{array}{c}
x  \\
y
\end{array} \right)
\,\, - \,\, \left(
\begin{array}{c}
a^{1}  \\
a^{2} 
\end{array}  \right) 
\end{equation}
represents the rotation and translation in $\, \mathbb{R}^{2} \, $.

 The potential $\, V_{2} (x, y) \, $ will then be given by the 
formula $\, V_{2} (x, y) \, = \, U \left( A (x,y) \right) \, $ 
for some fixed potential $\, U (x, y) \, $.

 As is easy to see, in cases (1) and (2), the function 
$\, F (z^{1}, z^{2}, z^{3}, z^{4}) \, $ is defined by the relations
\begin{equation}
\label{F1}
F (z^{1}, z^{2}, z^{3}, z^{4}) \,\,\, = \,\,\, 
V_{1} (z^{1}, z^{2}) \,\, + \,\, U (z^{3}, z^{4}) 
\end{equation}
and
\begin{equation}
\label{F2}
F (z^{1}, z^{2}, z^{3}, z^{4}) \,\,\, = \,\,\,  
Q \, \Big( V_{1} (z^{1}, z^{2}), \, U (z^{3}, z^{4}) \Big) 
\end{equation}
respectively.

 In case (3), the function $\, F \, $ is defined more complexly, 
based on the definition of $\, V (x, y) \, $ in each of the 
planes $\, \mathbb{R}^{2} \subset \mathbb{R}^{4} \, $. In this case, 
in fact, the function $\, F \, $ also acquires dependence on the 
angle $\, \alpha \, $: $\,\, F = F ({\bf z}, \alpha) \, $.

 In all cases, including case (3), the function 
$\, F = F ({\bf z}, \alpha) \, $ is a periodic function of
$\, {\bf z} \, $ with fixed periods determined by the periods 
of the potentials $\, V_{1} (x, y) \, $ and $\, U (x, y) \, $ 
(see e.g. \cite{AnnPhys}).
 
 In section 2 we consider in detail the case of rotational symmetry 
of order four, and in section 3 the cases of symmetry of order three 
and six. In section 4 we consider the case when $\, V_{1} \, $ and 
$\, U \, $ have the same rotational symmetry and incommensurable 
periods $\, T \, $ and $\, T^{\prime} \, $ in $\, \mathbb{R}^{2} \, $. 
(Like the case of equal periods, this case is also quite important 
in modern physics of two-dimensional systems.)

\section{The case of the fourth order symmetry.
Magic angles and generic rotations.}
\setcounter{equation}{0}
 
  Let us consider the situation when both potentials 
$\, V_{1} (x, y) \, $ and $\, U (x, y) \, $ have rotation symmetry 
by $90^{\circ}$ and identical periods of the length $T$.
As we have already said, the potential $\, V_{2} (x, y) \, $ 
is given by the potential $\, U (x, y) \, $ whose lattice of periods 
is rotated by a certain angle $\, \alpha \, $ relative to the 
lattice of periods of $\, V_{1} (x, y) \, $. Taking into account 
the rotational symmetry of the potential $\, U (x, y) \, $, 
it is natural to consider only the angles 
$\, \alpha \, \in \, (0, \pi / 2) \, $.
 
\begin{figure}[t]
\begin{center}
\includegraphics[width=\linewidth]{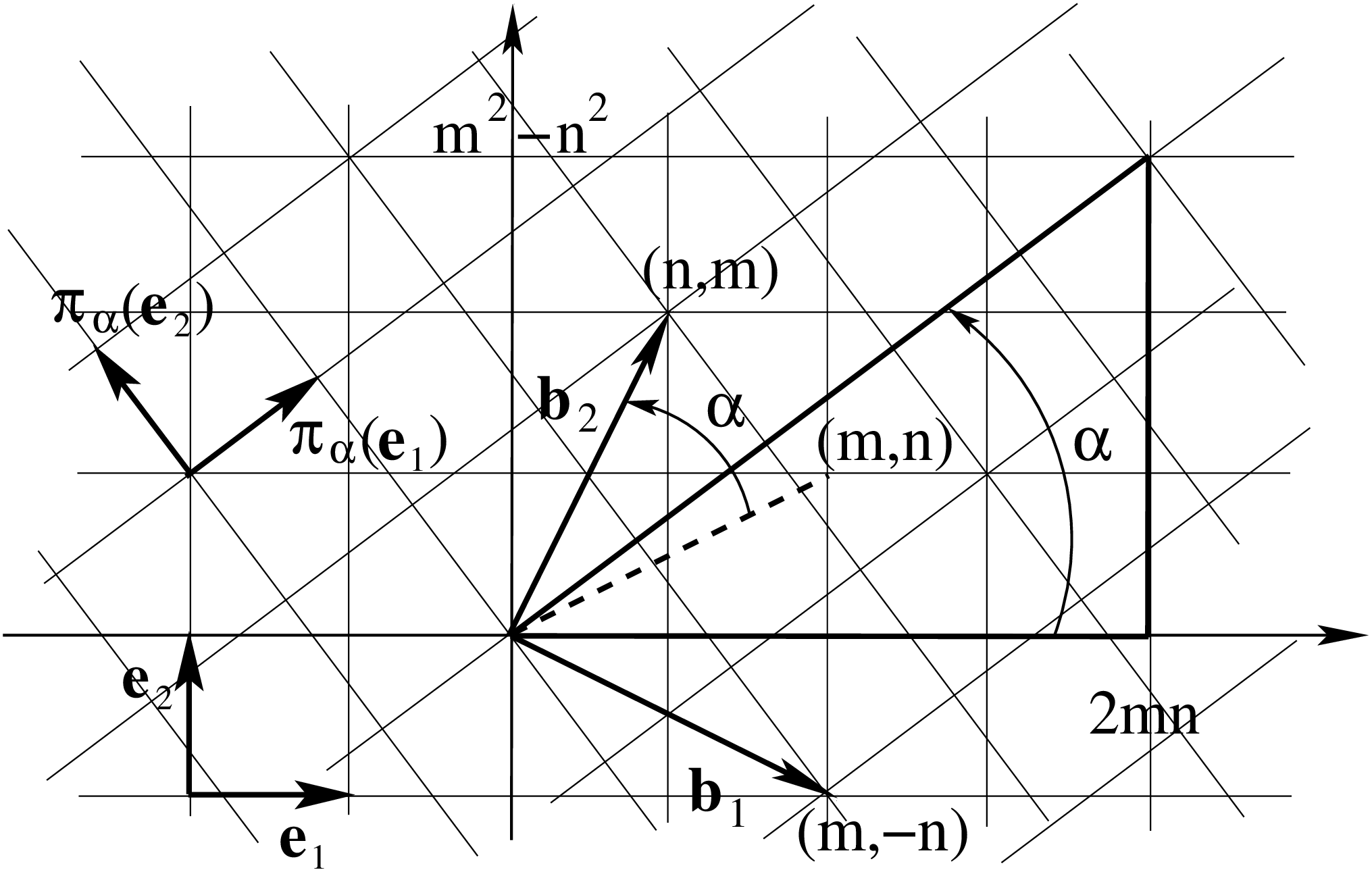}
\end{center}
\caption{Superposition corresponding to 
the ``magic'' angle $\, \alpha_{m,n} \, $ for the Pythagorean 
triple $\, (2 m n, m^{2} - n^{2}, m^{2} + n^{2}) \, $, and the 
periods of the resulting potential in the plane (schematically)}
\label{Periods1}
\end{figure}

\begin{figure}[t]
\begin{center}
\includegraphics[width=\linewidth]{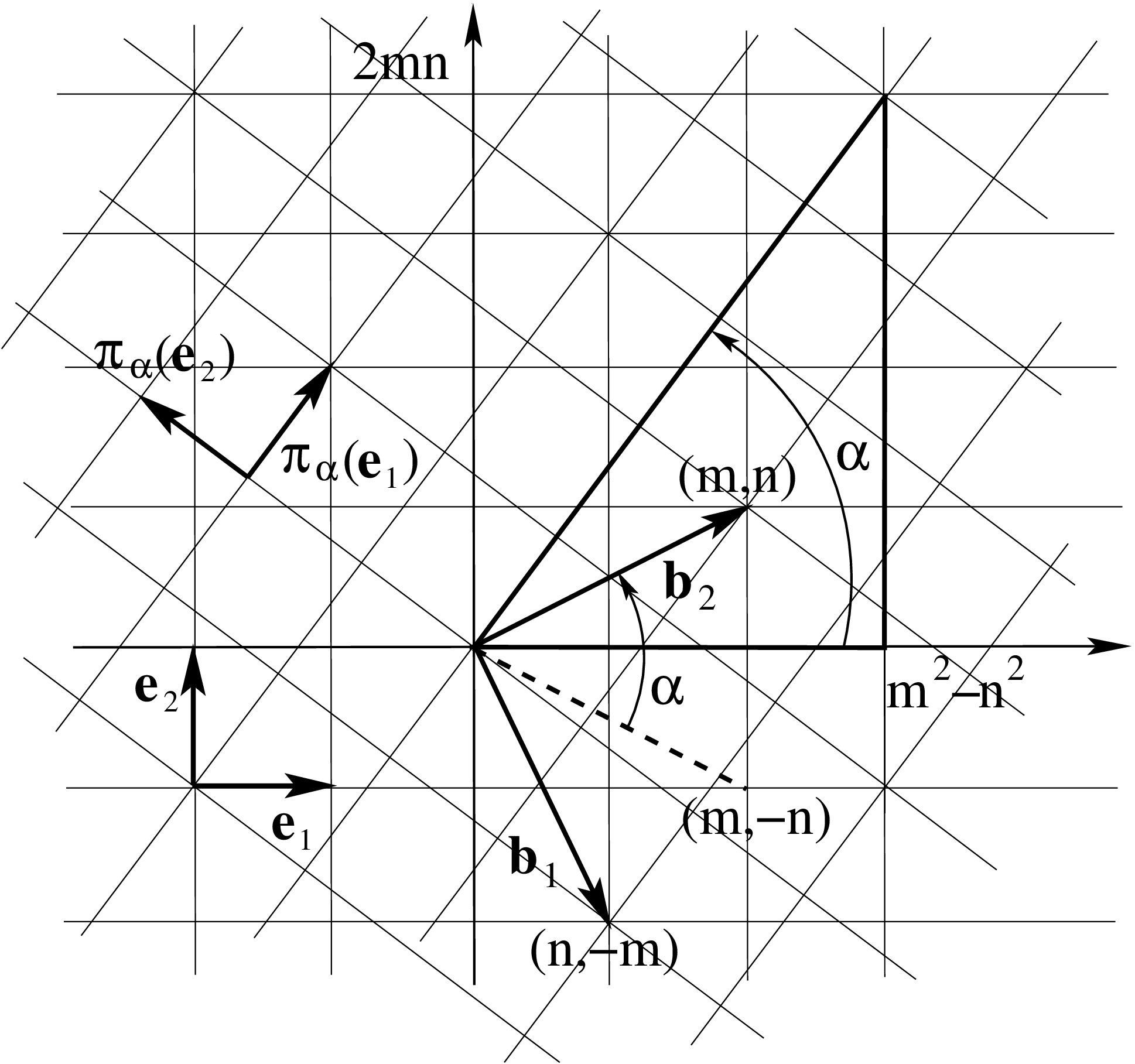}
\end{center}
\caption{Superposition corresponding to the ``magic'' angle 
$\, \bar{\alpha}_{m,n} \, $ for the Pythagorean triple 
$\, (m^{2} - n^{2}, 2 m n, m^{2} + n^{2}) \, $, and the 
periods of the resulting potential in the plane (schematically)}
\label{Periods2}
\end{figure}

 As is well known, the ``magic'' angles $\alpha$ are given in 
this case by (irreducible) Pythagorean triples
$$\Big( 2 m n \, , \,\, m^{2} - n^{2} , \,\, m^{2} + n^{2} \Big)  
\,\,\, , \quad m, n \in \mathbb{Z} $$
or 
$$\Big(m^{2} - n^{2}  \, , \,\, 2 m n , \,\, m^{2} + n^{2} \Big)  
\,\,\, , \quad m, n \in \mathbb{Z} $$
(see Fig. \ref{Periods1}, \ref{Periods2}). In this case,
$$\tan \alpha_{m,n} \,\,\, = \,\,\, {m^{2} - n^{2} \over 2 m n} 
\,\,\, = \,\,\, {1 \over 2} \left( {m \over n} - {n \over m} \right) $$
or
$$\tan \bar{\alpha}_{m,n} \,\,\, = \,\,\, {2 m n \over m^{2} - n^{2}} $$ 
 
 As is also well known, Pythagorean triples are in one-to-one 
correspondence with irreducible pairs $\, (m, n)\, $ (up to a sign) 
provided that the numbers $\, m\, $ and $\, n\, $ have different parities.

 In the first case, the periods of the superposition of 
$\, V_{1} (x, y) \, $ and $\, V_{2} (x, y) \, $ are given by 
the vectors $\, {\bf b}_{1} = (mT, - nT) \, $ and 
$\, {\bf b}_{2} = (nT, mT) \, $ (Fig. \ref{Periods1}).
Indeed, the vector $\, (n, m) \, $ is obtained by rotating
the vector $\, (m, n) \, $
$$ \left(
\begin{array}{cc}
{2 m n \over  m^{2} + n^{2}} &  
- {m^{2} - n^{2} \over m^{2} + n^{2}}   \\
{m^{2} - n^{2} \over m^{2} + n^{2}}  &  
{2 m n \over  m^{2} + n^{2}}
\end{array}  \right)    
\left(
\begin{array}{c}
m  \\ 
n
\end{array}  \right)
\,\,\, = \,\,\, 
\left(
\begin{array}{c}
n  \\ 
m
\end{array}  \right) $$
and, thus, is also integer in the rotated lattice 
(it has coordinates $[m, n]$ there). (Similarly for the vector
$\, (m, - n)$).

 In other words, the ``magic'' angles here are determined by 
rotations from the vector $\, (m, n) \, $ to the vector 
$\, (n, m) \, $ for all possible integer (irreducible) pairs 
$\, (m, n) \, $. As can be seen, the periods of superposition 
of potentials are reflections of the vector $\, (m, n) \, $ 
relative to the axes $\, (1, 0) \, $ and $\, (1, 1) \, $.

 In the second case, the rotation matrix has the form
$$ \left(
\begin{array}{cc}
{m^{2} - n^{2} \over  m^{2} + n^{2}} &  
- {2 m n \over m^{2} + n^{2}}   \\
{2 m n \over m^{2} + n^{2}}  &  
{m^{2} - n^{2} \over  m^{2} + n^{2}}
\end{array}  \right) $$

 The periods of superposition of $\, V_{1} (x, y) \, $ and
$\, V_{2} (x, y) \, $ are the vectors $\, {\bf b}_{1} = (nT, - mT) \, $
and $\, {\bf b}_{2} = (mT, nT) \, $ (Fig. \ref{Periods2}).
The ``magic'' angles $\, \bar{\alpha}_{m,n} \, $ are determined
here by the rotation from the vector $\, (m, - n) \, $ to the vector
$\, (m, n) \, $.

 In both cases, the periods $\, {\bf b}_{1} \, $ and 
$\, {\bf b}_{2} \, $ define a (minimal) fundamental region 
for the potential $\, V (x, y) \, $ provided that the numbers 
$\, m \, $ and $\, n \, $ have different parities.

 If $\, m \, $ and $\, n \, $ are both odd, the corresponding 
Pythagorean triple is reducible. The corresponding irreducible 
Pythagorean triple is determined by the numbers 
$\, m^{\prime} \, $ and $\, n^{\prime} \, $ of different parity, 
such that
\begin{multline*}
\Big( 2 m n \, , \,\, m^{2} - n^{2} , \,\, m^{2} + n^{2} \Big)  
\,\,\, =  \\
= \,\,\, 2 \, \Big((m^{\prime})^{2} - (n^{\prime})^{2}  \, , 
\,\, 2 m^{\prime} n^{\prime} , \,\, 
(m^{\prime})^{2} + (n^{\prime})^{2} \Big) 
\end{multline*}
(or vice versa)
\begin{multline*}
\Big( m^{2} - n^{2}  \, , \,\, 2 m n , \,\, 
m^{2} + n^{2} \Big) \,\,\, =  \\
= \,\,\,  2 \, \Big( 2 m^{\prime} n^{\prime} \, , \,\, 
(m^{\prime})^{2} - (n^{\prime})^{2} , \,\, 
(m^{\prime})^{2} + (n^{\prime})^{2} \Big)  
\end{multline*}
where
$$m \,\,\, = \,\,\, m^{\prime} \, + \, n^{\prime} \,\,\, , 
\quad n \,\,\, = \,\,\, m^{\prime} \, - \, n^{\prime} $$

 It is also easy to see that in this case we have a correspondence
$$\alpha_{m,n} \,\,\, = \,\,\, 
\bar{\alpha}_{m^{\prime}, n^{\prime}} \quad \quad \text{or} 
\quad \quad \alpha_{m^{\prime}, n^{\prime}} \,\,\, = \,\,\, 
\bar{\alpha}_{m, n} $$

 It will be convenient for us to use here only the angles
$\, \alpha_{m,n} \, $, where (coprime) numbers $\, m \, $ 
and $\, n \, $ can be of the same or different parity. 
It can be seen that for the complete set of ``magic'' angles 
$\, \alpha \in (0, \pi/2) \, $ we can restrict ourselves
to the values $\, m > n > 0 \, $. The limit 
$\, \alpha \rightarrow 0 \, $ corresponds to the limit 
$$m / n \,\,\, \rightarrow \,\,\, 1 \,\,\, , $$ 
and the limit $\, \alpha \rightarrow \pi / 2 \,\,\, $ 
corresponds to the limit 
$$m / n \,\,\, \rightarrow \,\,\, \infty $$

 Let us also define for each irreducible pair $\, (m, n) \, $ 
the numbers $\, (m_{0}, n_{0}) \, $:
\begin{equation}
\label{m0}
(m_{0}, n_{0}) \,\,\, = \,\,\, (m, n)
\end{equation}
if $\, m\, $ and $\, n\, $ have different parity, and
\begin{equation}
\label{n0}
(m_{0}, n_{0}) \,\,\, = \,\,\, \left( {m + n \over 2} , \,
{m - n \over 2} \right) 
\end{equation}
if $\, m \, $ and $\, n \, $ are both odd.

 Let $\, \Pi_{\bf b} \, $ be the parallelogram generated by 
the vectors $\, {\bf b}_{1} \, $ and $\, {\bf b}_{2} \, $ 
corresponding to the angle $\, \alpha_{m,n} \, $. As we have 
already said, $\, \Pi_{\bf b} \, $ determines the minimal 
fundamental domain for the potential $\, V (x, y) \, $ if 
$\, m \, $ and $\, n \, $ have different parities. In the 
case when $\, m \, $ and $\, n \, $ are both odd, 
$\, \Pi_{\bf b} \, $ corresponds to a ``doubled'' 
fundamental domain of the potential $\, V (x, y) \, $, and 
the minimal fundamental domain is determined by the vectors
$${\bf b}^{\prime}_{1} \,\, = \,\, 
{{\bf b}_{1} - {\bf b}_{2} \over 2} \,\,\, , \quad 
{\bf b}^{\prime}_{2} \,\, = \,\, 
{{\bf b}_{1} + {\bf b}_{2} \over 2} $$

 (Indeed, considering a rotation in $\, \mathbb{R}^{2} \, $ 
as a multiplication by a complex number $\, w_{\alpha} \, $ 
($| w_{\alpha} | = 1$), one can see that for ``magic'' angles 
the numbers $\, {\rm Re} \, w_{\alpha} \, $ and 
$\, {\rm Im} \, w_{\alpha} \, $ are represented by irreducible 
fractions with denominator $\, m_{0}^{2} + n_{0}^{2} \, $.
The existence of two complex numbers $\, u \, $ and $\, v \, $ 
with integer components such that
$\, u \, = \, w_{\alpha_{m,n}} v \, $ and
$\, |u| \, = \, |v| \, < \, \sqrt{m_{0}^{2} + n_{0}^{2}} \, $, 
would mean that
$${\rm Re} \,\, w_{\alpha_{m,n}}  = \,\, 
{\rm Re} \, {\bar u} v / |u|^{2} \quad \text{and} \quad 
{\rm Im} \,\, w_{\alpha_{m,n}}  = \,\, 
{\rm Im} \,\, {\bar u} v / |u|^{2} $$
can be represented as fractions with smaller denominators, 
which is impossible.)

 Further, it will be convenient for us to use the notation 
$\, \pi_{\alpha} (\boldmath{\xi}) \, $ for the rotation of an 
arbitrary vector $\, \boldmath{\xi} \, $ by an angle $\, \alpha \, $.
 
 As we have already said, we assume here that the potential 
$\, V_{2} ({\bf r}) = V_{2} (x,y) \, $ has the form
\begin{equation}
\label{V2U}
V_{2} ({\bf r}) \,\,\, = \,\,\, U \Big( 
\pi_{- \alpha} ({\bf r}) \, - \, {\bf a} \Big) \,\,\, , 
\quad {\bf a} = (a^{1}, a^{2}) \,\,\, , 
\end{equation}
where the potential $\, U (x,y) \, $ has the same lattice of 
periods as the potential $\, V_{1} (x,y) \, $. We will also 
assume that the point $\, (x, y) = (0, 0) \, $ is the center 
of rotational symmetry for both the potentials $\, V_{1} (x,y) \, $ 
and $\, U (x,y) \, $. It is easy to see then that all the points  
$$(x, y) \,\,\, = \,\,\, \left( {k + l \over 2} \, T \, , \,\, 
{k - l \over 2} \, T
\right) \,\,\, , \quad k, l \, \in \, \mathbb{Z} $$
also have the same property. 

 The potential $ \, V (x,y) \, $ is a function of the 
parameters $\, \alpha $ and $\, {\bf a} $
$$V (x, y) \,\,\, = \,\,\, V(x, y, \, \alpha , \, a^{1}, a^{2}) $$

 It is easy to see that
\begin{equation}
\label{Tozhd1}
V \big( 
{\bf r}, \, \alpha , \,\, {\bf a} + \pi_{\alpha} ({\bf e}_{1,2}) \big)
\,\,\, \equiv \,\,\, V ({\bf r}, \, \alpha, \, {\bf a} ) 
\end{equation}

 Besides that,
\begin{equation}
\label{Tozhd2}
V \big( 
{\bf r}, \, \alpha , \,\, {\bf a} + {\bf e}_{1,2} \big)
\,\,\, \equiv \,\,\, V \big( {\bf r} - {\bf e}_{1,2}, \,\, 
\alpha, \, {\bf a} \big) \,\,\, , 
\end{equation}
that is, in a certain sense, the potentials 
$\, V \big( {\bf r}, \, \alpha , \, {\bf a} + {\bf e}_{1,2} \big) $ 
are equivalent to the potential 
$\, V ({\bf r}, \, \alpha, \, {\bf a} ) \, $, namely, they are 
obtained from it by a shift in the plane $\, (x, y) $.
 
 For each potential $\, V ({\bf r}, \, \alpha, \, {\bf a} ) \, $ 
we can determine the corresponding class of equivalent potentials
\begin{multline}
\label{EquivPot}
V_{klpq} \big( {\bf r}, \, \alpha, \, {\bf a} \big) 
\,\,\, \cong   \\
\cong \,\,\, V \big( {\bf r}, \, \alpha, \,\, 
{\bf a} + k \pi_{\alpha} ({\bf e}_{1}) + 
l \pi_{\alpha} ({\bf e}_{2}) + p {\bf e}_{1} + q {\bf e}_{2} \big) 
\end{multline}
($k, l, p, q \, \in \, \mathbb{Z}$).

 It is not difficult to see that for generic angles $\, \alpha \, $ 
the set of the vectors
$${\bf a}_{0} \,\, + \,\, k \, \pi_{\alpha} ({\bf e}_{1}) \, + \, 
l \, \pi_{\alpha} ({\bf e}_{2}) \, + \, p \, {\bf e}_{1} 
\, + \, q \, {\bf e}_{2} $$
is everywhere dense in the space of parameters $\, {\bf a} \, $.

\begin{figure}[t]
\begin{center}
\includegraphics[width=\linewidth]{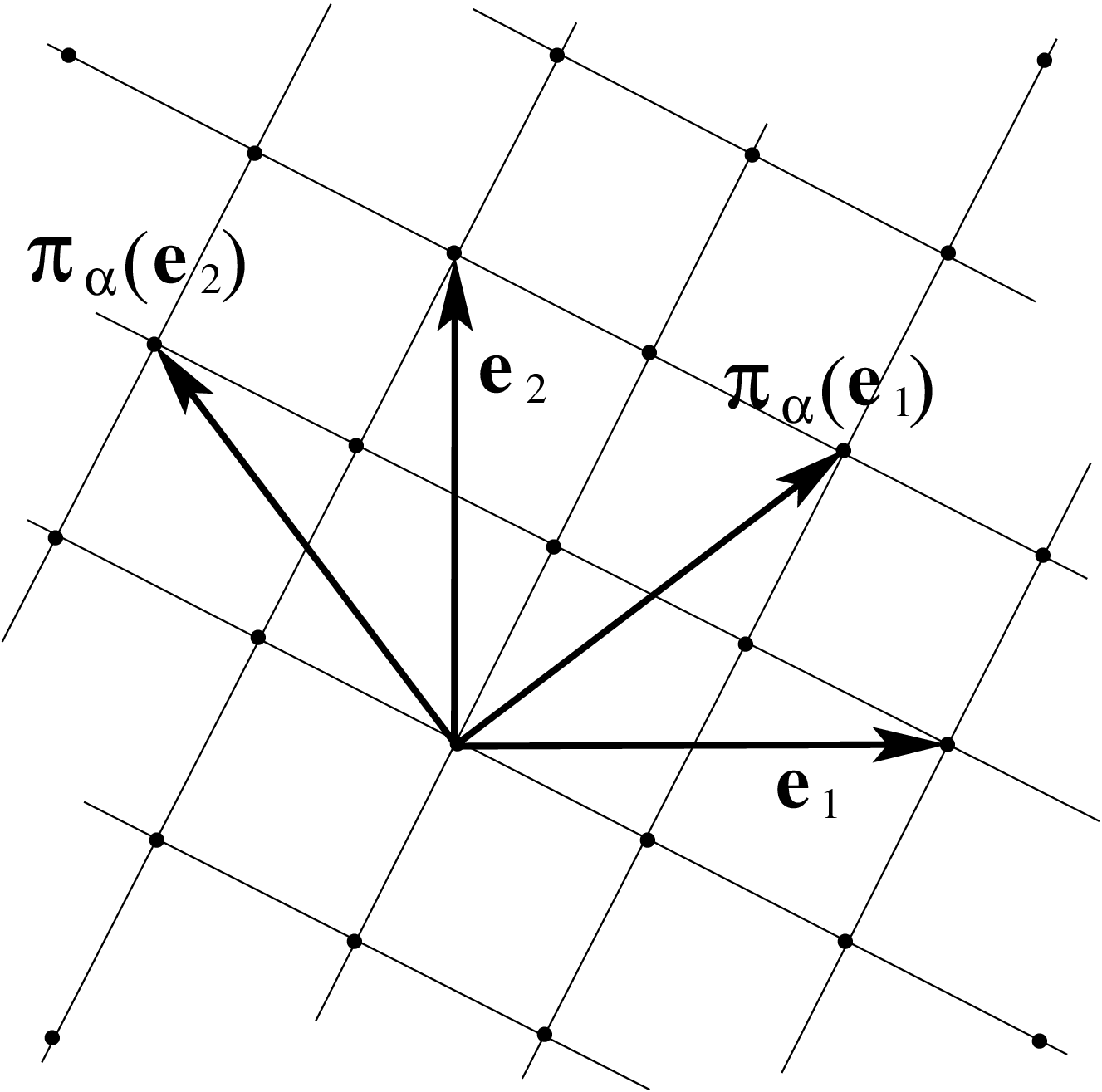}
\end{center}
\caption{Square lattice defined by vectors (\ref{DenseLat})}
\label{SmallLat}
\end{figure}

 For the ``magic'' angle $\, \alpha_{m,n} \, $ vectors
\begin{equation}
\label{DenseLat}
k \, \pi_{\alpha} ({\bf e}_{1}) \, + \, 
l \, \pi_{\alpha} ({\bf e}_{2}) \, + \, p \, {\bf e}_{1} 
\, + \, q \, {\bf e}_{2} 
\end{equation}
form a rotated (by the angle $\, - \arctan (n/m)$) square lattice 
with the step $\, T / \sqrt{m_{0}^{2} + n_{0}^{2}} \, $ 
(Fig. \ref{SmallLat}). It can be seen, therefore, that each 
potential $\, V (x, y, \, \alpha , \, {\bf a} ) \, $ is equivalent 
in this case to some potential
$\, V (x, y, \, \alpha, \, {\bf a}^{\prime} ) \, $, where
$\, | {\bf a}^{\prime} | \, \leq \, T / \sqrt{2(m_{0}^{2} + n_{0}^{2})} \, $.

 It is obvious that equivalent potentials have level lines of the 
same types at each of the levels $\, V (x, y) = c \, $.

\vspace{1mm}

 Here we will consider both regular and singular 
(containing singular points) level lines of the potential 
$\, V (x, y) \, $. We will assume that all singular points 
$\, V (x, y) \, $ are isolated minima or maxima or multiple saddles.

  As follows from \cite{DynMalNovUMN}, for a given $\, \alpha \, $, 
the set of levels $\, c \, $, containing open level lines of 
potentials $\, V (x, y, \, \alpha , \, {\bf a}) \, $ 
(at least for one $\, {\bf a}$), is always a closed segment 
$$ c \in [ c_{1} (\alpha), c_{2} (\alpha) ] \,\,\, , $$ 
which can be contracted to a single point 
$$ c_{1} (\alpha) = c_{2} (\alpha) = c_{0} (\alpha) $$

 If the angle $\, \alpha \, $ is ``magic'' and the potentials 
$\, V (x, y, \, \alpha , \, {\bf a}) \, $ are periodic, then 
each of the potentials $\, V (x, y, \, \alpha , \, {\bf a}) \, $ 
can have its own interval of open level lines
$$\left[ \hat{c}_{1} (\alpha_{m,n}, {\bf a}), \, 
\hat{c}_{2} (\alpha_{m,n}, {\bf a}) \right] \,\,\, \subset \,\,\, 
\left[ c_{1} (\alpha_{m,n}), c_{2} (\alpha_{m,n}) 
\right] $$

 It is easy to see that equivalent potentials 
$\, V (x, y, \, \alpha_{m,n} , \, {\bf a}) \, $ have  the same intervals 
$\, \left[ \hat{c}_{1} (\alpha_{m,n}, {\bf a}), 
\, \hat{c}_{2} (\alpha_{m,n}, {\bf a}) \right] \, $. 
 
\begin{figure}[t]
\begin{center}
\includegraphics[width=\linewidth]{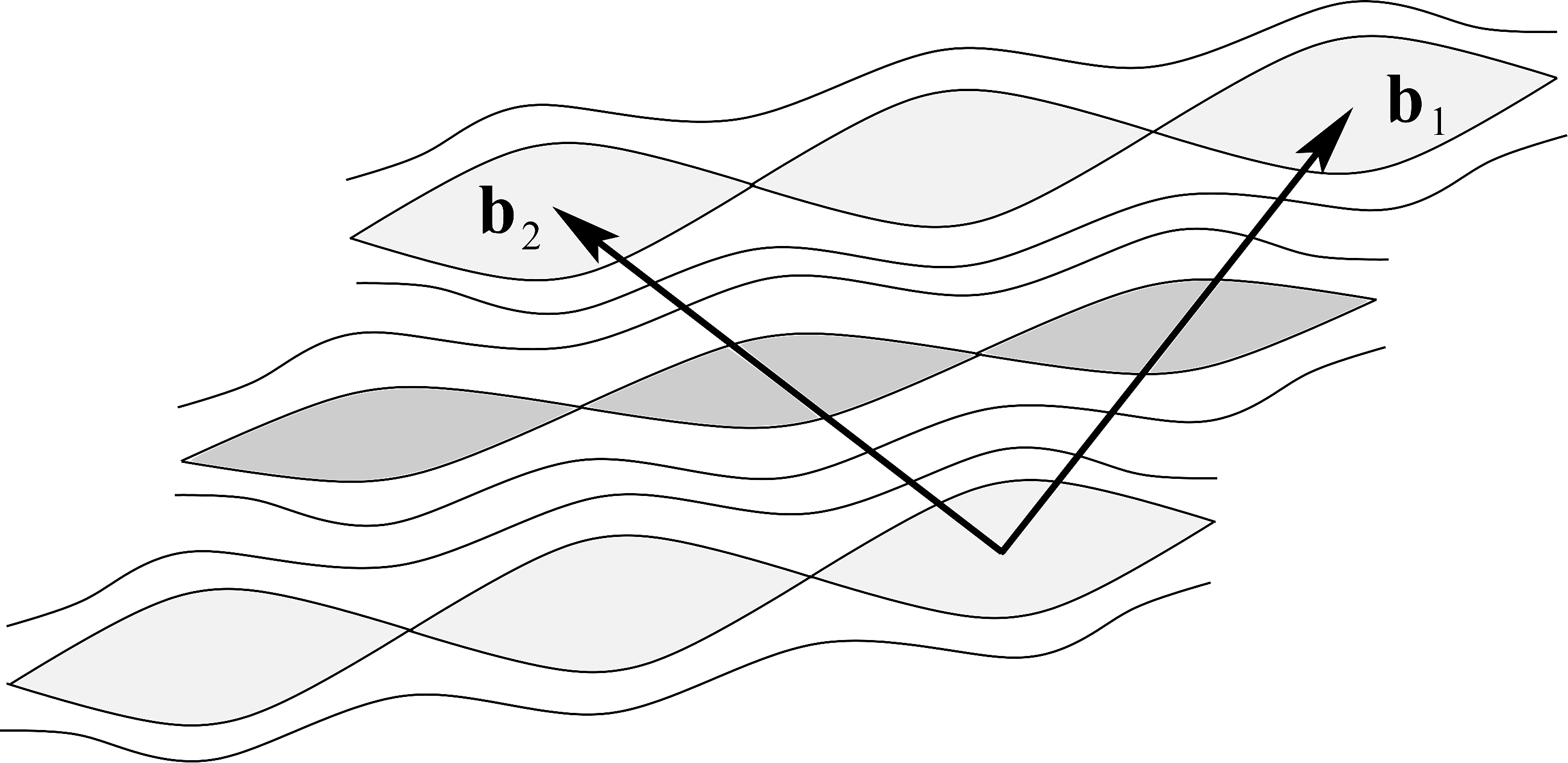}
\end{center}
\caption{Periodic level lines of potentials 
$\, V (x, y, \, \alpha_{m,n} , \, {\bf a}) \, $, 
corresponding to ``magic'' angles $\, \alpha_{m,n} \, $ 
(schematically)}
\label{PerLevLines}
\end{figure}

 Periodic potentials with finite intervals 
$\, \left[ \hat{c}_{1} (\alpha_{m,n}, {\bf a}), \, 
\hat{c}_{2} (\alpha_{m,n}, {\bf a}) \right] \, $ 
($\hat{c}_{2} > \hat{c}_{1}$) have periodic open level lines 
with some integer mean direction in the basis 
$\, \{ {\bf b}_{1}, \, {\bf b}_{2} \} \, $ or 
$\, \{ {\bf b}^{\prime}_{1}, \, {\bf b}^{\prime}_{2} \} \, $
(Fig. \ref{PerLevLines}). 
This direction, however, can be different for different values of 
the parameters $\, (a^{1}, a^{2}) \, $.

\begin{figure}[t]
\begin{center}
\includegraphics[width=\linewidth]{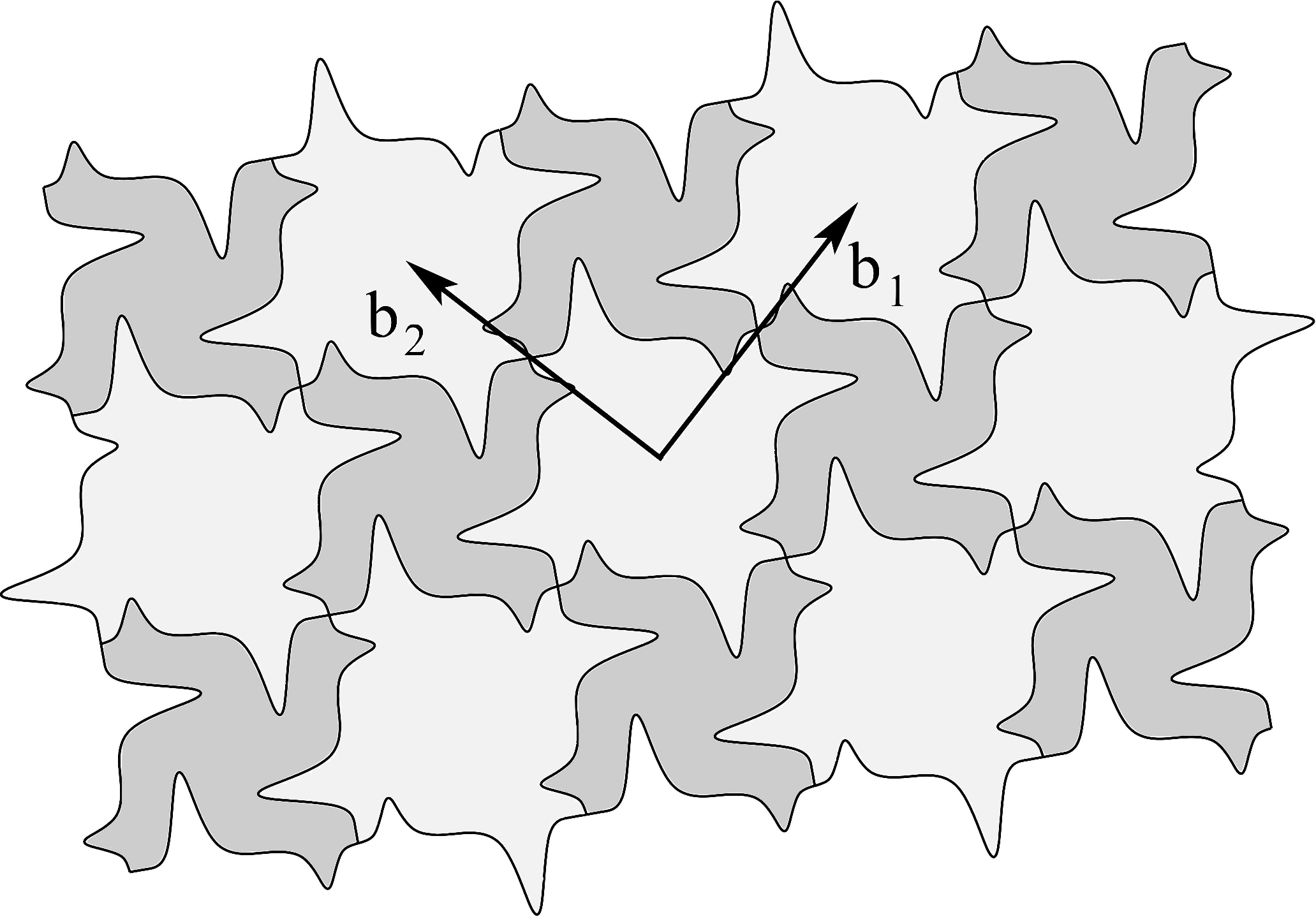}
\end{center}
\caption{A singular periodic ``net'' and regions of higher and 
lower energies for a periodic potential with $90^{\circ}$ rotation 
symmetry.}
\label{SingNet90Gen}
\end{figure}

  The potential $\, V (x, y, \, \alpha_{m,n} , \, 0, 0) \, $ has 
obvious rotational symmetry and cannot have level lines described 
above. For such potentials we have the relation
$$\hat{c}_{1} (\alpha_{m,n} , \, 0, 0) \,\,\, = \,\,\, 
\hat{c}_{2} (\alpha_{m,n} , \, 0, 0) \,\,\, = \,\,\,
\hat{c}_{0} (\alpha_{m,n} , \, 0, 0)  $$
and the open level line has the form of a singular 
periodic ``net'' arising at a single level 
$\, c = \hat{c}_{0} (\alpha_{m,n} , \, 0, 0) \, $.
It can be noted that for potentials 
$\, V (x, y, \, \alpha_{m,n} , \, 0, 0) \, $ with large periods, 
the shape of such a ``net'' can be quite complex 
(see, for example, Fig. \ref{SingNet90Gen}).

 The ``net'' shown in Fig. \ref{SingNet90Gen} is the simplest from 
the topological point of view and corresponds to the ``generic'' case 
(in particular, all singular points on it are non-degenerate and the 
level $\, V (x, y) = c_{0} \, $ contains exactly two nonequivalent 
singular points $\, V (x, y) $). As is easy to see, all ``cells'' 
of such a ``net'' have rotational symmetry. For our purposes, in fact, 
we can restrict ourselves to the generic case (but we will not impose 
this restriction here). More generally, a singular ``net'' may contain 
saddle points of higher multiplicity and have a more complex topology 
(in particular, it may contain ``cells'' that do not have rotational 
symmetry and transform into each other under the rotation by $90^{\circ}$).

 A potential $\, V (x, y, \, \alpha , \, {\bf a}) \, $ 
has exact rotational symmetry (has a center of rotational symmetry) 
if the translation vector $\, {\bf a} \, $ satisfies the relation
\begin{equation}
\label{SuperDenseLat}
{\bf a} \,\, = \,\, {k + l \over 2} \, {\bf e}_{1} \, + \,
{k - l \over 2} \, {\bf e}_{2} \, + \, 
{p + q \over 2} \, \pi_{\alpha} ({\bf e}_{1}) \, + \, 
{p - q \over 2} \, \pi_{\alpha} ({\bf e}_{2}) 
\end{equation}
($k, l, p, q \, \in \, \mathbb{Z}$). 

 For the angles $\, \alpha = \alpha_{m,n} \, $ the vectors 
(\ref{SuperDenseLat}) form a (rotated) square lattice with the 
step $\, T / 2 \sqrt{(m_{0}^{2} + n_{0}^{2})} \, $. The lattice 
(\ref{SuperDenseLat}) contains the sublattice (\ref{DenseLat}) 
and is denser. For each of the potentials 
$\, V (x, y, \, \alpha_{m,n} , \, {\bf a}) \, $, there thus 
exists a potential 
$\, V (x, y, \, \alpha_{m,n} , \, {\bf a}^{\prime}) \, $ with 
exact rotational symmetry and such that 
$\, | {\bf a} - {\bf a}^{\prime} | \, 
\leq T / 2 \sqrt{2(m_{0}^{2} + n_{0}^{2})} \, $.

\vspace{1mm} 

 For each potential $\, V (x, y, \, \alpha , \, {\bf a}) \, $ 
in $\, \mathbb{R}^{2} \, $ and each level $\, c \, $ one can define 
the sets
$$\Omega^{-}_{c} [V] \,\,\, = \,\,\,
\Omega^{-}_{c} (\alpha, \, {\bf a}) \,\,\, = \,\,\,
\left\{ (x, y): \,\, V (x, y) < c \right\} $$
and 
$$\Omega^{+}_{c} [V] \,\,\, = \,\,\, 
\Omega^{+}_{c} (\alpha, \, {\bf a}) \,\,\, = \,\,\,
\left\{ (x, y): \,\, V (x, y) > c \right\} $$
 
\begin{figure}[t]
\begin{center}
\includegraphics[width=\linewidth]{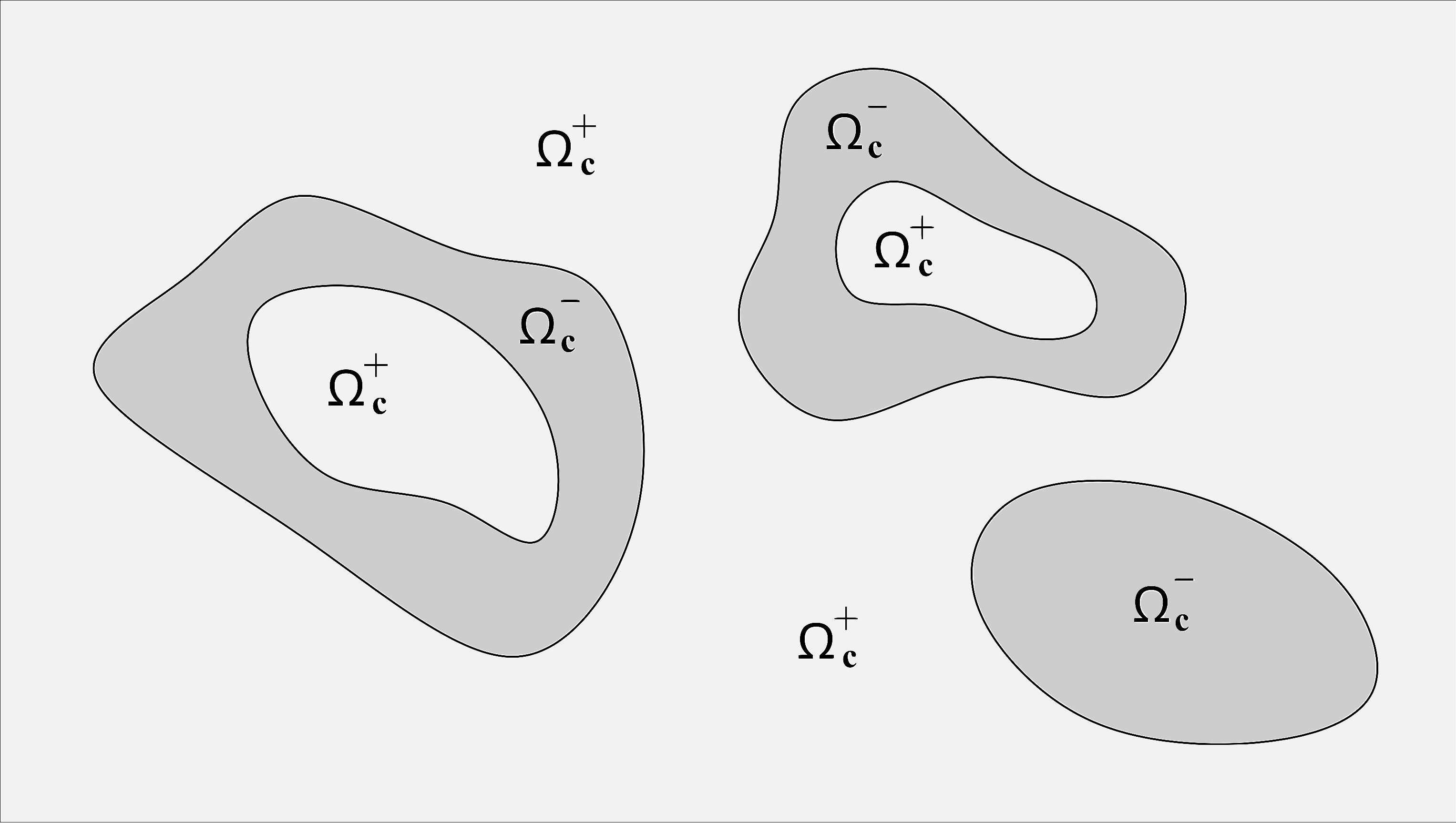}
\end{center}
\caption{Situation of type $\, A(-) \, $ in the 
plane $\, \mathbb{R}^{2} \, $}
\label{SituationAMinus}
\end{figure}

 For all values $\, c < c_{1} (\alpha) \, $ the set 
$\, \Omega^{+}_{c} (\alpha, \, {\bf a}) \, $ 
(for all values of $\, {\bf a}$) has a single unbounded component,
while all other (connected) components of the sets 
$\, \Omega^{-}_{c} (\alpha, \, {\bf a}) \, $ and 
$\, \Omega^{+}_{c} (\alpha, \, {\bf a}) \, $ are bounded 
(Fig. \ref{SituationAMinus}). As follows from \cite{DynMalNovUMN}, 
the sizes of all bounded components of
$\, \Omega^{-}_{c} (\alpha, \, {\bf a}) \, $ and 
$\, \Omega^{+}_{c} (\alpha, \, {\bf a}) \, $ 
(for all $\, {\bf a}$) for a given value $\, c < c_{1} (\alpha) \, $ 
are limited by one constant, which, however, can tend to infinity 
as $\, c \, $ approaches $\, c_{1} (\alpha) \, $.

 For ``magic'' angles $\, \alpha_{m,n} \, $ a similar situation 
also arises in the case $\, c < \hat{c}_{1} (\alpha, \, {\bf a}) \, $. 
We will call the described situation in $\, \mathbb{R}^{2} \, $
a situation of type $\, A(-) \, $.

 Conversely, the occurrence of the situation $\, A(-) \, $ in 
the plane $\, \mathbb{R}^{2} \, $ for generic (not ``magic'') 
angles $\, \alpha \, $ implies $\, c < c_{1} (\alpha) \, $ 
(see \cite{dynn3,DynMalNovUMN,BigQuas}). For ``magic'' angles 
$\, \alpha_{m,n} \, $ the occurrence of the situation $\, A(-) \, $ 
means $\, c < \hat{c}_{1} (\alpha, \, {\bf a}) \, $ 
(but not necessarily $\, c < c_{1} (\alpha)$).

\vspace{1mm}

\begin{figure}[t]
\begin{center}
\includegraphics[width=\linewidth]{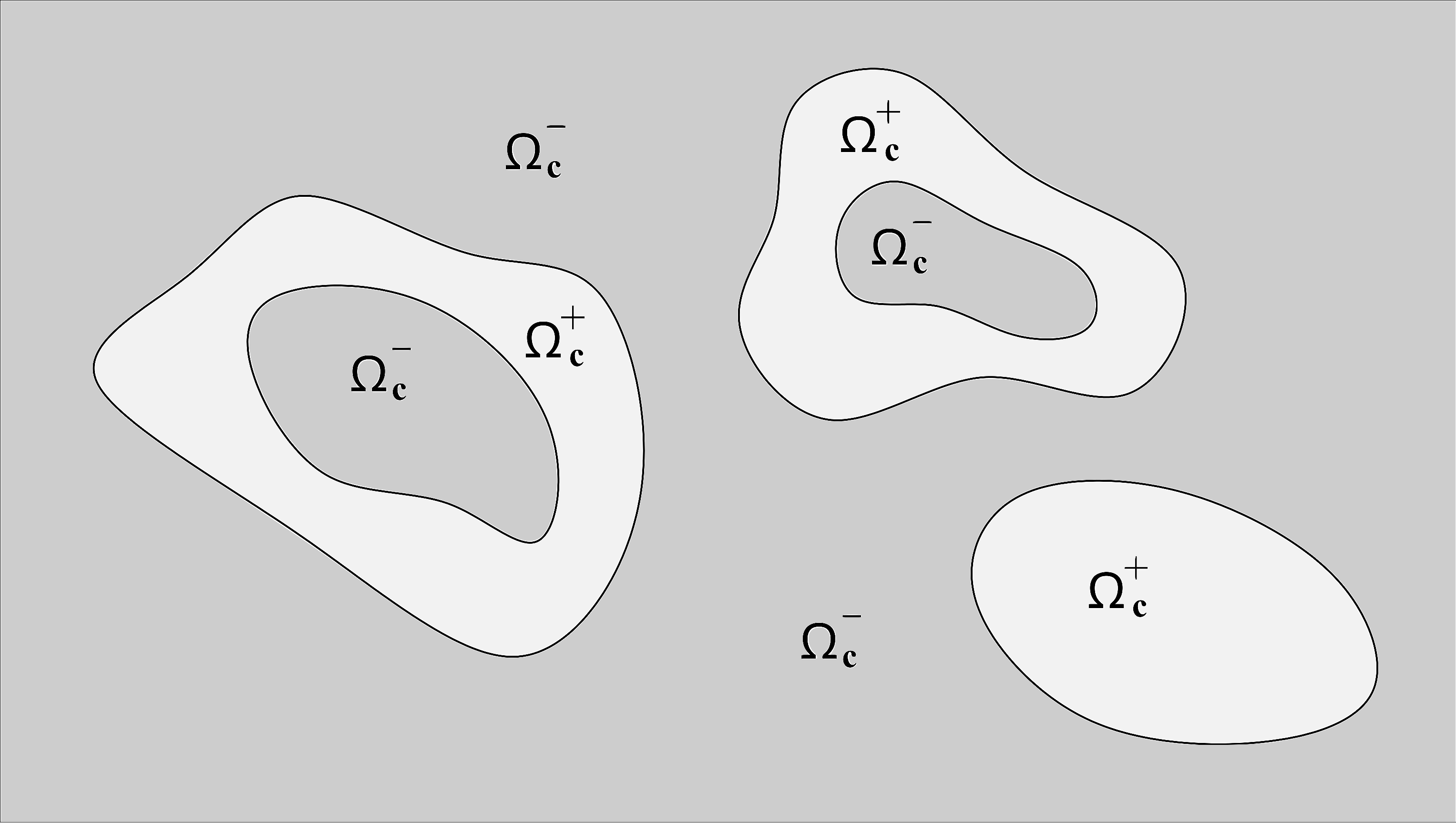}
\end{center}
\caption{Situation of type $\, A(+) \, $ in the 
plane $\, \mathbb{R}^{2} \, $}
\label{SituationAPlus}
\end{figure}

 In the same way, for all values $\, c > c_{2} (\alpha) \, $ the 
set $\, \Omega^{-}_{c} (\alpha, \, {\bf a}) \, $ (for all values of
$\, {\bf a}$) has a single unbounded component, while all other 
(connected) components of the sets 
$\, \Omega^{-}_{c} (\alpha, \, {\bf a}) \, $ and
$\, \Omega^{+}_{c} (\alpha, \, {\bf a}) \, $ are bounded
(Fig. \ref{SituationAPlus}). The sizes of all bounded components
of $\, \Omega^{-}_{c} (\alpha, \, {\bf a}) \, $ and
$\, \Omega^{+}_{c} (\alpha, \, {\bf a}) \, $ (for all $\, {\bf a}$)
for a given value $\, c > c_{2} (\alpha) \, $ are limited 
by one constant, which, however, can tend to infinity 
as $\, c \, $ approaches $\, c_{2} (\alpha) \, $.

 For ``magic'' angles $\, \alpha_{m,n} \, $ a similar situation 
also arises in the case $\, c > \hat{c}_{2} (\alpha, \, {\bf a}) \, $. 
We will call the described situation in $\, \mathbb{R}^{2} \, $ 
a situation of type $\, A(+) \, $.

 Conversely, the occurrence of the situation $\, A(+) \, $ in 
the plane $\, \mathbb{R}^{2} \, $ for generic (not ``magic'') 
angles $\, \alpha \, $ implies $\, c > c_{2} (\alpha) \, $. 
For ``magic'' angles $\, \alpha_{m,n} \, $ the occurrence of 
the situation $\, A(+) \, $ means 
$\, c > \hat{c}_{2} (\alpha, \, {\bf a}) \, $ 
(but not necessarily $\, c > c_{2} (\alpha)$).

\vspace{1mm}

 As we have already said, the potentials 
$\, V (x, y, \, \alpha , \, {\bf a}) \, $ are quasiperiodic 
functions defined by some (affine) embedding 
$\, \mathbb{R}^{2} \subset \mathbb{R}^{4} \, $ and a 4-periodic 
function $\, F (z^{1}, z^{2}, z^{3}, z^{4}, \, \alpha) \, $. 
A shift along the coordinates $\, x \, $ and $\, y \, $, 
as well as a change in the parameters $\, (a^{1}, a^{2}) \, $, 
correspond in this case to parallel shifts of the plane 
$\, \mathbb{R}^{2} \, $ in the space $\, \mathbb{R}^{4} \, $. 
In cases (1) and (2) described in the introduction, the function 
$$F (z^{1}, z^{2}, z^{3}, z^{4}, \, \alpha) \,\,\, = \,\,\, 
F (z^{1}, z^{2}, z^{3}, z^{4}) $$ 
is obviously a smooth function of the variable $ \, {\bf z} \, $. 
In the more general case (3), we will also assume that the 
functional dependence of $\, V \, $ on $\, V_{1} \, $ and 
$\, V_{2} \, $ is sufficiently regular, and the function 
$\, F (z^{1}, z^{2}, z^{3}, z^{4}, \, \alpha) \, $ is a smooth 
function of all its arguments. In particular, this implies the 
existence of constants $\, C_{1} \, $, $\, C_{2} \, $, such that
\begin{equation}
\label{NablaZRel}
\left| \nabla_{\bf z} F ({\bf z}, \, \alpha ) \right| 
\,\,\, \leq \,\,\, C_{1} 
\end{equation}
\begin{equation}
\label{DerAlphaRel} 
\left| \partial_{\alpha} F ({\bf z}, \, \alpha ) 
\right| \,\,\, \leq \,\,\, C_{2} 
\end{equation}
for all values of the variables $\, {\bf z} \, $ and $\, \alpha \, $.  

 For the potentials $\, V (x, y, \, \alpha , \, {\bf a}) \, $ this,
in particular, implies the inequalities
\begin{equation}
\label{VaRel}
\left| V (x, y, \, \alpha , \, {\bf a}) \, - \, 
V (x, y, \, \alpha , \, {\bf a}^{\prime}) \right| \,\,\,\,\, \leq 
\,\,\,\,\, C_{1} \, \left| {\bf a} - {\bf a}^{\prime} \right| 
\end{equation}
\begin{equation}
\label{VCoordRel}
\Big| V ({\bf r}, \, \alpha , \, {\bf a}) \, - \, 
V ({\bf r}^{\prime}, \, \alpha , \, {\bf a}) \Big| \,\,\,\,\, \leq 
\,\,\,\,\, C_{1} \, \left| {\bf r} - {\bf r}^{\prime} \right| 
\end{equation}

\vspace{1mm}

 If the condition (\ref{NablaZRel}) is satisfied, the following 
lemma can be formulated for the potentials 
$\, V (x, y, \, \alpha_{m,n} , \, {\bf a}) \, $.

\vspace{1mm}

\noindent
{\bf Lemma 2.1}

 For a function $\, F ({\bf z}, \, \alpha ) \, $ satisfying 
condition (\ref{NablaZRel}), the width of the interval 
$\, [ c_{1} (\alpha_{m,n}) , \, c_{2} (\alpha_{m,n}) ] \, $ 
(for each ``magic'' angle $\, \alpha_{m,n} $) does not exceed 
$\, \sqrt{2} C_{1} T / \sqrt{m_{0}^{2} + n_{0}^{2}} \, $.

\vspace{1mm}

\noindent
Proof.

 To prove the Lemma, we need to show that the union of the intervals 
$\, [ \hat{c}_{1} (\alpha_{m,n} , {\bf a}), \, 
\hat{c}_{2} (\alpha_{m,n} , {\bf a}) ] \, $ 
for all potentials $\, V (x, y, \, \alpha_{m,n} , \, {\bf a}) \, $ 
for a given $\, \alpha_{m,n} \, $ belongs to some interval whose 
width does not exceed 
$\, \sqrt{2} C_{1} T / \sqrt{m_{0}^{2} + n_{0}^{2}} \, $.

 As we have already mentioned, each of the potentials
$\, V (x, y, \, \alpha_{m,n} , \, {\bf a}) \, $ is equivalent 
in our case to some potential 
$\, V (x, y, \, \alpha_{m,n} , \, {\bf a}^{\prime} ) \, $, where 
$\, | {\bf a}^{\prime} | \, \leq \, T / \sqrt{2(m_{0}^{2} + n_{0}^{2})} \, $.

 The potential $\, V (x, y, \, \alpha_{m,n} , \, 0, 0) \, $, 
as we have already noted, has an open level line in the form 
of a singular periodic ``net'' (Fig. \ref{SingNet90Gen}), arising 
at a single level $\, c = \hat{c}_{0} (\alpha_{m,n} , \, 0, 0) \, $.

 From inequality (\ref{VaRel}) we have
\begin{multline*}
V (x, y, \, \alpha_{m,n} , \, {\bf a}^{\prime} ) \,\,\, \geq  \\
\geq \,\,\, V (x, y, \, \alpha_{m,n} , \, 0, 0 ) \,\, - \,\, 
C_{1} T / \sqrt{2(m_{0}^{2} + n_{0}^{2})}  \,\,\, , 
\end{multline*}
which implies
$$\Omega^{-}_{c - C_{1} T / \sqrt{2(m_{0}^{2} + n_{0}^{2})}} \, 
\left( \alpha_{m,n}, \, {\bf a}^{\prime} \right) \,\,\, \subset \,\,\, 
\Omega^{-}_{c} \left( \alpha_{m,n}, \, 0, 0 \right) $$
for any value of $\, c \, $.

 In particular, this means that the region
$$ \Omega^{-}_{\hat{c}_{0} (\alpha_{m,n} , \, 0, 0) - 
C_{1} T / \sqrt{2(m_{0}^{2} + n_{0}^{2})}} \, 
\left( \alpha_{m,n}, \, {\bf a}^{\prime} \right) $$
lies entirely in the region
$\, \Omega^{-}_{\hat{c}_{0} (\alpha_{m,n} , \, 0, 0)} 
\left( \alpha_{m,n}, \, 0, 0 \right) \, $,
which corresponds to a situation of type $\, A(-) \, $ for the 
potential $\, V (x, y, \, \alpha_{m,n} , \, {\bf a}^{\prime} ) \, $ 
at the level
$$c \,\,\, = \,\,\, \hat{c}_{0} (\alpha_{m,n} , \, 0, 0) 
\,\, - \,\, C_{1} T / \sqrt{2(m_{0}^{2} + n_{0}^{2})} $$ 

 Thus, we get
$$\hat{c}_{1}  (\alpha_{m,n} , {\bf a}) \,\,\, > \,\,\, 
\hat{c}_{0} (\alpha_{m,n} , \, 0, 0) \,\, - \,\, 
C_{1} T / \sqrt{2(m_{0}^{2} + n_{0}^{2})} $$

 The inequality
$$\hat{c}_{2}  (\alpha_{m,n} , {\bf a}) \,\,\, < \,\,\, 
\hat{c}_{0} (\alpha_{m,n} , \, 0, 0) \,\, + \,\,
C_{1} T / \sqrt{2(m_{0}^{2} + n_{0}^{2})} $$
is proved in a completely similar way. 
Thus, we obtain the statement of the Lemma.

\hfill{Lemma 2.1 is proven.}

\vspace{1mm}

 Similar to Lemma 2.1, we can formulate the following lemma.

\vspace{1mm}

\noindent
{\bf Lemma 2.2}

 For a function $\, F ({\bf z}, \, \alpha ) \, $ satisfying 
condition (\ref{NablaZRel}), the width of the intervals 
$\, [ \hat{c}_{1} (\alpha_{m,n}, \, {\bf a}) , \, 
\hat{c}_{2} (\alpha_{m,n}, \, {\bf a}) ] \, $ 
(for any $\, \alpha_{m,n} \, $ and $\, {\bf a}$) does not exceed 
$\, C_{1} T / \sqrt{2(m_{0}^{2} + n_{0}^{2})} \, $.

\vspace{1mm}

 The proof of Lemma 2.2 is similar to the proof of Lemma 2.1, 
since each potential 
$\, V (x, y, \, \alpha_{m,n} , \, {\bf a}) \, $ is equivalent to 
a potential $\, V (x, y, \, \alpha_{m,n} , \, {\bf a}^{\prime}) \, $ 
that has exact rotational symmetry and is such that 
$\, | {\bf a} - {\bf a}^{\prime} | \, 
\leq T / 2 \sqrt{2(m_{0}^{2} + n_{0}^{2})} \, $.

\vspace{1mm}

 As can be seen, the set of ``magic'' angles is everywhere dense 
in the set of all angles $\, \alpha \, $. It is easy to show 
that ``magic'' angles can also be used for ``good'' approximations 
of generic angles $\, \alpha \, $, according to Dirichlet's 
theorem on the approximation of irrational numbers by rational ones. 
Let us present the corresponding arguments here.

\vspace{1mm}

 Let $\, m \geq n \, $. For
$${1 \over 2} \left( {m \over n} - {n \over m} \right) \,\,\, = \,\,\,
\tan \alpha_{m,n} $$
we have 
\begin{multline*}
{m \over n}  \,\,\, =  \,\,\, \sqrt{\tan^{2} \alpha_{m,n} + 1} \,\, + \,\, 
\tan \alpha_{m,n} \,\,\, ,  \\
{n \over m} \,\,\, = \,\,\, \sqrt{\tan^{2} \alpha_{m,n} + 1} \,\, - \,\, 
\tan \alpha_{m,n} 
\end{multline*}
 
 Let us have (without loss of generality) for some angle 
$\, \alpha \, $:
$\, \tan \alpha > 0 \, $ ($\alpha \in (0, \pi / 2)$),  
$\,\, m > n \, $ and 
$$\left| {m \over n} \, - \, \sqrt{\tan^{2} \alpha + 1} 
\, - \, \tan \alpha \right| \,\,\, < \,\,\, \delta $$
 
 Then 
$$\left|  {1 \over 2} \left( {m \over n} - {n \over m} \right)
\, - \, \tan \alpha \right| \,\,\, < \,\,\, \delta $$
(according to Lagrange's theorem for
$\, f (x) = {1 \over 2} \left( x - {1 \over x} \right) $). That is,
\begin{multline*}
\left| {m \over n} \, - \, \sqrt{\tan^{2} \alpha + 1} 
\, - \, \tan \alpha \right| \,\,\, < \,\,\, \delta
\quad \Rightarrow  \\
\Rightarrow \quad \left| \tan \alpha_{m,n} \, - \, \tan \alpha
\right| \,\,\, < \,\,\, \delta \quad \Rightarrow  \quad 
\left| \alpha_{m,n} \, - \, \alpha \right| \,\,\, < \,\,\, \delta
\end{multline*}

 In particular, if we have for a ``good'' approximation of 
the number $\, \sqrt{\tan^{2} \alpha + 1} \, + \, \tan \alpha \, $ 
by a rational fraction
$$\left| {m \over n} \, - \, \sqrt{\tan^{2} \alpha + 1} 
\, - \, \tan \alpha \right| \,\,\, < \,\,\, {1 \over n^{2}} 
\,\,\, , $$
then
$$\left|  {1 \over 2} \left( {m \over n} - {n \over m} \right)
\, - \, \tan \alpha \right| \,\,\, < \,\,\, {1 \over n^{2}} $$
and
$$\left| \alpha_{m,n} \, - \, \alpha \right| \,\,\, < \,\,\, 
{1 \over n^{2}} $$
 
\vspace{1mm}

 This circumstance, in particular, allows us to prove the 
following theorem.

\vspace{1mm}

\noindent
{\bf Theorem 2.1}

 Let $\, V_{1} (x, y) \, $ and $\, U (x, y) \, $ be periodic 
potentials with the same periods, symmetric with respect to 
rotation by $90^{\circ}$. Then for all potentials 
$\, V (x, y, \, \alpha, \, {\bf a} ) $ corresponding to generic 
(not ``magic'') angles $\, \alpha \, $ the following relation holds
$$c_{1} (\alpha) \,\,\, = \,\,\, c_{2} (\alpha) \,\,\, = \,\,\, 
c_{0} (\alpha) $$

\vspace{1mm}

 To prove Theorem 2.1 we will also need the following lemma.

 \vspace{1mm}

\noindent
{\bf Lemma 2.3}

 Let a potential $\, V (x, y) \, $ have rotation symmetry at 
the angle $\, \alpha_{0} = 90^{\circ} \, $ and periods 
$\, \{ {\bf e}_{1} \, , \, {\bf e}_{2} \} \, $ such that
$$\left| {\bf e}_{1} \right| \,\, = \,\, \left| {\bf e}_{2} \right|
\,\, = \,\, L \,\,\, , \quad 
\pi_{\alpha_{0}} \left[ {\bf e}_{1} \right] \,\, = \,\, 
{\bf e}_{2} $$

 Let
$$\left| \nabla \, V (x, y) \right| \,\, < \,\, C_{1} $$
and $\, V (x, y) \, $ have a ``singular periodic net'' at a level
$$V (x, y) \,\,\, = \,\,\, c_{0} $$

 Then for any $\, \Delta c > 0 $ the diameter of any connected region 
of the set $\, \Omega^{-}_{c_{0} - \Delta c} \left[ V \right] \, $ 
(as well as $\, \Omega^{+}_{c_{0} + \Delta c} \left[ V \right]$) 
does not exceed
$$\sqrt{ D L \over  \Delta c} \,\, L \,\,\, , $$
where $\, D \, = \, \sqrt{5} C_{1} / 2 \, $.

\vspace{1mm}

 We give a proof of Lemma 2.3 in the Appendix.

 \vspace{1mm}
 
 \noindent
 Proof of Theorem 2.1.
 
  As is easy to see, the presence of a non-singular non-closed 
level line $\, V (x, y) \, = \, c \, $ implies the presence of 
unbounded components both for the set 
$\, \Omega^{-}_{c} \left[ V \right] \, $ and for the set 
$\, \Omega^{+}_{c} \left[ V \right] \, $ 
(adjacent to $\, V (x, y) \, = \, c \, $ from different sides). 
The same is true for an unbounded level line 
$\, V (x, y) \, = \, c \, $ containing a finite number of 
singular points (which we assume here for level lines 
of non-periodic potentials).

 Let a potential $\, V (x, y, \, \alpha, \, {\bf a} ) $ have 
an unbounded connected component of 
$\, \Omega^{-}_{c} \left[ \alpha, \, {\bf a} \right] \, $ 
containing a point $\, (x_{0}, y_{0}) \, $.

 Let us take an angle $\, \alpha_{m,n} \, $, such that
$$\left| \alpha_{m,n} - \alpha \right| \,\,\, < \,\,\, 
{1 \over n^{2}} $$

 Consider the potential $\, V^{\prime} (x, y) \, $ created by 
the superposition of $\, V_{1} (x, y) \, $ and the potential 
$\, V_{2} (x, y) \, $ which is rotated by the angle
$$\delta \alpha \,\,\, = \,\,\, \alpha_{m,n} - \alpha $$
around the point $\, (x_{0}, y_{0}) \, $. 

 Obviously
$$V^{\prime} (x, y) \,\,\, = \,\,\, 
V (x, y, \, \alpha_{m,n}, \, {\bf a}_{m,n}^{\prime} ) $$
for some $\, {\bf a}_{m,n}^{\prime} \, $.

 According to (\ref{NablaZRel}) - (\ref{DerAlphaRel}) 
(taking into account a change in the values of $\, z^{3} \, $
and $\, z^{4} \, $ as well as the value 
of $\, \alpha \, $ in embedding 
$\, \mathbb{R}^{2} \subset \mathbb{R}^{4} $) in the circle 
of radius 
$$(m_{0}^{2} + n_{0}^{2})^{5/6} \,\, T $$
with the center $\, (x_{0}, y_{0}) \, $ we have the following
relation
\begin{multline*}
\Big| V (x, y, \, \alpha, \, {\bf a} )  \, - \,   
V (x, y, \, \alpha_{m,n}, \, {\bf a}_{m,n}^{\prime} ) \Big| 
\,\,\, \leq  \\
\leq \,\,\, C_{1} \,\, 
(m_{0}^{2} + n_{0}^{2})^{5/6} \,\, T \,\, \left| \delta \alpha \right|
\,\,\, + \,\,\, C_{2} \, \left| \delta \alpha \right| \,\,\, <  \\ 
< \,\,\, C_{1} \,\, {(m_{0}^{2} + n_{0}^{2})^{5/6} \over n^{2}} \,\, T 
\,\,\, + \,\,\, {C_{2} \over n^{2}}  
\end{multline*}

 As we have already noted, there is also a vector 
$\, {\bf a}_{m,n}^{\prime\prime} \, $, such that
$$\left| {\bf a}_{m,n}^{\prime} \, - \, {\bf a}_{m,n}^{\prime\prime} 
\right| \,\,\, \leq \,\,\, T / \sqrt{2 (m_{0}^{2} + n_{0}^{2})} $$
and the potential
$$V^{m,n} (x, y) \,\,\, = \,\,\, 
V (x, y, \, \alpha_{m,n}, \, {\bf a}_{m,n}^{\prime\prime} ) $$
is equivalent to the potential
$\, V (x, y, \, \alpha_{m,n}, \, 0, 0 ) \, $. 

 The potential $\, V^{m,n} (x, y) \, $ has exact rotational 
symmetry, and, according to (\ref{VaRel})
$$\Big| V (x, y, \, \alpha_{m,n}, \, {\bf a}_{m,n}^{\prime} )
\, - \, V^{m,n} (x, y) \Big| \,\,\, \leq \,\,\, 
{C_{1} \, T \over \sqrt{2 (m_{0}^{2} + n_{0}^{2})}} $$

 Thus, we have in the circle of radius 
$\, (m_{0}^{2} + n_{0}^{2})^{5/6} \,\, T \, $ with center at 
$\, (x_{0}, y_{0}) \, $
\begin{multline*}
\Big| V (x, y, \, \alpha, \, {\bf a} )
\, - \, V^{m,n} (x, y) \Big| \quad <  \\
< \quad  C_{1} \,\,
{(m_{0}^{2} + n_{0}^{2})^{5/6} \over n^{2}} \,\, T
\,\,\, + \,\,\, {C_{2} \over n^{2}} 
\,\,\, + \,\,\, {C_{1} \, T \over \sqrt{2 (m_{0}^{2} + n_{0}^{2})}}
\end{multline*}

 The potential $\, V^{m,n} (x, y) \, $ has an unbounded level 
line (singular periodic net) at a single level
$$ c_{0} (\alpha_{m,n}, \, {\bf a}_{m,n}^{\prime\prime}) 
\,\,\, = \,\,\, c_{0} (\alpha_{m,n}, 0, 0) $$

  Let 
$\, c \, < \, c_{0} (\alpha_{m,n}, \, {\bf a}_{m,n}^{\prime\prime}) \, $. 
By assumption, the set $\, \Omega^{-}_{c} (\alpha, \, {\bf a}) \, $ has an 
unbounded connected component containing the point $\, (x_{0}, y_{0}) \, $. 
It follows then 
\begin{multline}
\label{FirstRel}
c \,\,\,\,\, > \,\,\,\,\, 
c_{0} (\alpha_{m,n}, \, {\bf a}_{m,n}^{\prime\prime}) \,\, - \,\, 
{D T \over (m_{0}^{2} + n_{0}^{2})^{1/6}} \,\, -  \\ 
- \,\, C_{1} \,\, {(m_{0}^{2} + n_{0}^{2})^{5/6} \over n^{2}} \,\, T
\,\,\, - \,\,\, {C_{2} \over n^{2}} \,\,\, - \,\,\, 
{C_{1} \, T \over \sqrt{2 (m_{0}^{2} + n_{0}^{2})}} 
\,\,\,\,\, \equiv  \\
\equiv \,\,\,\,\, 
c_{0} (\alpha_{m,n}, 0, 0) \,\, - \,\, \Delta_{m,n} 
\end{multline}

 Indeed, if (\ref{FirstRel}) is wrong, then we have in the circle 
of radius $\, (m_{0}^{2} + n_{0}^{2})^{5/6} \, T \, $ with center 
$\, (x_{0}, y_{0}) \, $
\begin{multline*}
\Omega^{-}_{c} (\alpha, \, {\bf a}) \,\, \subset \,\,
\Omega^{-}_{c_{0} (\alpha_{m,n}, {\bf a}_{m,n}^{\prime\prime})
- \overline{\Delta c}} \,\, (\alpha_{m,n}, {\bf a}_{m,n}^{\prime\prime}) 
\,\,  \equiv  \\
\equiv \,\, \Omega^{-}_{c_{0} (\alpha_{m,n}, 0, 0)
- \overline{\Delta c}} \, \left[ V^{m,n} \right] \,\,\, ,
\end{multline*}
where
$$\overline{\Delta c} \,\,\, = \,\,\, 
D T \big/ (m_{0}^{2} + n_{0}^{2})^{1/6} $$

 The potential $\, V^{m,n} (x, y) \, $, however, has the required 
rotational symmetry and period
$$L \,\,\, = \,\,\,  T_{m,n} \,\,\, = \,\,\, 
(m_{0}^{2} + n_{0}^{2})^{1/2} \, T $$
  
  Using Lemma 2.3, we can then conclude that all connected 
components of the set 
$\, \Omega^{-}_{c_{0}(\alpha_{m,n}, 0, 0) - 
\overline{\Delta c}} \, \left[ V^{m,n} \right] \, $ 
that contain the point $\, (x_{0}, y_{0}) \, $ 
lie in the circle of radius 
$\, (m_{0}^{2} + n_{0}^{2})^{5/6} \, T \, $ centered at 
$\, (x_{0}, y_{0}) \, $, which then implies the same 
property for any connected component of 
$\, \Omega^{-}_{c} (\alpha, \, {\bf a}) \, $
containing the point $\, (x_{0}, y_{0}) \, $.

 Absolutely the same
\begin{equation}
\label{SecondRel}
c_{2} (\alpha) \,\,\, < \,\,\, c_{0} (\alpha_{m,n}, 0, 0) 
\,\, + \,\, \Delta_{m,n} 
\end{equation}
and thus
\begin{equation}
\label{2Deltamn}
c_{2} (\alpha) \, - \, c_{1} (\alpha) 
\,\,\,\,\, < \,\,\,\,\, 2 \Delta_{m,n}
\end{equation}

 In the limit
$$n \, \rightarrow \, \infty \, , \,\,\, m \, \rightarrow \, \infty \, ,
\quad  {m \over n} \,\, \rightarrow \,\,
\sqrt{\tan^{2} \alpha + 1} \,\,\, + \,\,\, \tan \alpha $$
we get 
$$\Delta_{m,n} \,\,\, \rightarrow \,\,\, 0 $$
and thus
$\, c_{1} (\alpha) \, = \, c_{2} (\alpha) \, = \, c_{0} (\alpha) \, $.

{\hfill Theorem 2.1 is proven.}

\vspace{1mm}

 From the proof of Theorem 2.1, in fact, it also follows that 
the relations (\ref{FirstRel}) - (\ref{SecondRel}) hold for all 
angles $\, \alpha^{\prime} \, $ such that
$$\left| \alpha^{\prime} - \alpha_{m,n} \right| \,\,\, < \,\,\,
1 / n^{2} $$

 Let us define open sets
$$O_{m,n} \,\,\, = \,\,\, \left\{ \alpha^{\prime} \, : \,\, 
\left| \alpha^{\prime} - \alpha_{m,n} \right| \,\,\, < \,\,\,
1 / n^{2} \right\} $$

 We can state that
 
\vspace{1mm} 
 
\noindent
1) All sets $\, O_{m,n} \, $ contain the angle $\, \alpha \, $;

\vspace{1mm} 

\noindent
2) Each set $\, O_{m,n} \, $ corresponds to a segment
$$I_{m,n} \, = \, \Big[ c_{0} (\alpha_{m,n}, 0, 0) - 
\Delta_{m,n} \, , \,\, c_{0} (\alpha_{m,n}, 0, 0) + \Delta_{m,n}
\Big] \, , $$ 
such that
$$\left[ c_{1} (\alpha^{\prime}) , \, c_{2} (\alpha^{\prime}) 
\right] \,\,\, \subset I_{m,n} $$
for all $\, \alpha^{\prime} \in O_{m,n} \, $;

\vspace{1mm}

\noindent
3) The length of $\, I_{m,n} \, $ tends to zero as
$\, n \rightarrow \infty \, $. 

\vspace{1mm}

 It follows from the above statements that the functions 
$\, c_{1} (\alpha) \, $ and $\, c_{2} (\alpha) \, $ are 
continuous at generic angles $\, \alpha \, $ (where they
both coincide with $\, c_{0} (\alpha)$). In particular, 
the function $\, c_{0} (\alpha) \, $, considered on the 
set of generic angles $\, \alpha \, $ (not ``magic''), 
is continuous on this set.

\section{The cases of symmetry of the third and sixth order}
\setcounter{equation}{0}

  The cases where the potentials $\, V_{1} (x, y) \, $ and 
$\, V_{2} (x, y) \, $ have rotation symmetry by $60^{\circ}$ 
or $120^{\circ}$ play an even more important role in the physics 
of two-layer systems than the case of the symmetry of order 
four. A detailed description of the set of ``magic'' angles, 
as well as the corresponding periods of superposition of 
$\, V_{1} (x, y) \, $ and $\, V_{2} (x, y) \, $, for symmetry 
of order six, is given in \cite{Shallcross1,Shallcross2}. For our 
purposes here we use a somewhat ``simplified'' representation of 
the ``magic'' angles in both cases, similar to that used in 
the previous chapter.

 In both the case of the third-order and sixth-order symmetry, 
the lattice of periods of potentials $\, V_{1} \, $ and 
$\, V_{2} \, $ is triangular. In the case of sixth-order symmetry, 
the ``complete'' set of rotation angles is given by the interval 
$\, \alpha \in (0, \, \pi / 3 ) \, $. For the third-order symmetry, 
we will consider here the interval 
$\, \alpha \in ( - \pi / 3 , \, \pi / 3 ) \, $. Now let
\begin{equation}
\label{3OrdBas}
{\bf e}_{1} \,\,\, = \,\,\, \big(T, \, 0 \big) \,\,\, , \quad
{\bf e}_{2} \,\,\, = \,\,\, \big( T/2 , \, \sqrt{3} T / 2 \big) 
\end{equation}

\begin{figure}[t]
\begin{center}
\includegraphics[width=\linewidth]{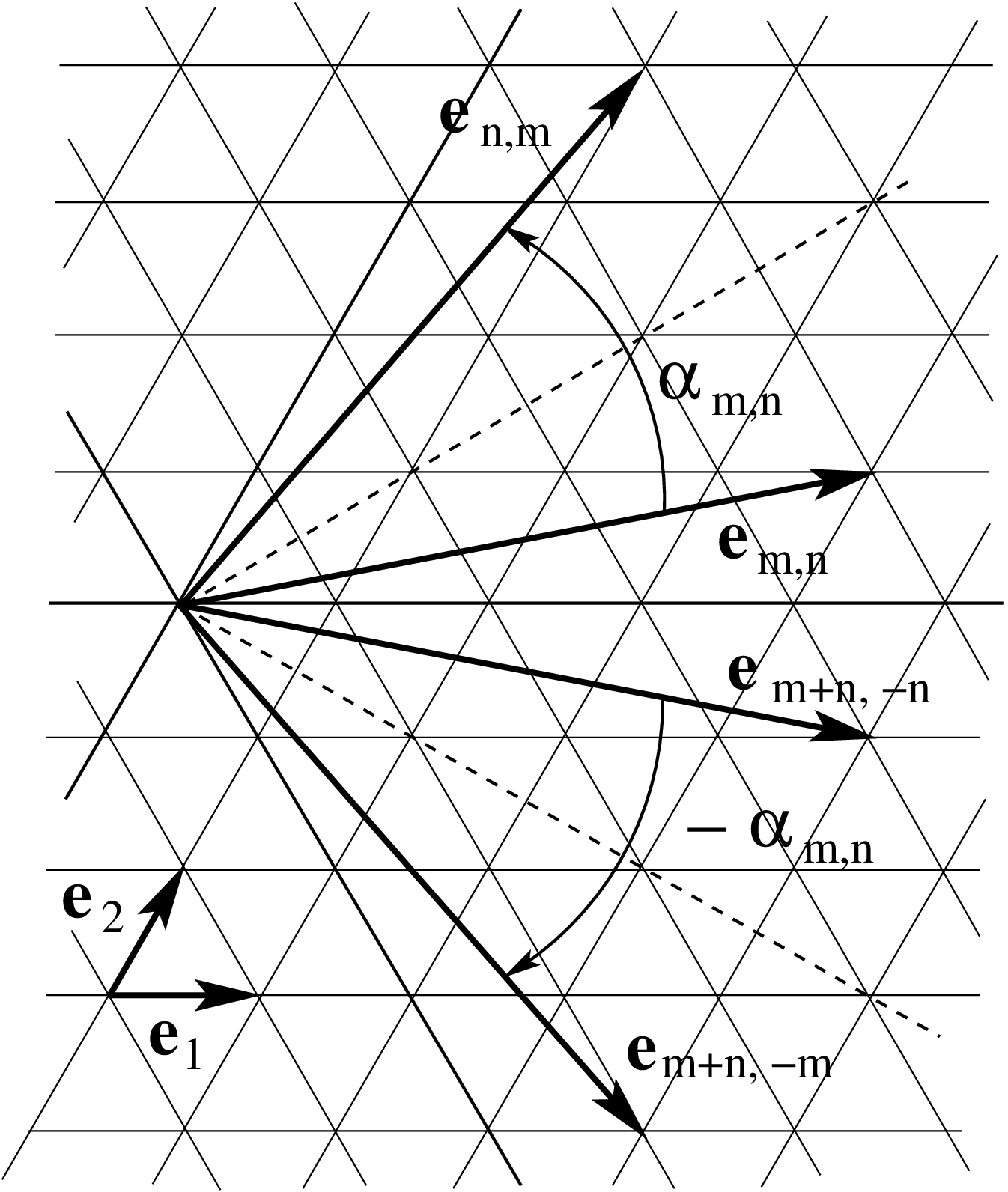}
\end{center}
\caption{Rotations from the vector 
$m \, {\bf e}_{1}\, + \, n \, {\bf e}_{2}$ to the vector 
$n \, {\bf e}_{1}\, + \, m \, {\bf e}_{2}$, and also from the 
vector $\, (m + n) \, {\bf e}_{1}\, - \, n \, {\bf e}_{2} \, $ 
to the vector 
$\, (m + n) \, {\bf e}_{1}\, - \, m \, {\bf e}_{2} \, $ 
in a triangular lattice.}
\label{ThirdOrdePovmn}
\end{figure}

 Similar to the previous section, we define a series 
of ``magic'' angles $\, \alpha_{m,n} \, $ using the rotation from 
the vector
$${\bf e}_{m,n} \,\,\, = \,\,\, m \, {\bf e}_{1} \,\, + \,\, 
n \, {\bf e}_{2} $$
to the vector
$${\bf e}_{n,m} \,\,\, = \,\,\, n \, {\bf e}_{1} \,\, + \,\, 
m \, {\bf e}_{2} \,\,\, , $$
where $\, m > n > 0 \, $ (Fig. \ref{ThirdOrdePovmn}).

 The angles $\, \alpha_{m,n} \, $ lie in the interval
$\, (0, \, \pi / 3 ) \, $ and give us the desired family 
of ``magic'' angles in the case of sixth-order symmetry. 
The periods of the potential $\, V (x, y) \, $ 
(not necessarily minimal) here are the vectors
$${\bf b}_{1} \,\,\, = \,\,\, {\bf e}_{m+n, -n} \,\,\, , \quad
{\bf b}_{2} \,\,\, = \,\,\, {\bf e}_{n,m} \,\,\, , $$
having length $\, \sqrt{m^{2} + n^{2} + mn} \, \cdot \, T \, $.

\vspace{1mm}

\begin{figure}[t]
\begin{center}
\includegraphics[width=\linewidth]{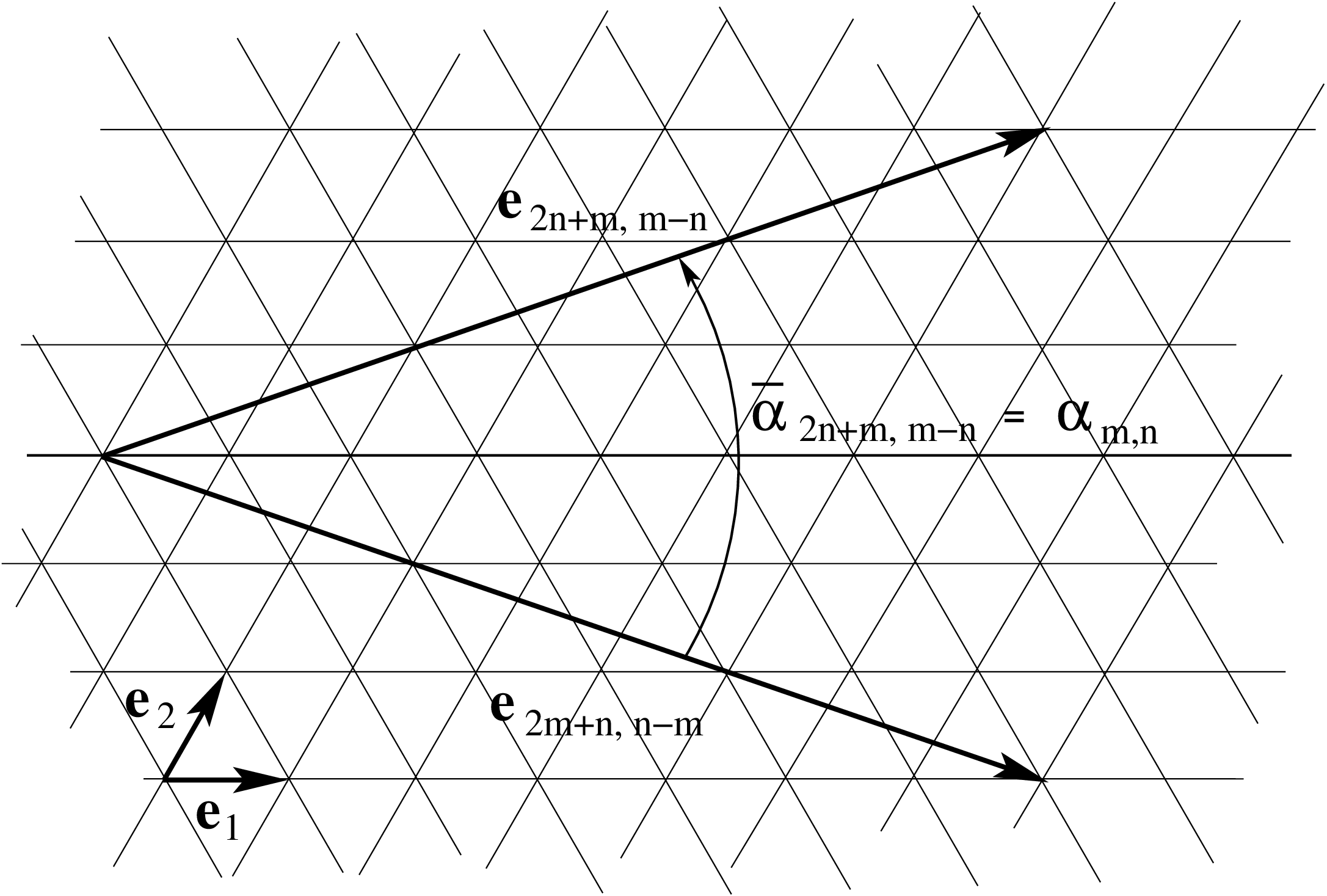}
\end{center}
\caption{``Magic'' angles 
$\, \bar{\alpha}_{2n+m, m-n} \, = \, {\alpha}_{m,n} \, $, 
defined by the rotation from the vector 
$\, {\bf e}_{2m+n, n-m} \, $ to the vector 
$\, {\bf e}_{2n+m, m-n} \, $ in a triangular lattice.} 
\label{ThirdOrdePovbar}
\end{figure}

 The same set of ``magic'' angles $\, {\alpha}_{m,n} \,$ can be 
represented by the angles $\, \bar{\alpha}_{2n+m, m-n} $ defined 
by rotations from the vector
$${\bf e}_{2m+n, n-m} \,\,\, = \,\,\, (2m + n) \, {\bf e}_{1}
\,\, + \,\, (n - m) {\bf e}_{2} $$
to the vector
$${\bf e}_{2n+m, m-n} \,\,\, = \,\,\, (2n + m) \, {\bf e}_{1}
\,\, + \,\, (m - n) {\bf e}_{2} $$
(Fig. \ref{ThirdOrdePovbar}).

 The minimal periods of the potential $\, V (x, y) \, $ 
are either the vectors $\, \{ {\bf b}_{1}, {\bf b}_{2} \} \, $, 
or
$${\bf b}^{\prime}_{1} \,\,\, = \,\,\, {1 \over \gamma} 
\left(
\begin{array}{c}
n + 2 m   \\
- 2 n - m 
\end{array} \right) \,\,\, , \quad
{\bf b}^{\prime}_{2} \,\,\, = \,\,\, {1 \over \gamma} 
\left(
\begin{array}{c}
2 n + m   \\
m - n  
\end{array} \right) \,\,\, ,  $$
where $\, \gamma \, = \, \text{gcd} \, (2n + m, \, m - n) \, $. 

 It is also easy to see that
$$ \gamma \, = \, \text{gcd} \, (3n , \, m - n) $$

 Since 
$\, \text{gcd} \, (n , \, m - n) \, = \,
\text{gcd} \, (n , \, m) \, = \, 1 \, $, 
$\, \gamma \, $ is different from 1 (and is equal to 3) only if 
$\, m - n \, $ is divisible by 3. It can thus be seen that 
$\, \{ {\bf b}^{\prime}_{1}, {\bf b}^{\prime}_{2} \} \, $ 
are minimal periods if 
$\, m - n = 3 k\, $, $\,\, k \in \mathbb{N} \, $.

 Similar to the previous section, we introduce here the 
numbers $\, (m_{0}, n_{0}) \, $:
$$(m_{0}, n_{0}) \,\,\, = \,\,\, (m, n) \,\,\, , \quad
m - n \, \neq \, 3 k \,\,\, , \,\,\,\,\, k \in \mathbb{N} $$
\begin{multline*}
(m_{0}, n_{0}) \,\,\, = \,\,\, \left( 
{2n + m \over 3} \, , \,\, {m - n \over 3} \right) 
\,\,\, = \,\,\, (n + k, \, k ) \,\,\, ,  \\
m - n \, = \, 3 k \,\,\, , \,\,\,\,\, k \in \mathbb{N} 
\end{multline*}

 As in the previous section, we will use here the series 
of angles $\, \alpha_{m,n} \, $ for all irreducible pairs 
$\, (m, n) \, $, $\,\, m > n > 0 \, $. 

\vspace{1mm}

 In the case of third-order symmetry, we also need to add 
the series of ``magic'' angles $\, - \, \alpha_{m,n} \, $ 
for $\, m > n \geq 0 \, $ (Fig. \ref{ThirdOrdePovmn}). 
It can be seen that the minimal periods of the corresponding 
potentials $\, V (x, y) \, $ are the vectors
$${\bf b}_{1} \,\,\, = \,\,\, {\bf e}_{m+n, -m} \,\,\, , \quad
{\bf b}_{2} \,\,\, = \,\,\, {\bf e}_{m, n} \,\,\, , $$
or 
$${\bf b}^{\prime}_{1} \,\,\, = \,\,\, {1 \over \gamma} 
\left(
\begin{array}{c}
2 m + n   \\
n - m
\end{array} \right) \,\,\, , \quad
{\bf b}^{\prime}_{2} \,\,\, = \,\,\, {1 \over \gamma} 
\left(
\begin{array}{c}
m - n   \\
2 n + m
\end{array} \right) \,\,\, ,  $$

\vspace{1mm}

 The value of the angle $\, \alpha_{m,n} \, $
is determined by the expression
$$\tan \alpha_{m,n} \,\, = \,\, 
{\sqrt{3} (m^{2} - n^{2}) \over m^{2} + n^{2} + 4 m n} 
\,\, = \,\, {\sqrt{3} ( (m/n)^{2} - 1) \over 
(m/n)^{2} + 4 (m/n) + 1} $$

 For the derivative of the function
$$f (x) \,\,\ = \,\,\,  {\sqrt{3} ( x^{2} - 1) \over 
x^{2} + 4 x + 1} $$
that is
$${d f \over d x}  \,\,\ = \,\,\, 4 \sqrt{3} \,\, 
{x^{2} + 3 x / 2 + 1 \over (x^{2} + 4 x + 1)^{2}} $$
we have in the region $\, x \geq 1 \, $
$${ d f \over d x} \,\,\, < \,\,\, 
{4 \sqrt{3} \over x^{2} + 4 x + 1} \,\,\, \leq \,\,\, 
{2 \over \sqrt{3}} $$

 Let $\, f(x) \, = \, \tan \, \alpha \, $ for  
$\, 0 < \alpha < 60^{\circ} \, $, i.e.
$$x \,\,\, = \,\,\, {\sqrt{3} \, \sqrt{\tan^{2} \alpha + 1} \,\, + \,\, 
2 \, \tan \, \alpha \over \sqrt{3} \,\, - \,\, \tan \, \alpha} 
\,\,\, > \,\,\, 1 \,\,\, , $$

 Let us consider a ``good'' approximation of the 
value $\, x \, $ by a value $\, m/n \, $. From Lagrange's 
theorem we can then conclude
$$\left| {m \over n} - x \right| \,\, < \,\, {1 \over n^{2}} 
\,\,\, \Rightarrow \,\,\, 
\left| \tan \, \alpha_{m,n} \, - \, \tan \, \alpha \right| 
\,\, < \,\, {2 \over \sqrt{3}} \, {1 \over n^{2}} \,\,\, \Rightarrow  $$
$$\Rightarrow \,\,\, \left| \alpha_{m,n} \, - \, \alpha \right| 
\,\, < \,\, {2 \over \sqrt{3}} \, {1 \over n^{2}} $$
which gives sequences of ``good'' approximations for 
angles $\, \alpha \in (0, \, \pi / 3 ) \, $ by the ``magic'' angles 
$\, \alpha_{m,n} \, $ (similarly, by the angles 
$\, - \, \alpha_{m,n} \, $ for $\, \alpha \in ( - \pi / 3 , 0 )$).

\vspace{1mm}

 As in the previous section, we assume here that the 
potential $\, V_{2} (x, y) \, $ has the form (\ref{V2U}), 
and the point $\, (0, 0) \, $ is the center of rotational 
symmetry for both the potentials $\, V_{1} (x, y) \, $ and  
$\, U (x, y) \, $.

 As before, the identities (\ref{Tozhd1}) - (\ref{Tozhd2}) 
hold here, and we can also define classes of equivalent 
potentials given by the formula (\ref{EquivPot}).

 For ``magic'' angles $\, \alpha_{m,n} \, $ the vectors 
(\ref{DenseLat}) here form a lattice with step 
$\, T / \sqrt{m_{0}^{2} + n_{0}^{2} + m_{0} n_{0}} \, $, 
which now has triangular symmetry. In particular, each 
potential $\, V (x, y, \, \alpha_{m,n}, \, {\bf a}) \, $ 
is now equivalent to some potential 
$\, V (x, y, \, \alpha_{m,n}, \, {\bf a}^{\prime}) \, $ 
such that 
$\, |{\bf a}^{\prime}| \, \leq \, 
T / \sqrt{3 (m_{0}^{2} + n_{0}^{2} + m_{0} n_{0})} \, $.

 Besides that, every potential 
$\, V (x, y, \, \alpha_{m,n}, \, {\bf a}) \, $ is equivalent 
to a potential 
$\, V (x, y, \, \alpha_{m,n}, \, {\bf a}^{\prime\prime}) \, $ 
that has exact rotational symmetry and is such that
$$\left| {\bf a} - {\bf a}^{\prime\prime} \right| \,\,\, \leq \,\,\,
T \big/ 2 \sqrt{3 (m_{0}^{2} + n_{0}^{2} + m_{0} n_{0})} $$

\begin{figure}[t]
\begin{center}
\includegraphics[width=\linewidth]{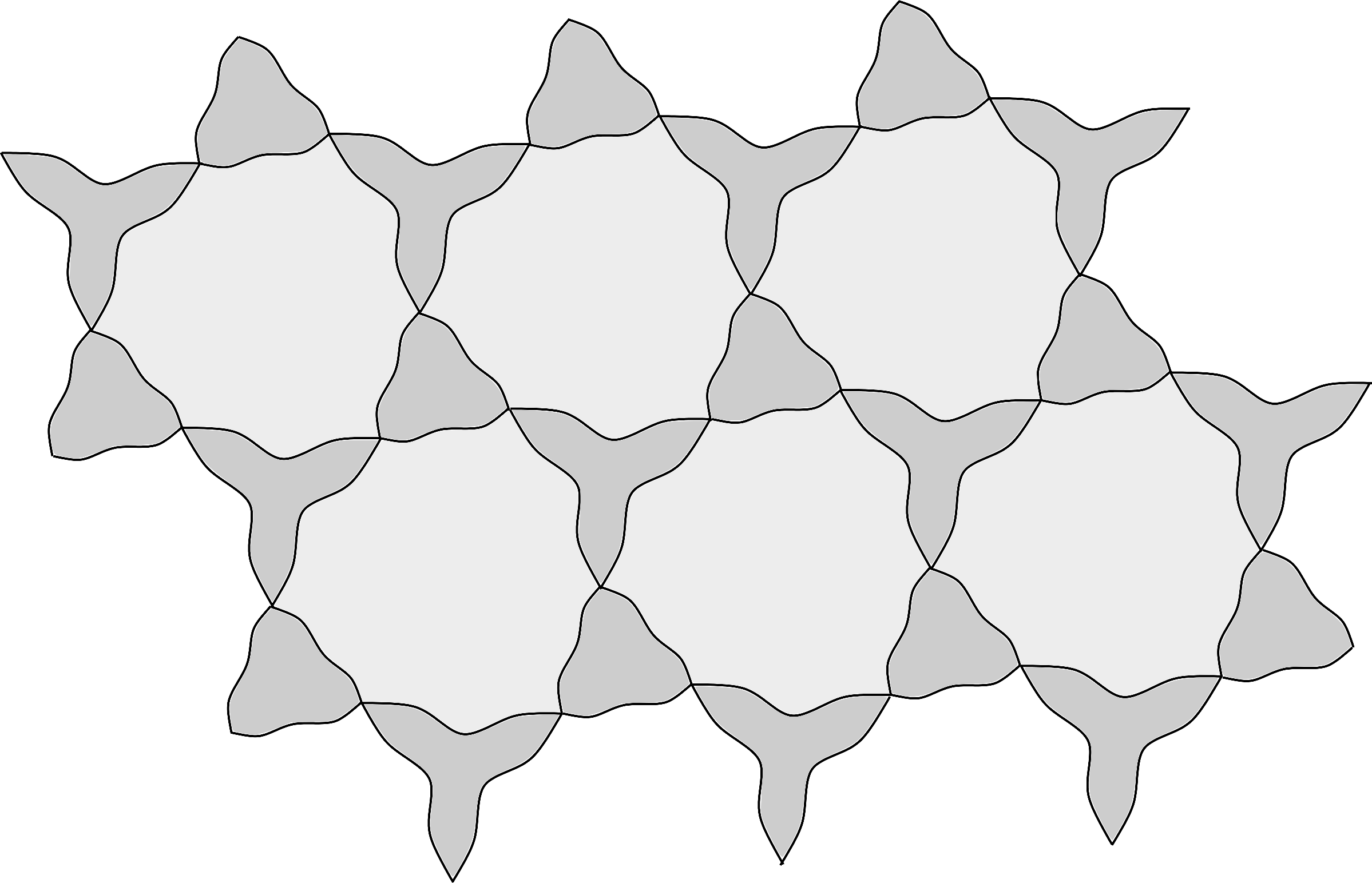}
\end{center}
\caption{An example of a singular kagome-type ``net'' 
for a third-order rotational symmetry potential with 
non-degenerate critical points.}
\label{kagome}
\end{figure}

 Periodic potentials with exact rotational symmetry of 
the third or sixth order also cannot have non-singular open 
level lines. The situations $\, A(-) \, $ and $\, A(+) \, $ 
for them are also separated by a singular periodic ``net'' 
arising at a single level $\, V (x, y) = c_{0} \, $ 
(Fig. \ref{kagome}).

 Periodic potentials $\, V (x, y, \, \alpha_{m,n}, \, {\bf a}) \, $, 
which do not have exact rotational symmetry, here also have intervals 
$\, \left[ \hat{c}_{1} (\alpha_{m,n}, {\bf a}), \, 
\hat{c}_{2} (\alpha_{m,n}, {\bf a}) \right] \, $, where they have 
open level lines. Here we can also define the interval 
$\, \left[ c_{1} (\alpha_{m,n}), \, c_{2} (\alpha_{m,n}) \right] \, $, 
where open level lines of potentials 
$\, V (x, y, \, \alpha_{m,n}, \, {\bf a}) \, $ arise for 
at least one value of the parameters $ {\bf a} \, $. All non-singular 
open level lines of $\, V (x, y, \, \alpha_{m,n}, \, {\bf a}) \, $ 
are periodic, however, they can have different directions for 
different values of $\, {\bf a} \, $. As before, we have the embeddings
$$\left[ \hat{c}_{1} (\alpha_{m,n}, {\bf a}), \,
\hat{c}_{2} (\alpha_{m,n}, {\bf a}) \right] \,\,\, \subset \,\,\, 
\left[ c_{1} (\alpha_{m,n}), \, c_{2} (\alpha_{m,n}) \right] $$

 In exactly the same way as in the previous section, the following 
statements can also be formulated here.

\vspace{1mm}

\noindent
{\bf Lemma 3.1}

 For a function $F ({\bf z}, \alpha) \, $ satisfying condition 
(\ref{NablaZRel}), the width of the interval 
$\, \left[ c_{1} (\alpha_{m,n}), \, c_{2} (\alpha_{m,n}) \right] \, $ 
for the ``magic'' angle $\, \alpha_{m,n} \, $ does not exceed 
$\, 2 C_{1} T / \sqrt{3 (m_{0}^{2} + n_{0}^{2} + m_{0} n_{0})} \, $.

\vspace{1mm}

\noindent
{\bf Lemma 3.2}

 For a function $F ({\bf z}, \alpha) \, $ satisfying condition 
(\ref{NablaZRel}), the width of the intervals 
$\, \left[ \hat{c}_{1} (\alpha_{m,n}, {\bf a}), \, 
\hat{c}_{2} (\alpha_{m,n}, {\bf a}) \right] \, $ does not exceed 
$\, C_{1} T / \sqrt{3 (m_{0}^{2} + n_{0}^{2} + m_{0} n_{0})} \, $.

\vspace{1mm}

 Similar to the case of the 4th order symmetry, the following 
statement can also be proven here.

\vspace{1mm}
 
\noindent
{\bf Lemma 3.3}

 Let the potential $\, V (x, y) \, $ have rotational symmetry 
at the angle $\, \alpha_{0} = 60^{\circ} \, $ or $\, 120^{\circ}$ 
and periods $\, {\bf e}_{1} \, $ and $\, {\bf e}_{2} \, $ such that
$$\left| {\bf e}_{1} \right| \,\, = \,\, \left| {\bf e}_{2} \right|
\,\, = \,\, L \,\,\, , \quad 
\pi_{\alpha_{0}} \left[ {\bf e}_{1} \right] \,\, = \,\, 
{\bf e}_{2} $$

 Let 
$$\left| \nabla \, V (x, y) \right| \,\, < \,\, C_{1} $$
and $\, V (x, y) \, $ have a ``singular periodic net'' at the level
$$V (x, y) \,\,\, = \,\,\, c_{0} $$

 Then (for any $\, \Delta c > 0 $) the diameter of any connected region 
of the set $\, \Omega^{-}_{c_{0} - \Delta c} \left[ V \right] \, $ 
(as well as $\, \Omega^{+}_{c_{0} + \Delta c} \left[ V \right]$) 
does not exceed
$$\sqrt{C_{1} L \over \Delta c} \,\, L $$

\vspace{1mm}

 As with the proof of Lemma 2.3, we give the proof of this lemma 
in the Appendix.

 \vspace{1mm}

\vspace{1mm} 

\noindent
{\bf Theorem 3.1}

  Let $\, V_{1} (x, y) \, $ and $\, U (x, y) \, $ be periodic 
potentials with the same periods, symmetric with respect to 
rotation by $60^{\circ}$ or $120^{\circ}$. Then

\vspace{1mm} 

\noindent
1) For all potentials $\, V (x, y, \, \alpha, \, {\bf a} ) $, 
corresponding to generic (not ``magic'') angles $\, \alpha \, $, 
we have the relation 
$$c_{1} (\alpha) \,\,\, = \,\,\, c_{2} (\alpha) \,\,\, = \,\,\, 
c_{0} (\alpha) $$

\vspace{1mm} 

\noindent
2) The functions $\, c_{1} (\alpha) \, $ and $\, c_{2} (\alpha) \, $ 
are continuous at generic angles $\, \alpha \, $ (where they both 
coincide with $\, c_{0} (\alpha)$). In particular, the function 
$\, c_{0} (\alpha) \, $, considered on the set of generic 
$\, \alpha \, $ (not ``magic''), is continuous on this set.

\vspace{1mm} 

 The proofs of Lemmas 3.1 and 3.2, as well as Theorem 3.1, 
almost literally repeat the arguments given in the previous section.

\section{Potentials with incommensurate periods}
\setcounter{equation}{0}

\begin{figure}[t]
\begin{center}
\includegraphics[width=\linewidth]{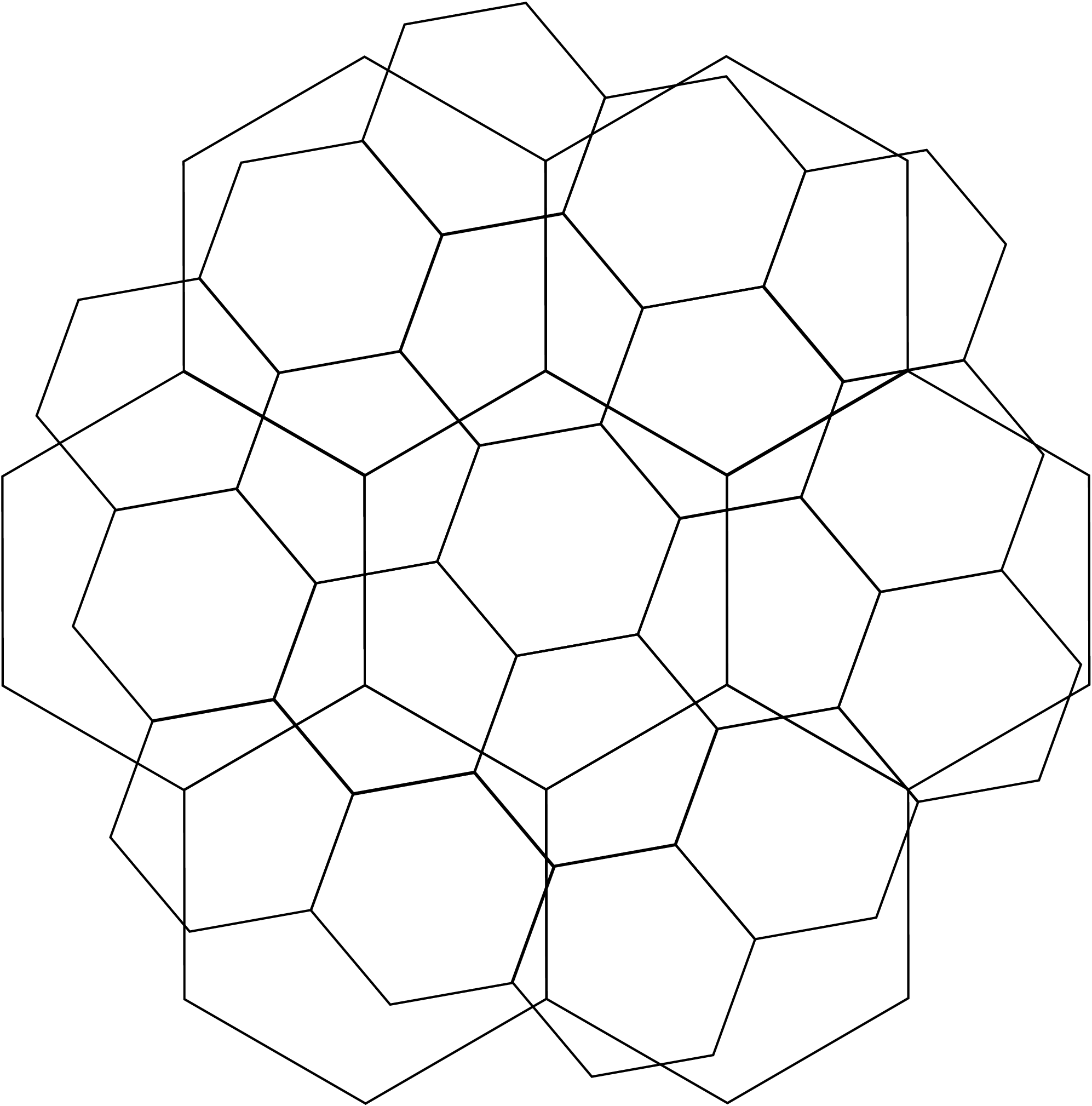}
\end{center}
\caption{Incommensurate lattices of potentials 
$\, V_{1} \, $ and $\, V_{2} \, $, possessing the same 
rotational symmetry (schematically).}
\label{DifPeriods}
\end{figure}

 Potentials $\, V_{1} \, $, $\, V_{2} \, $, with the 
same rotational symmetry and incommensurate periods 
($T$ and $T^{\prime}$) in $\, \mathbb{R}^{2} \, $ 
(Fig. \ref{DifPeriods}) are also very important in the  
physics of two-dimensional systems.

 Superpositions of such potentials correspond to two-layer 
systems containing layers of different substances with the 
same rotational symmetry. In this case, the ``magic'' angles 
no longer play such an important role as in the case of 
commensurate periods, and none of the potentials 
$\, V (x, y, \, \alpha, \, {\bf a}) \, $ is periodic.

 For quasi-periodic potentials, however, we can also formulate 
here analogs of the statements given in the previous sections, 
since the approximation of potentials 
$\, V (x, y, \, \alpha, \, {\bf a}) \, $ by periodic potentials 
using ``magic'' angles is not the only possible one. 
For simplicity, we consider here the case of rotational symmetry 
of the fourth order (other cases are considered in a completely 
similar way). Let, for definiteness, $\, T^{\prime} < T \, $.

\begin{figure}[t]
\begin{center}
\includegraphics[width=\linewidth]{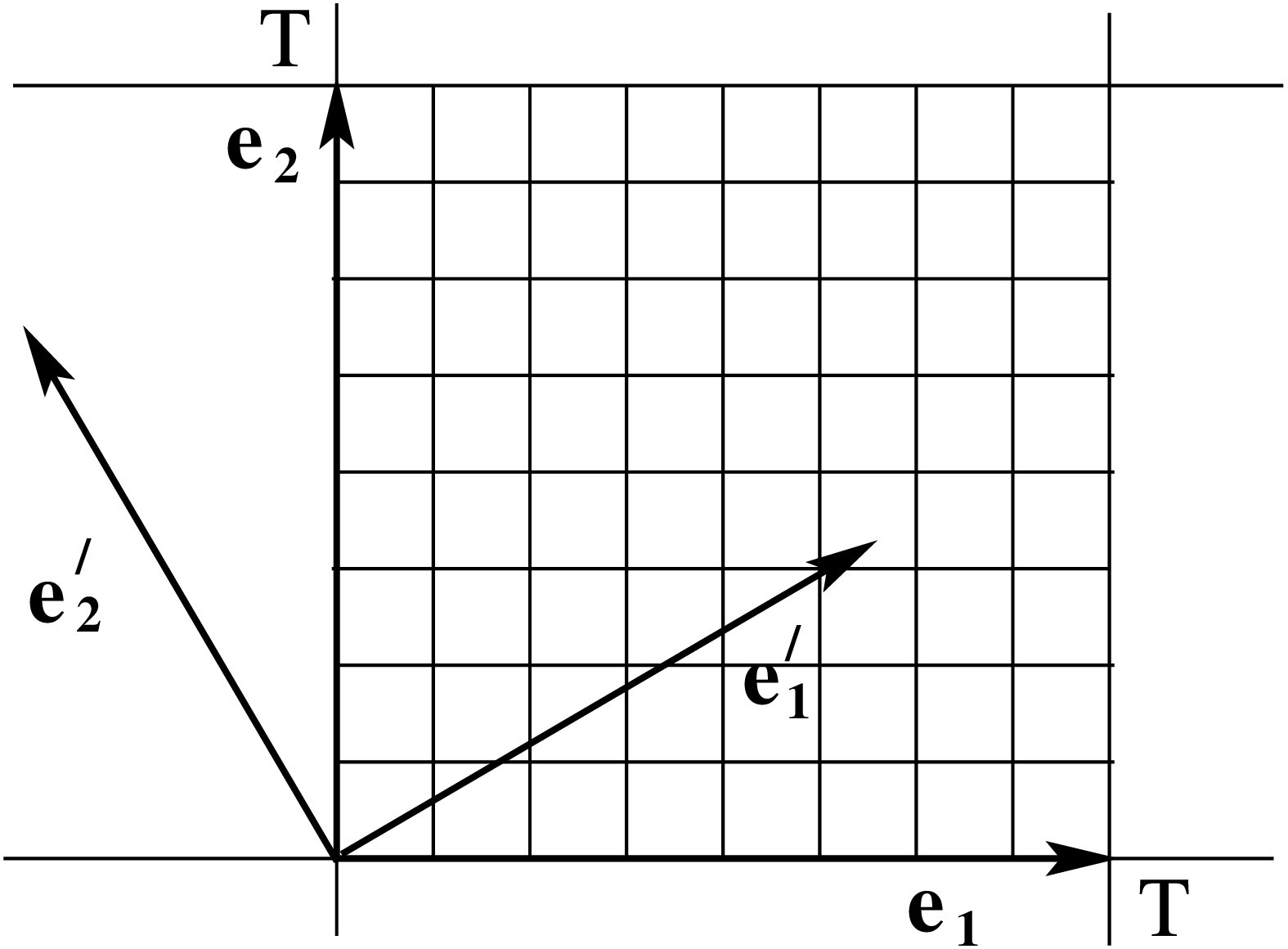}
\end{center}
\caption{Periods of potentials 
$\, V_{1} \, $ ($\{ {\bf e}_{1}, {\bf e}_{2} \}$) and 
$\, V_{2} \, $ ($\{ {\bf e}^{\prime}_{1}, {\bf e}^{\prime}_{2} \}$) 
and the fundamental domain of potential $\, V_{1} \, $, divided 
into $\, q^{2} \, $ cells with side $\, T/q \, $.}
\label{Appr}
\end{figure}

 Indeed, following the proof of Dirichlet's theorem, let us 
divide the fundamental domain for $\, V_{1} \, $ into 
$\, q^{2} \, $ cells with side $\, T/q \, $ (Fig. \ref{Appr}) 
and consider all integer linear combinations of the periods of 
the potential $\, V_{2} \, $
$$ \tilde{m}^{1} {\bf e}^{\prime}_{1} \, + \, 
\tilde{m}^{2} {\bf e}^{\prime}_{2} $$
modulo the periods of $\, V_{1} \, $.
The circle of radius $\, q \, $ certainly contains more than 
$\, q^{2} \, $ different integer vectors 
$\, \tilde{\bf m} \, $, $\,\, | \tilde{\bf m} | \leq q \, $, 
so that for any $\, q \in \mathbb{N} \, $ there exist 
$\, m^{1}, m^{2}, n^{1}, n^{2} \, \in \, \mathbb{Z} \, $, such that
\begin{multline*}
\left| m^{1} {\bf e}^{\prime}_{1} \, + \, m^{2} {\bf e}^{\prime}_{2} 
\, - \, n^{1} {\bf e}_{1} \, - \, n^{2} {\bf e}_{2} \right| 
\,\,\, < \,\,\, \sqrt{2} T / q \,\,\, ,  \\
\left| {\bf m} \right| \,\, = \,\, \sqrt{(m^{1})^{2} + (m^{2})^{2}} 
\,\,\, \leq \,\,\, 2 q   
\end{multline*}

 For $\, T^{\prime} < T \, $ and sufficiently large $\, q \, $ 
we can also write
$$\left| {\bf n} \right| \,\,\, = \,\,\, \sqrt{(n^{1})^{2} + (n^{2})^{2}}
\,\,\, < \,\,\, \left| {\bf m} \right| \,\,\, \leq \,\,\,  2 q $$

 This immediately implies the existence of a sequence of (integer) 
vectors $\, {\bf m}_{(s)} \, $, $\, {\bf n}_{(s)} \, $, such that
\begin{multline*}
\left| m_{(s)}^{1} {\bf e}^{\prime}_{1} \, + \, 
m_{(s)}^{2} {\bf e}^{\prime}_{2} 
\, - \, n_{(s)}^{1} {\bf e}_{1} \, - \, 
n_{(s)}^{2} {\bf e}_{2} \right|  \,\,\, < \,\,\, 
{ 2 \sqrt{2} T \over  | {\bf n}_{(s)} | } \,\,\, ,  \\
| {\bf m}_{(s)} | \, \rightarrow \, \infty \,\,\, , \quad
| {\bf n}_{(s)} | \, \rightarrow \, \infty
\end{multline*}

 Let us now fix an arbitrary point 
$\, (x_{0}, y_{0}) \in \mathbb{R}^{2} \, $ and consider 
the rotation 
$$(x, y) \,\,\, \rightarrow \,\,\, 
\hat{\pi}_{\delta \alpha_{(s)}} (x, y) $$ 
centered at $\, (x_{0}, y_{0}) \, $, by the minimum angle 
$\, \delta \alpha_{(s)} \, $ that maps the direction 
$\, n_{(s)}^{1} {\bf e}_{1} \, + \, n_{(s)}^{2} {\bf e}_{2} \, $ 
to the direction 
$\, m_{(s)}^{1} {\bf e}^{\prime}_{1} \, + \, 
m_{(s)}^{2} {\bf e}^{\prime}_{2} \, $ 
(here we use a ``hat'' over the operator to distinguish it 
from a rotation centered at the origin).

 Based on the above relations, for sufficiently large 
$\, | {\bf n}_{(s)} | \, $, we can, for example, use the estimate
$$\left| \delta \alpha_{(s)} \right| \,\,\, < \,\,\, 
{3 \over | {\bf n}_{(s)} |^{2}} $$ 

 Then, consider the minimal stretching (compression) 
in $\, \mathbb{R}^{2} \, $ centered at $\, (x_{0}, y_{0}) \, $
\begin{multline*}
\hat{\sigma}_{\delta \lambda_{(s)}} \, : \,\, (x, y) \,\,\, \rightarrow  \\
\rightarrow \,\,\,
\Big( x_{0} \, + \, \left(1 + \delta \lambda_{(s)}\right) (x - x_{0}) \, , \,\, 
\, y_{0} \, + \left(1 + \delta \lambda_{(s)}\right) (y - y_{0}) \Big) 
\end{multline*}
such that
$$\left(1 + \delta \lambda_{(s)}\right) \,\, | {\bf n}_{(s)} | \,\, T 
\,\,\,\,\, = \,\,\,\,\, |{\bf m}|_{(s)} \,\, T^{\prime} $$
($| \delta \lambda_{(s)} | \, < \, 2 \sqrt{2} / | {\bf n}_{(s)} |^{2}$). 

\vspace{1mm}

 The potential
$$V^{(s)}_{2} (x, y) \,\,\, = \,\,\, 
V_{2} \Big( \hat{\sigma}_{\delta \lambda_{(s)}} \circ 
\hat{\pi}_{\delta \alpha_{(s)}} (x, y) \Big) \,\,\, , $$
is obviously periodic with periods
$$\sigma_{\delta \lambda_{(s)}}^{-1} \circ 
\pi_{\delta \alpha_{(s)}}^{-1} ({\bf e}^{\prime}_{1}) \quad \text{and}
\quad \sigma_{\delta \lambda_{(s)}}^{-1} \circ 
\pi_{\delta \alpha_{(s)}}^{-1} ({\bf e}^{\prime}_{2}) $$

 Let us now consider the potential 
$\, V^{(s)}_{(x_{0}, y_{0})} (x, y) \, $,
defined by the superposition of $\, V_{1} (x, y) \, $ 
and $\, V^{(s)}_{2} (x, y) \, $. Before doing so, however, 
we make one remark.

 Considering the potentials $\, V (x, y, \, \alpha, \, {\bf a}) \, $ 
as quasiperiodic functions, we used the embedding 
$\, \mathbb{R}^{2} \subset \mathbb{R}^{4} \, $ given by the 
formulas (\ref{Emb}) - (\ref{Axy}). Adding the stretching operation, 
we now define the operator $\, A \, $ by the more general formula
\begin{equation}
\label{newAxy}
A (x, y) \,\,\, = \,\,\, \lambda^{-1} \,\, 
\pi_{- \alpha} (x, y) \,\, - \,\, {\bf a} 
\end{equation}

 In cases (1) and (2), the function $\, F ({\bf z}) \, $, as before, 
is defined by formulas (\ref{F1}) and (\ref{F2}), respectively. 
In case (3), however, the function $\, F \, $ was defined more 
complexly and could also contain a dependence on the angle $\, \alpha \, $.

 The situation when the layers undergo additional stretching and 
the superposition of the corresponding potentials is determined 
by a complex functional may be also important in physical systems
(see e.g. \cite{PaulCrowleyFu} and references therein).
As is easy to see, in this case only isotropic stretching 
corresponds to the presence of rotational symmetry. In this 
situation, the function $\, F \, $ may also depend on the 
parameter $\, \lambda \, $ and be of independent interest. 
Its periods, as before, are determined by the periods of the 
potentials $\, V_{1} (x, y) \, $ and $\, U (x, y) \, $. For 
sufficiently regular functionals defining the superposition 
of $\, V_{1} \, $ and $\, V_{2} \, $, in addition to the 
relations (\ref{NablaZRel}) - (\ref{DerAlphaRel}), the 
condition
\begin{equation}
\label{C3Usl}
\left| \partial_{\lambda} F ({\bf z}, \, \alpha , \, \lambda ) 
\right| \,\,\, \leq \,\,\, C_{3} 
\end{equation}
is also satisfied. 
 
 If, however, this situation is not interesting from the 
physical point of view, and we use only a formal construction 
for approximating the potentials 
$\, V (x, y, \, \alpha, \, {\bf a}) \, $ by periodic potentials, 
we can consider the same function 
$\, F \, = \, F ({\bf z}, \, \alpha ) \, $ and the 
operator (\ref{newAxy}) for all values of $\, \lambda \, $.

\vspace{1mm}

 It is easy to see that the potential
$\, V^{(s)}_{(x_{0}, y_{0})} (x, y) \, $ is periodic with periods
\begin{equation}
\label{nPeriods}
n_{(s)}^{1} {\bf e}_{1} \, + \, n_{(s)}^{2} {\bf e}_{2} \,\,\,
\quad \text{and} \quad
- n_{(s)}^{2} {\bf e}_{1} \, + \, n_{(s)}^{1} {\bf e}_{2} 
\end{equation}

 Without loss of generality, we can assume that the periods 
(\ref{nPeriods}) are minimal for the potentials 
$\, V^{(s)}_{(x_{0}, y_{0})} (x, y) \, $.

 Indeed, the existence of common periods for potentials 
$\, V_{1} (x, y) \, $ and $\, V^{(s)}_{2} (x, y) \, $, 
smaller than (\ref{nPeriods}), implies the existence of 
integer pairs 
$\, (\widetilde{\bf m}_{(s)}, \widetilde{\bf n}_{(s)}) \, $, 
satisfying the conditions
\begin{multline}
\label{newcond}
\left| \widetilde{m}_{(s)}^{1} {\bf e}^{\prime}_{1} \, + \, 
\widetilde{m}_{(s)}^{2} {\bf e}^{\prime}_{2} 
\, - \, \tilde{n}_{(s)}^{1} {\bf e}_{1} \, - \, 
\tilde{n}_{(s)}^{2} {\bf e}_{2} \right|  \,\,\, < \,\,\, 
{ 2 \sqrt{2} T \over  | {\bf n}_{(s)} | } \,\,\, ,  \\
| \widetilde{\bf m}_{(s)} | \, < \, | {\bf m}_{(s)} |  \,\,\, , \quad
| \widetilde{\bf n}_{(s)} | \, < \, | {\bf n}_{(s)} | 
\end{multline}

 Taking such a pair with the smallest value of 
$\, \widetilde{\bf n}_{(s)} \, $, we can use it instead of 
$\, ({\bf m}_{(s)}, {\bf n}_{(s)}) \, $, while the first of 
the conditions (\ref{newcond}) also implies the property
$$| \widetilde{\bf m}_{(s)} | \, \rightarrow \, \infty \,\,\, , \quad
| \widetilde{\bf n}_{(s)} | \, \rightarrow \, \infty $$

\vspace{1mm}

 From the relations (\ref{NablaZRel}) - (\ref{VCoordRel}), 
we get in the circle of radius 
$\, \lambda \, T^{(s)} \, = \, \lambda \, |{\bf n}_{(s)}| \, T \, $ 
centered in $\, (x_{0}, y_{0}) \, $ 
\begin{equation}
\label{VVsrel}
\left| V (x, y, \, \alpha, \, {\bf a}) \, - \, 
V^{(s)}_{(x_{0}, y_{0})} (x, y) \right| \,\,\,\,\, \leq \,\,\,\,\,
\widehat{C} \lambda \,\,  {T \over |{\bf n}_{(s)}| } 
\end{equation}
($\forall \lambda > 0$) with some (universal for the family 
$\, V (x, y, \, \alpha, \, {\bf a})$) constant $\, \widehat{C} > 0 \, $.

 In the ``formal'' definition of potentials 
$\, V^{(s)}_{(x_{0}, y_{0})} (x, y) \, $ the constant 
$\, \widehat{C} \, $ is determined from the relations 
(\ref{NablaZRel}) - (\ref{VCoordRel}). Thus, for
$$\left| \delta \alpha_{(s)} \right| \,\, < \,\, 
{3 \over | {\bf n}_{(s)} |^{2}} \,\,\, , \quad 
\left| \delta \lambda_{(s)} \right| \,\, < \,\, 
{2 \sqrt{2} \over | {\bf n}_{(s)} |^{2}} \,\,\, , $$
we can put
$$\widehat{C} \,\, = \,\, \left( 3 + 2 \sqrt{2} \right) C_{1} $$

 If we are also interested in the possibility of additional 
stretching of layers in physical systems, it is natural to 
include also the constant $\, C_{3} \, $ in the constant 
$\, \widehat{C} \, $.

 The potentials $ V (x, y) \, $ formed by superpositions of 
the potential $\, V_{1} (x, y) \, $ and 
$\, V^{(s)}_{2} (x - a^{1}, \, y - a^{2}) \, $ define a family 
of periodic potentials 
$ \, V^{(s)}_{(x_{0}, y_{0})} ({\bf r}, \, {\bf a}) \, $, 
which is in many ways similar to those considered earlier. 
As before, we have here the identities
$$V^{(s)}_{(x_{0}, y_{0})} \left( {\bf r}, \,\, 
{\bf a} + \sigma_{\delta \lambda_{(s)}}^{-1} \circ 
\pi_{\delta \alpha_{(s)}}^{-1} ({\bf e}^{\prime}_{i}) \right) 
\,\,\, \equiv \,\,\, V^{(s)}_{(x_{0}, y_{0})} 
\left( {\bf r}, \, {\bf a} \right) $$
$$V^{(s)}_{(x_{0}, y_{0})} \left( {\bf r}, \,\, 
{\bf a} + {\bf e}_{i} \right) \,\,\, \equiv \,\,\, 
V^{(s)}_{(x_{0}, y_{0})} \left( {\bf r} - {\bf e}_{i}, \,\, 
{\bf a} \right) $$

 Thus, we can also introduce classes of equivalent potentials 
on the set of 
$ \, V^{(s)}_{(x_{0}, y_{0})} ({\bf r}, \, {\bf a}) \, $, 
which are defined by the conditions
$$V^{(s)}_{(x_{0}, y_{0})} ({\bf r}, \, {\bf a}) 
\,\,\,\,\, \cong \,\,\,\,\, V^{(s)}_{(x_{0}, y_{0})} 
\left( {\bf r}, \,\, {\bf a} + {\bf A}_{klpq} \right) \,\,\, , $$
where 
\begin{multline}
\label{Avectors}
{\bf A}_{klpq} \,\, =   \\ 
= \,\, k \, \sigma_{\delta \lambda_{(s)}}^{-1} \! \circ \, 
\pi_{\delta \alpha_{(s)}}^{-1} ({\bf e}^{\prime}_{1}) \,\, + \,\, 
l \, \sigma_{\delta \lambda_{(s)}}^{-1} \! \circ \, 
\pi_{\delta \alpha_{(s)}}^{-1} ({\bf e}^{\prime}_{2}) \,\, + \,\,
p \, {\bf e}_{1} \,\, + \,\, q \, {\bf e}_{2} 
\end{multline}
($k, l, p, q \, \in \, \mathbb{Z}$). 

 In particular, we can distinguish a class 
$\, Q^{(s)} (\alpha) \, $ of equivalent potentials that have exact 
rotational symmetry and a ``singular net'' of level lines at the 
same energy level $\, c^{(s)}_{0} (\alpha) \, $.

 Vectors (\ref{Avectors}) form a (rotated) square lattice in 
the space of parameters $\, {\bf a} \, $ with the step
$${T \over \left| {\bf m}_{(s)} \right|}  \quad \leq \quad
{T \over \left| {\bf n}_{(s)} \right|} $$

 In particular, together with the potential 
$\, V^{(s)}_{(x_{0}, y_{0})} ({\bf r}) \, $, there also 
exists a potential
$${\widetilde V}^{(s)}_{(x_{0}, y_{0})} ({\bf r}) \,\,\, = \,\,\, 
V^{(s)}_{(x_{0}, y_{0})} ({\bf r}, \, {\bf a}_{(s)}) 
\,\,\, \in \,\,\, Q^{(s)} (\alpha) \,\,\, , $$
with exact rotational symmetry, such that
$$\left| {\bf a}_{(s)} \right| \quad \leq \quad
{T \over \sqrt{2} \left| {\bf n}_{(s)} \right|} $$
(we note here that the center of rotational symmetry of 
the potential 
$\, {\widetilde V}^{(s)}_{(x_{0}, y_{0})} ({\bf r}) \, = \, 
V^{(s)}_{(x_{0}, y_{0})} ({\bf r}, \, {\bf a}_{(s)}) \, $ 
can be significantly distant from the point $\, (x_{0}, y_{0}) $).

 Using the relations (\ref{NablaZRel}) - (\ref{VCoordRel}), 
we can write in this case
$$\left|  V^{(s)}_{(x_{0}, y_{0})} ({\bf r}) \, - \, 
{\widetilde V}^{(s)}_{(x_{0}, y_{0})} ({\bf r}) \right| 
\quad \leq \quad {C_{1} T \over \sqrt{2} \, |{\bf n}_{(s)}|} $$

 Combining this relation with the relation (\ref{VVsrel}), 
and assuming that $\, \lambda \, $ and $\, |{\bf n}_{(s)}| \, $ 
are sufficiently large, we can formulate the following lemma.
 
\vspace{1mm}

\noindent
{\bf Lemma 4.1} 

 For every potential $\, V (x, y, \, \alpha, \, {\bf a}) \, $ 
and every fixed point $\, (x_{0}, y_{0}) \in \mathbb{R}^{2} \, $ 
there exists a sequence of integer vectors
$${\bf n}_{(s)} (\alpha) \,\,\, = \,\,\, 
\left( n^{1}_{(s)}, \, n^{2}_{(s)} \right) \,\,\, , \quad
\left| {\bf n}_{(s)} \right| \rightarrow \infty $$
and potentials
$\, {\widetilde V}^{(s)}_{(x_{0}, y_{0}, \alpha, {\bf a})} (x, y) \, $,
periodic with periods (\ref{nPeriods}) and having exact rotational 
symmetry, such that: 

\vspace{1mm}

\noindent
1) In the circle of radius
$$\lambda \, T^{(s)} \,\,\, = \,\,\, 
\lambda \, \left| {\bf n}_{(s)} \right| \, T \,\,\, , $$
centered at $\, (x_{0}, y_{0}) \, $, we have the relation
\begin{multline*}
\left| V (x, y, \, \alpha, \, {\bf a}) \, - \, 
{\widetilde V}^{(s)}_{(x_{0}, y_{0}, \alpha, {\bf a})} (x, y) 
\right| \quad \leq  \\
\leq \quad {\widetilde C} \lambda \,\, 
{T^{(s)} \over \left| {\bf n}_{(s)} \right|^{2} } \quad = \quad 
{\widetilde C} \lambda \,\, {T \over \left| {\bf n}_{(s)} \right| } 
\end{multline*}
(with some constant $\, {\widetilde C} \, $, universal for the 
family $\, V (x, y, \, \alpha, \, {\bf a})$).

\noindent
2) All potentials 
$\, {\widetilde V}^{(s)}_{(x_{0}, y_{0}, \alpha, {\bf a})} (x, y) \, $ 
have a ``singular periodic net'' at the level 
$\, c \, = \, c^{(s)}_{0} (\alpha) \, $.

\vspace{1mm}

 All the above reasoning can be reproduced literally also in 
the case of rotational symmetry of order 3 or 6. The fundamental 
domain of the potential $\, V_{1} \, $ now has the form 
of a parallelogram, and the lattice of periods is a regular 
triangular one. In other details, the construction of a sequence of 
potentials 
$\, {\widetilde V}^{(s)}_{(x_{0}, y_{0}, \alpha, {\bf a})} (x, y) \, $, 
approximating a given potential $\, V (x, y, \, \alpha, \, {\bf a}) \, $, 
is no different from the scheme given above. The periods of 
$\, {\widetilde V}^{(s)}_{(x_{0}, y_{0}, \alpha, {\bf a})} (x, y) \, $ 
are now given by the vectors
$$n_{(s)}^{1} {\bf e}_{1} \, + \, n_{(s)}^{2} {\bf e}_{2} \,\,\,
\quad \text{and} \quad
- n_{(s)}^{2} {\bf e}_{1} \, + \,
\left( n_{(s)}^{1} + n_{(s)}^{2} \right) {\bf e}_{2} \,\,\, , $$
where the vectors $\, {\bf e}_{1} \, $ and $\, {\bf e}_{2} \, $ 
are determined by the relations (\ref{3OrdBas}).

 The value of the period $\, T^{(s)} \, $ is equal here
$$\sqrt{\left( n_{(s)}^{1} \right)^{2} \, + \, 
\left( n_{(s)}^{2} \right)^{2} \, + \, 
n_{(s)}^{1} n_{(s)}^{2}} \,\, \cdot \,\, T \,\,\, , $$
so it is natural to put now
$$\left| {\bf n}_{(s)} \right| \,\,\, = \,\,\, 
\sqrt{\left( n_{(s)}^{1} \right)^{2} \, + \, 
\left( n_{(s)}^{2} \right)^{2} \, + \, n_{(s)}^{1} n_{(s)}^{2}} $$
and $\, T^{(s)} \, = \, \left| {\bf n}_{(s)} \right| \, \cdot \, T \, $.

 Here we can also use the same estimates for $\, \widehat{C} \, $ 
and $\, \widetilde{C} \, $, taking the same estimates for 
$\, \left| \delta \alpha_{(s)} \right| \, $ and 
$\, \left| \delta \lambda_{(s)} \right| \, $
(although in the case of $\, \alpha_{0} = 60^{\circ} \, $ 
or $\, 120^{\circ} \, $ it is easy to write better estimates 
for these quantities).

\vspace{1mm}

 Each of the potentials 
$\, {\widetilde V}^{(s)}_{(x_{0}, y_{0}, \alpha, {\bf a})} (x, y) \, $ 
has a ``singular periodic net'' at the level 
$\, c_{0}^{(s)} (\alpha) \, $. According to Lemma 2.3 or Lemma 3.3, 
the diameter of any connected component of the set
$$\Omega^{-}_{c_{0}^{(s)} (\alpha) - \Delta c} \left[ 
{\widetilde V}^{(s)}_{(x_{0}, y_{0}, \alpha, {\bf a})} 
\right] \,\,\, , $$
as well as the set
$$\Omega^{+}_{c_{0}^{(s)} (\alpha) + \Delta c} \left[ 
{\widetilde V}^{(s)}_{(x_{0}, y_{0}, \alpha, {\bf a})} 
\right] \,\,\, , $$
does not exceed
\begin{equation}
\label{Ds}
\sqrt{C T^{(s)} \over \Delta c} \,\, T^{(s)} \,\,\, = \,\,\, 
\sqrt{C T \over \Delta c} \,\, \left| {\bf n}_{(s)} \right|^{3/2} 
\,\, T
\end{equation}
($\Delta c > 0$) with some constant $\, C \, $, which is universal 
for the family $\, V (x, y, \, \alpha, \, {\bf a}) \, $.

\vspace{1mm}

 Similar to the previous sections, the above statements imply 
the relations
\begin{equation}
\label{c12Rel}
\left| c_{1,2} (\alpha) - c_{0}^{(s)} (\alpha) \right| 
\,\,\, \leq \,\,\, 
{(C + {\widetilde C}) \, T \over | {\bf n}_{(s)} |^{1/3}} 
\,\,\, \equiv \,\,\, \Delta_{(s)} 
\end{equation}
for the set of potentials
$$V (x, y, \, \alpha, \, {\bf a}) \,\,\, , \quad 
{\bf a} \, \in \, \mathbb{R}^{2} $$

\vspace{1mm}

 To prove them, we assume again that some potential $\, V \, $ 
from this set has an unbounded level line (singular or regular) 
at a level $\, c \, $. We need to show that
$$\left| c - c_{0}^{(s)} (\alpha) \right| \,\,\, \leq \,\,\, 
{(C + {\widetilde C}) \, T \over | {\bf n}_{(s)} |^{1/3}} $$

 Let, for example,
$$\Delta c \,\,\, = \,\,\, c_{0}^{(s)} (\alpha) \, - \, c 
\,\,\, > \,\,\, 
{(C + {\widetilde C}) \, T \over | {\bf n}_{(s)} |^{1/3}}
\,\,\, > \,\,\, 0 $$

 Consider an unbounded connected component of the set 
$\, \Omega^{-}_{c} \left[ V \right] \, $ adjacent to the 
unbounded level line $\, V = c \, $. Let this component 
contain some point $\, (x_{0}, y_{0}) \, $.

 Setting in Lemma 4.1 
$\,\, \lambda \, = \, \left| {\bf n}_{(s)} \right|^{2/3} \, $, 
we then have in the circle of radius 
$\, \left| {\bf n}_{(s)} \right|^{5/3} \, T \, $ 
with center $\, (x_{0}, y_{0}) \, $
$$\Omega^{-}_{c} \left[ V \right] \,\,\, \subset \,\,\, 
\Omega^{-}_{c_{0}^{(s)} (\alpha) - \overline{\Delta c}} \left[ 
{\widetilde V}^{(s)}_{(x_{0}, y_{0}, \alpha, {\bf a})} \right] 
\,\,\, , $$
where
$$\overline{\Delta c} \,\,\, = \,\,\, C T \Big/ 
| {\bf n}_{(s)} |^{1/3} $$

 However, as follows from (\ref{Ds}), any connected component 
of the set 
$$ \Omega^{-}_{c_{0}^{(s)} (\alpha) - \overline{\Delta c}} \left[ 
{\widetilde V}^{(s)}_{(x_{0}, y_{0}, \alpha, {\bf a})} \right] 
\,\,\, , $$
containing the point $\, (x_{0}, y_{0}) \, $, lies entirely in 
the circle of radius $\, \left| {\bf n}_{(s)} \right|^{5/3} \, T \, $ 
centered at this point, which implies the same property for the 
connected component of $\, \Omega^{-}_{c} \left[ V \right] \, $
containing the point $\, (x_{0}, y_{0}) \, $. The obtained 
contradiction proves the relation (\ref{c12Rel}) 
(similarly for the case $\, c \, > \, c_{0}^{(s)} (\alpha) \, $).

\vspace{1mm}

 From relations (\ref{c12Rel}), it obviously follows
\begin{equation}
\label{c2c1ns}
c_{2} (\alpha) \, - \, c_{1} (\alpha) \,\,\, \leq \,\,\, 
{2 (C + {\widetilde C}) \, T \over | {\bf n}_{(s)} |^{1/3}}
\end{equation}

 In the limit 
$\,\, \left| {\bf n} \right|_{(s)} \rightarrow \infty \,\, $ 
we thus obtain
$$c_{1} (\alpha) \,\,\, = \,\,\, c_{2} (\alpha) \,\,\, = \,\,\,
c_{0} (\alpha) $$

\vspace{1mm}

 As in the previous cases, for each $\, {\bf n}_{(s)} \, $ 
there also exists an open subset $\, O_{(s)} (\alpha) \, $ 
(in the space of angles) such that the relations (\ref{c12Rel}) 
are in fact satisfied for all 
$\, \alpha^{\prime} \in O_{(s)} (\alpha) \, $. 
Subsets $\, O_{(s)}  (\alpha) \, $ can, for example, 
be defined by the condition
$$\left| \alpha^{\prime} - \alpha_{(s)} \right| \,\,\, < \,\,\,
3 \Big/ \left| {\bf n}_{(s)} \right|^{2} \,\,\, , $$
where 
$$\alpha_{(s)} \,\,\, = \,\,\, \alpha \,\, + \,\, 
\delta \alpha_{(s)} $$

 The sets $\, O_{(s)} (\alpha) \, $ have then the following 
properties

\vspace{1mm} 
 
\noindent
1) All sets $\, O_{(s)} (\alpha) \, $ contain the angle 
$\, \alpha \, $;

\vspace{1mm} 

\noindent
2) Each set $\, O_{(s)} (\alpha) \, $ is associated with a segment
$$I_{(s)} (\alpha) \, = \, \Big[ c_{0}^{(s)} (\alpha) - \Delta_{(s)} 
\, , \,\, c_{0}^{(s)} (\alpha) + \Delta_{(s)} \Big] \, , $$
such that
$$\left[ c_{1} (\alpha^{\prime}) , \, c_{2} (\alpha^{\prime})
\right] \,\,\, \subset I_{(s)} (\alpha) $$
for all $\, \alpha^{\prime} \in O_{(s)} (\alpha) \, $;

\vspace{1mm}

\noindent
3) The length of $\, I_{(s)} (\alpha) \, $ tends to zero as 
$\, |{\bf n}_{(s)}| \rightarrow \infty \, $.

\vspace{1mm}

 Repeating the arguments from the previous sections and 
taking into account that now all potentials 
$\, V (x, y, \, \alpha, \, {\bf a}) \, $ are not periodic, 
we can formulate the following theorem here.

\vspace{1mm}

\noindent
{\bf Theorem 4.1} 

 Let $\, V_{1} (x, y) \, $ and $\, U (x, y) \, $ be potentials 
with incommensurate periods that have the same type of rotational 
symmetry. Then

\vspace{1mm}

\noindent
1) For all potentials $\, V (x, y, \, \alpha, \, {\bf a}) \, $ 
we have the relation
$$c_{1} (\alpha) \,\,\, = \,\,\, c_{2} (\alpha) \,\,\, = \,\,\,
c_{0} (\alpha) $$

\noindent
2) The function $\, c_{0} (\alpha) \, $ is a continuous function 
of the angle $\, \alpha \, $.

\vspace{1mm}

 As we have already noted earlier, the condition 
$\, c = c_{0} (\alpha) \, $ does not necessarily imply 
the existence of open level lines
\begin{equation}
\label{c0levels}
V (x, y, \, \alpha, \, {\bf a}) \,\,\, = \,\,\, c
\,\,\, = \,\,\, c_{0} (\alpha)
\end{equation}
for all values of $\, {\bf a} \, $. Open level lines
(\ref{c0levels}), however, must arise in this case at
least for some values of $\, {\bf a} \, $. It can also be 
shown that if open level lines do not arise for all 
$\, {\bf a} \, $, then each level (\ref{c0levels}) must 
also contain arbitrarily large closed components 
(see \cite{DynNov,DynMalNovUMN,BigQuas}). The set of
values $\, {\bf a} \, $ corresponding to the occurrence
of open level lines (\ref{c0levels}) then forms an 
everywhere dense subset in the space of parameters 
$\, {\bf a} \, $.

 The open level lines arising for the potentials considered 
here are (in our terminology) of the chaotic type. Their 
geometric description, however, requires further serious study. 
Certainly, the potentials $\, V (x, y, \, \alpha, \, {\bf a}) \, $ 
arising in real physical systems are of particular interest here 
(here we would especially like to cite the work 
\cite{TitovKatsnelson}, where such trajectories were studied 
from the point of view of percolation theory and where their 
geometry plays a very important role).

\section{Appendix}
\setcounter{equation}{0}

 In this section we give the proof of Lemmas 2.3 and 3.3, 
formulated in Sections 2 and 3. We will give here a detailed 
proof for the case of the fourth-order symmetry; other cases are 
treated in a completely similar way (and even somewhat simpler).

 In our situation, we have the relation
$$\left| F_{\max} - F_{\min} \right| \,\,\, \leq \,\,\,
C_{1} \, L \,\,\, , $$
so that for $\, \Delta c > C_{1} L \, $ the sets 
$\, \Omega^{-}_{c_{0} - \Delta c} \, [V] \, $ 
(and also $\, \Omega^{+}_{c_{0} + \Delta c} \, [V] $) are empty. 
We can thus immediately imply the relation 
$\, \Delta c \leq C_{1} L \, $ 
(in fact, we will be interested here in the limit 
$\, \Delta c / C_{1} L \, \rightarrow \, 0 $).

 First of all, we note that, due to the periodicity of the 
potential $\, V (x, y) \, $, each of the connected regions 
$\, \Omega \, $ of the set 
$\, \Omega^{-}_{c_{0} - \Delta c} \, [V] \, $ does not intersect 
its integer shifts in $\, \mathbb{R}^{2} \, $.

 The region $\, \Omega \, $ can either have rotational symmetry 
or pass into another region by rotating by 
$\, \alpha_{0} \, $ around any of the rotational symmetry centers. 
In the latter case, we obviously have 
(for $\, \alpha_{0} = 90^{\circ}$) either two or four families 
$$\Omega^{(i)} \,\,\, + \,\,\, k \, {\bf e}_{1} 
\,\,\, + \,\,\, l \, {\bf e}_{2} \,\,\, , \quad \quad
k, l \, \in \, \mathbb{Z} \,\,\, , $$
that are disjoint from each other (we use the notation 
$\, {\cal M} + {\bf a} \, $ for the shift of any set 
${\cal M} \subset \mathbb{R}^{2} \, $ by a vector $\, {\bf a}$).

\begin{figure}[t]
\begin{center}
\includegraphics[width=\linewidth]{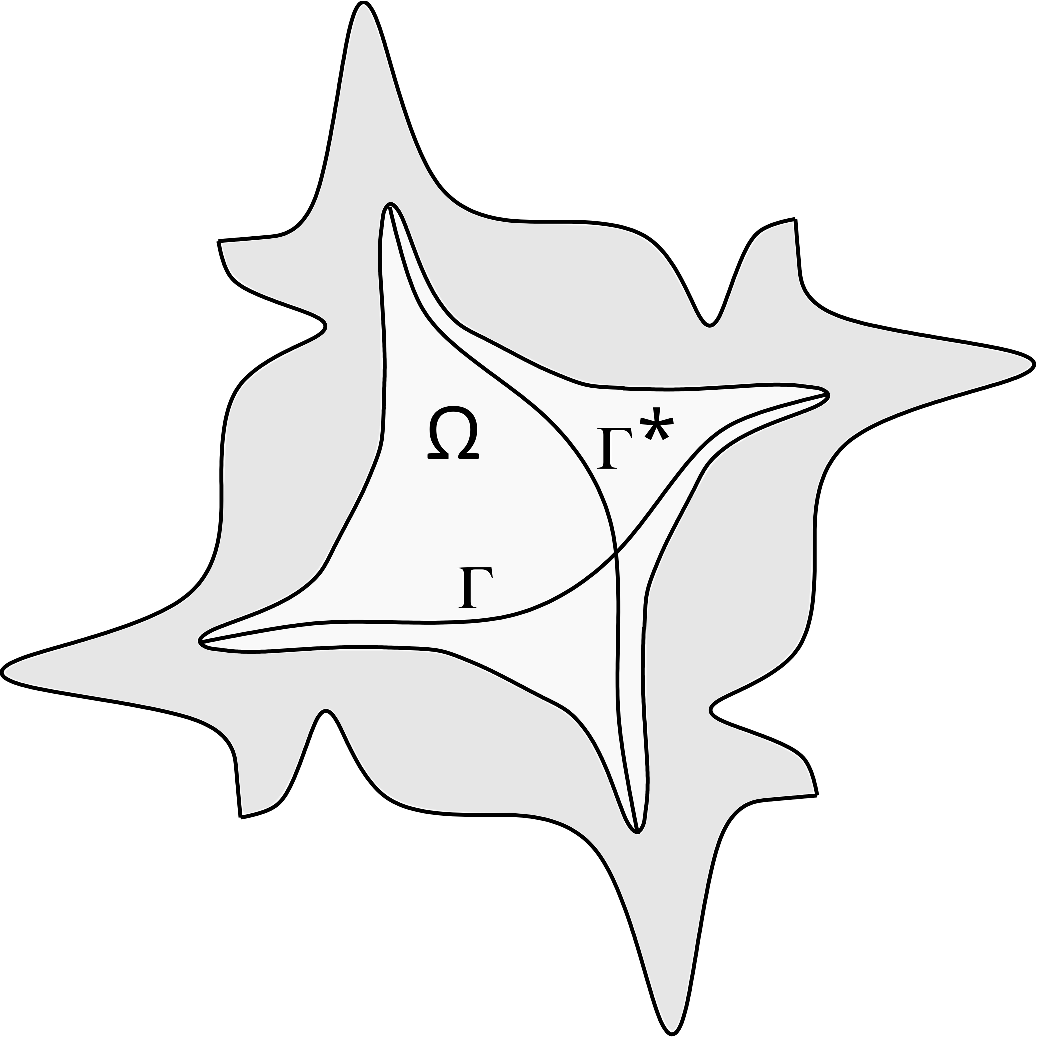}
\end{center}
\caption{A graph $\, G = \Gamma \cup \Gamma^{*} \, $ lying 
in a region $\, \Omega \, $ inside a symmetric cell 
of a ``singular net''.}
\label{Graph}
\end{figure}

 Let us consider the first of the above cases. Let us 
consider 4 points on the boundaries of the region 
$\, \Omega \, $, the most distant from its center and 
passing into each other under the rotation 
by $\, 90^{\circ} \, $. Let $\, \Gamma \, $ be a path 
connecting 2 diametrically opposite points, lying entirely 
in the domain $\, \Omega \, $, and $\, \Gamma^{*} \, $ be 
its rotation by $\, 90^{\circ} \, $ relative to the 
center of $\, \Omega \, $ (Fig. \ref{Graph}). It is easy 
to show that $\, \Gamma \, $ and $\, \Gamma^{*} \, $ can be 
chosen to be non-self-intersecting and intersecting each 
other at only one point. The distances between the ends 
of $ \, \Gamma \, $ and $\, \Gamma^{*} \, $ are obviously 
equal to the diameter $\, d \, $ of the domain $\, \Omega \, $.

 It can be seen that integer shifts of the graph 
$\, G = \Gamma \cup \Gamma^{*} \, $ 
do not intersect each other. In particular, the mapping of $\, G \, $ 
onto its image in the torus $\, \mathbb{T}^{2} \, $ under the 
factorization
$$\mathbb{R}^{2} \big/ \left\{ k \, {\bf e}_{1} 
\, + \, l \, {\bf e}_{2} \right\} \,\,\, = \,\,\, 
\mathbb{T}^{2} $$
is one-to-one.

\begin{figure}[t]
\begin{center}
\includegraphics[width=\linewidth]{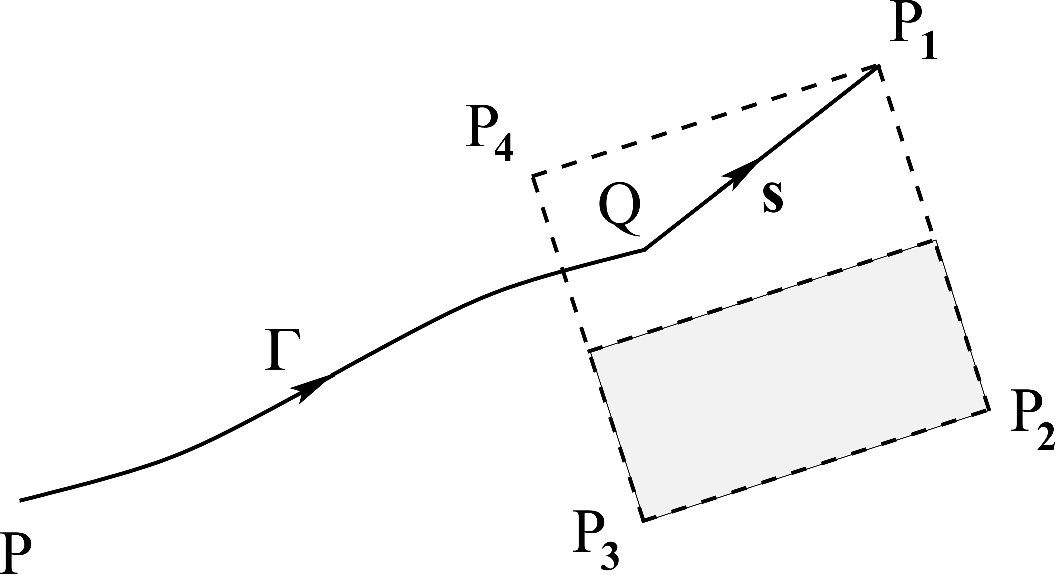}
\end{center}
\caption{The oriented path $\, \Gamma \, $ and the cycle
$\, \widehat{\Gamma} \, = \, \Gamma \cup {\bf s} \, $.}
\label{Gamma}
\end{figure}

 Consider the path $\, \Gamma \, $, 
oriented as shown in Fig. \ref{Gamma}. Let the point $\, P \, $ be the 
beginning of $\, \Gamma \, $, and the point $\, Q \, $ be its end.

 Let the point $\, Q \, $ belong to a fundamental domain 
$\, \square \, P_{1} P_{2} P_{3} P_{4} \, $, defined by integer 
shifts of the point $\, P \, $. It is easy to see that the 
point $\, P_{1} \, $ can be chosen in such a way that
\begin{equation}
\label{PP1}
\left| Q P_{1} \right| \, < \, \sqrt{5} L / 2 \quad  \text{and} \quad 
\left| P P_{1} \right| \, \geq \, \left| P Q \right|
\end{equation}
 
 More precisely, the domain 
$\, \square \, P_{1} P_{2} P_{3} P_{4} \, $ can be divided 
in half, so that the segment $\, Q P_{1} \, $ is entirely contained 
in one of the halves, while its projections onto the basis vectors 
do not exceed $\, L \, $ and $\, L / 2 \, $ (Fig. \ref{Gamma}).
 
 Let us connect the point $\, Q \, $ with the point $\, P_{1} \, $ 
by the oriented segment 
$\, \overrightarrow{Q P_{1}} \, = \, {\bf s} \, $ and define 
an oriented cycle $\, \widehat{\Gamma} \, = \, \Gamma \cup {\bf s} \, $ 
(Fig. \ref{Gamma}).
 
 Let, for any figure $\, {\cal F} \subset \mathbb{R}^{2} \, $, 
$\, \pi \left[ {\cal F} \right] \, $ denote its rotation by the 
angle $\, 90^{\circ} \, $ around the origin. We will also use 
the same notation for any figure
$\, {\cal F} \subset \mathbb{T}^{2} \, $.

 We have the well-known relations for the intersection indices 
of the basis cycles $\, {\bf e}_{1} \, $, $\, {\bf e}_{2} \, $ 
in the torus $\, \mathbb{T}^{2} \, $ 
$${\bf e}_{1} \circ {\bf e}_{1} \,\,\, = \,\,\, 
{\bf e}_{2} \circ {\bf e}_{2} \,\,\, = \,\,\, 0 \,\,\, , \quad
{\bf e}_{1} \circ {\bf e}_{2} \,\,\, = \,\,\, 1 $$

 It is easy to see that for any cycle
$${\bf a} \,\,\, = \,\,\, m \, {\bf e}_{1} \,\, + \,\, 
n \, {\bf e}_{2}  \quad \subset \quad \mathbb{T}^{2} $$
we then have the relations
$${\bf a} \circ \pi [{\bf a}] \,\,\, = \,\,\, 
m^{2} \, {\bf e}_{1} \circ {\bf e}_{2} \,\, - \,\, 
n^{2} \, {\bf e}_{2} \circ {\bf e}_{1} \,\,\, = \,\,\, 
m^{2} \, + \, n^{2} $$

 In other words, the intersection index 
$\, {\bf a} \, \circ \, \pi [{\bf a}] \, $ is equal to the square 
of the distance between the beginning and end of the path 
$\, {\bf a} \, $ representing the cycle $\, {\bf a} \, $ in 
$\, \mathbb{R}^{2} \, $, divided by $\, L^{2} \, $.

 Let us use here the notation $\, \widehat{\circ} \, $ for the 
algebraic (taking into account the sign) number of intersection 
points of any two oriented curves in $\, \mathbb{R}^{2} \, $.

 Now let in Fig. \ref{Gamma}
$$\overrightarrow{P P_{1}} \,\,\, = \,\,\, 
m \, {\bf e}_{1} \,\, + \,\, n \, {\bf e}_{2} \,\,\, , $$
where $\, d^{2} \, \leq \, (m^{2} + n^{2}) \, L^{2} \, $. 

 For the cycles $\, \widehat{\Gamma} \, $ and 
$\, \pi [ \widehat{\Gamma} ] \, $ in the torus 
$\, \mathbb{T}^{2} \, $ we have
$$\widehat{\Gamma} \circ \pi \big[ \widehat{\Gamma} \big] 
\,\,\, = \,\,\, m^{2} \, + \, n^{2} $$

 On the other hand, we can calculate the value 
$\, \widehat{\Gamma} \circ \pi \big[ \widehat{\Gamma} \big] \, $ 
by summing the algebraic intersection numbers of 
$\, \widehat{\Gamma} \, $ with all integer shifts of 
$\, \pi [ \widehat{\Gamma} ] \, $ in $\, \mathbb{R}^{2} \, $
\begin{multline}
\label{FullIndex}
\widehat{\Gamma} \circ \pi \big[ \widehat{\Gamma} \big] 
\,\,\, = \,\,\, \widehat{\Gamma} \,\,\, \widehat{\circ} \, 
\bigcup_{(k,l)} \Big( \pi \big[ \widehat{\Gamma} \big] 
\,\, + \,\, k \, {\bf e}_{1} \, + \, l \, {\bf e}_{2} \Big) 
\,\,\, =  \\
= \,\,\, \Gamma \,\,\, \widehat{\circ} \, 
\bigcup_{(k,l)} \Big( \pi \big[ \Gamma \big] 
\,\, + \,\, k \, {\bf e}_{1} \, + \, l \, {\bf e}_{2} \Big) 
\,\,\, +  \\ 
+ \,\,\, {\bf s} \,\,\, \widehat{\circ} \, 
\bigcup_{(k,l)} \Big( \pi \big[ \Gamma \big] 
\,\, + \,\, k \, {\bf e}_{1} \, + \, l \, {\bf e}_{2} \Big) 
\,\,\, +  \\ 
+ \,\,\, \Gamma \,\,\, \widehat{\circ} \, 
\bigcup_{(k,l)} \Big( \pi [ {\bf s} ] 
\,\, + \,\, k \, {\bf e}_{1} \, + \, l \, {\bf e}_{2} \Big) 
\,\,\, +  \\ 
+ \,\,\, {\bf s} \,\,\, \widehat{\circ} \, 
\bigcup_{(k,l)} \Big( \pi [ {\bf s} ] 
\,\, + \,\, k \, {\bf e}_{1} \, + \, l \, {\bf e}_{2} \Big) 
\end{multline}
($k, l \in \mathbb{Z}$). 

\begin{figure}[t]
\begin{center}
\includegraphics[width=\linewidth]{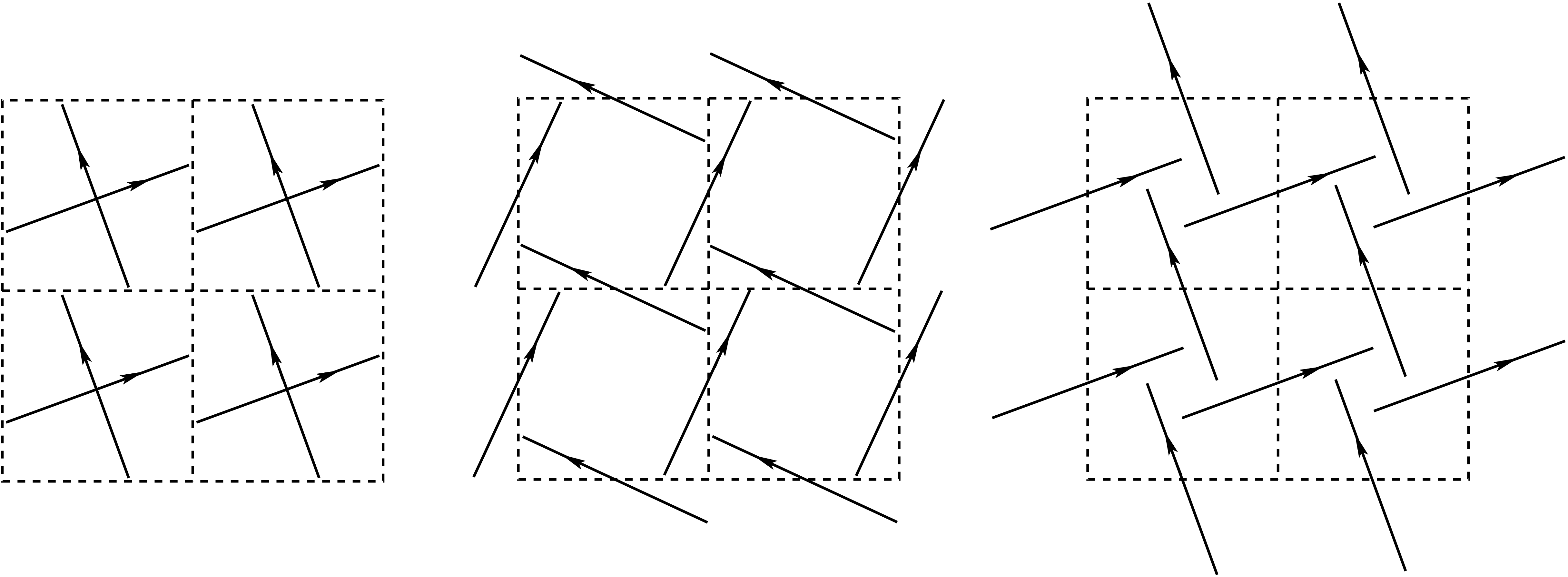}
\end{center}
\caption{Integer shifts of the segments $\, {\bf s} \, $ and 
$\, \pi [ {\bf s} ] \, $ in the case when their midpoints are 
centers of rotational symmetry, and in the cases when they 
are only inversion centers of the potential $\, V (x, y) \, $.}
\label{SpisSym}
\end{figure}

 By construction, the first term in this sum is equal to 1, 
and the next two terms are equal to each other. The midpoint 
of the segment $\, {\bf s} \, $, as well as $\, \pi [ {\bf s} ] \, $, 
in our case is an inversion center of the potential 
$\, V (x, y) \, $. Depending on whether these points are also 
centers of rotational symmetry of $\, V (x, y) \, $ or not, 
the segment $\, {\bf s} \, $ either intersects only one of 
the integer shifts of $\, \pi [ {\bf s} ] \, $ or does not 
intersect any of them (Fig. \ref{SpisSym}).

 Let $\, d \, = \, q L \, $. By (\ref{PP1}) we then have
$$\widehat{\Gamma} \circ \pi \big[ \widehat{\Gamma} \big] 
\,\,\, \geq \,\,\, q^{2} $$
and thus
$${\bf s} \,\,\, \widehat{\circ} \, 
\bigcup_{(k,l)} \Big( \pi \big[ \Gamma \big] 
\,\, + \,\, k \, {\bf e}_{1} \, + \, l \, {\bf e}_{2} \Big) 
\,\,\, \geq \,\,\, {q^{2} - 2 \over 2} $$

\begin{figure}[t]
\begin{center}
\includegraphics[width=\linewidth]{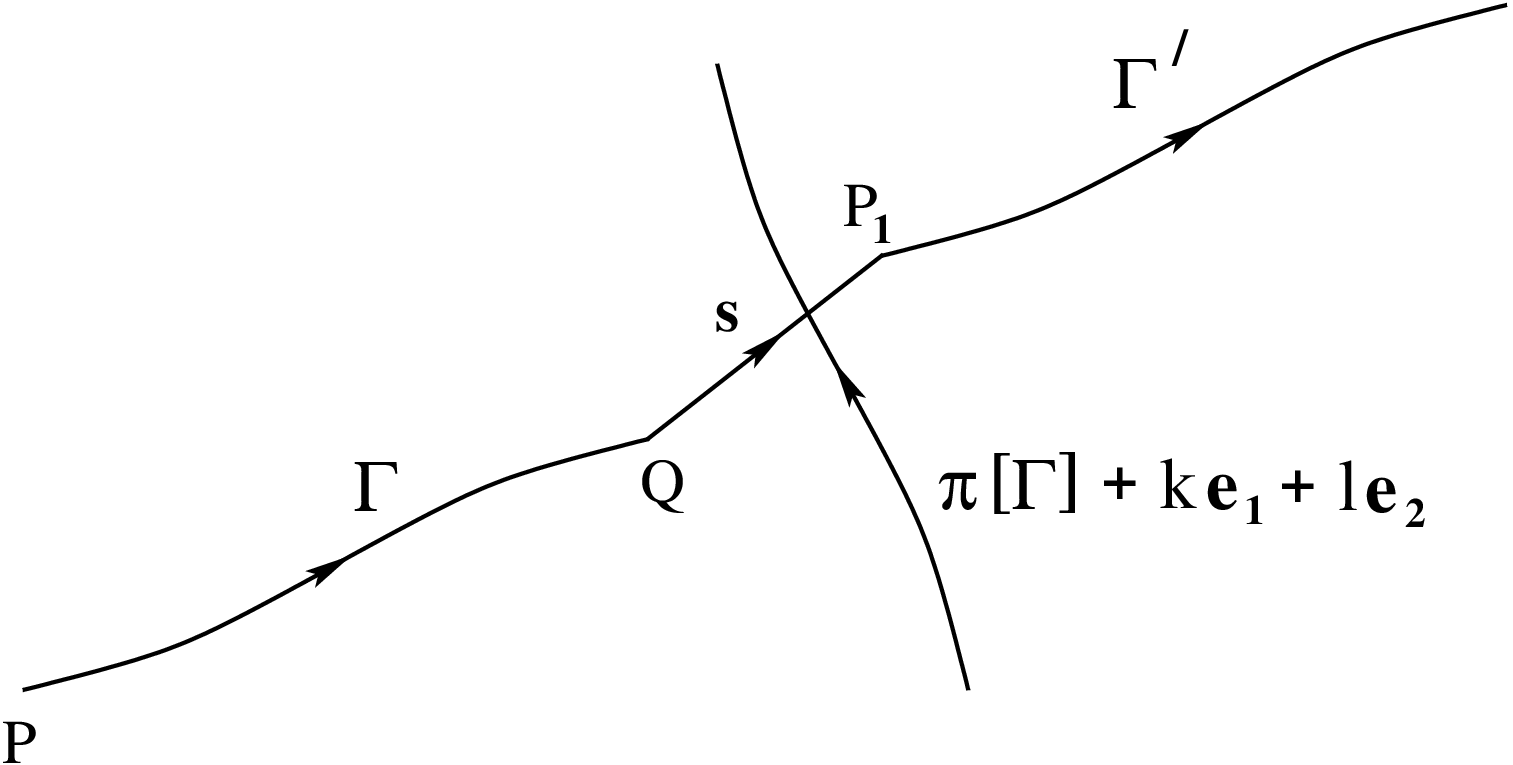}
\end{center}
\caption{Paths $\, \Gamma \, $, $\, \Gamma^{\prime} \, $, and a path 
$\, \pi \big[ \Gamma \big] \, + \, k \, {\bf e}_{1} 
\, + \, l \, {\bf e}_{2} \, $
intersecting segment $\, {\bf s} \, $.}
\label{TwoGamma}
\end{figure}

 In addition to the path $\, \Gamma \, $, consider also its 
integer shift $\, \Gamma^{\prime} \, $ with the beginning at 
the point $\, P_{1} \, $ (Fig. \ref{TwoGamma}). 
The path $\, \Gamma^{\prime} \, $, like the path $\, \Gamma \, $, 
intersects only one of the integer shifts of the path 
$\, \pi \big[ \Gamma \big] \, $, forming a symmetric graph 
in $\, \mathbb{R}^{2} \, $. Both the corresponding shifts of 
the path $\, \pi \big[ \Gamma \big] \, $, obviously, do not 
intersect the segment $\, {\bf s} \, $. The segment 
$\, {\bf s} \, $, however, can be intersected by other integer 
shifts of $\, \pi \big[ \Gamma \big] \, $.

 By construction, the distance between the ends ($P$ and $Q$) 
of the path $\, \Gamma \, $ is the greatest distance between 
points in the region $\, \Omega \, $. As a consequence, 
none of the remaining paths
\begin{equation}
\label{piGammaSdvig}
\pi \big[ \Gamma \big] \,\, + \,\, k \, {\bf e}_{1} \, + \,
l \, {\bf e}_{2}
\end{equation}
can ``go around'' the path $\, \Gamma \, $ or $\, \Gamma^{\prime} \, $ 
and its algebraic intersection number with the segment 
$\, {\bf s} \, $ can only be $\, -1 \, $, $\, 0 \, $ or $\, 1 \, $.

 It can be seen, therefore, that the segment $\, {\bf s} \, $ 
must be intersected by at least $\, (q^{2} - 2) / 2 \, $ 
different curves (\ref{piGammaSdvig}).

 By construction, the value of the potential $\, V (x, y) \, $ 
on each curve (\ref{piGammaSdvig}) does not exceed 
$\, c_{0} - \Delta c \, $. At the same time, each curve 
(\ref{piGammaSdvig}) lies in its ``own'' cell of the ``singular net'' 
$\, V (x, y) \, = \, c_{0} \, $ (Fig. \ref{Graph}). Each intersection 
of $\, {\bf s} \, $ with a curve (\ref{piGammaSdvig}) is thus contained 
in some segment $\, I_{p} \subset {\bf s} \, $ ($p = 1, \dots, N$), 
lying in the same cell of the ``singular net''. The boundaries of 
$\, I_{p} \, $ are the closest intersections of 
$\, {\bf s} \, $ with the boundaries of the cell, where the value 
of $\, V (x, y) \, $ is equal to $ \, c_{0} \, $. It is easy to see 
that different segments $\, I_{p} \, $ do not intersect each other, 
and their total number is at least $\, q^{2}/2 \, - \, 1 \, $.

\begin{figure}[t]
\begin{center}
\includegraphics[width=\linewidth]{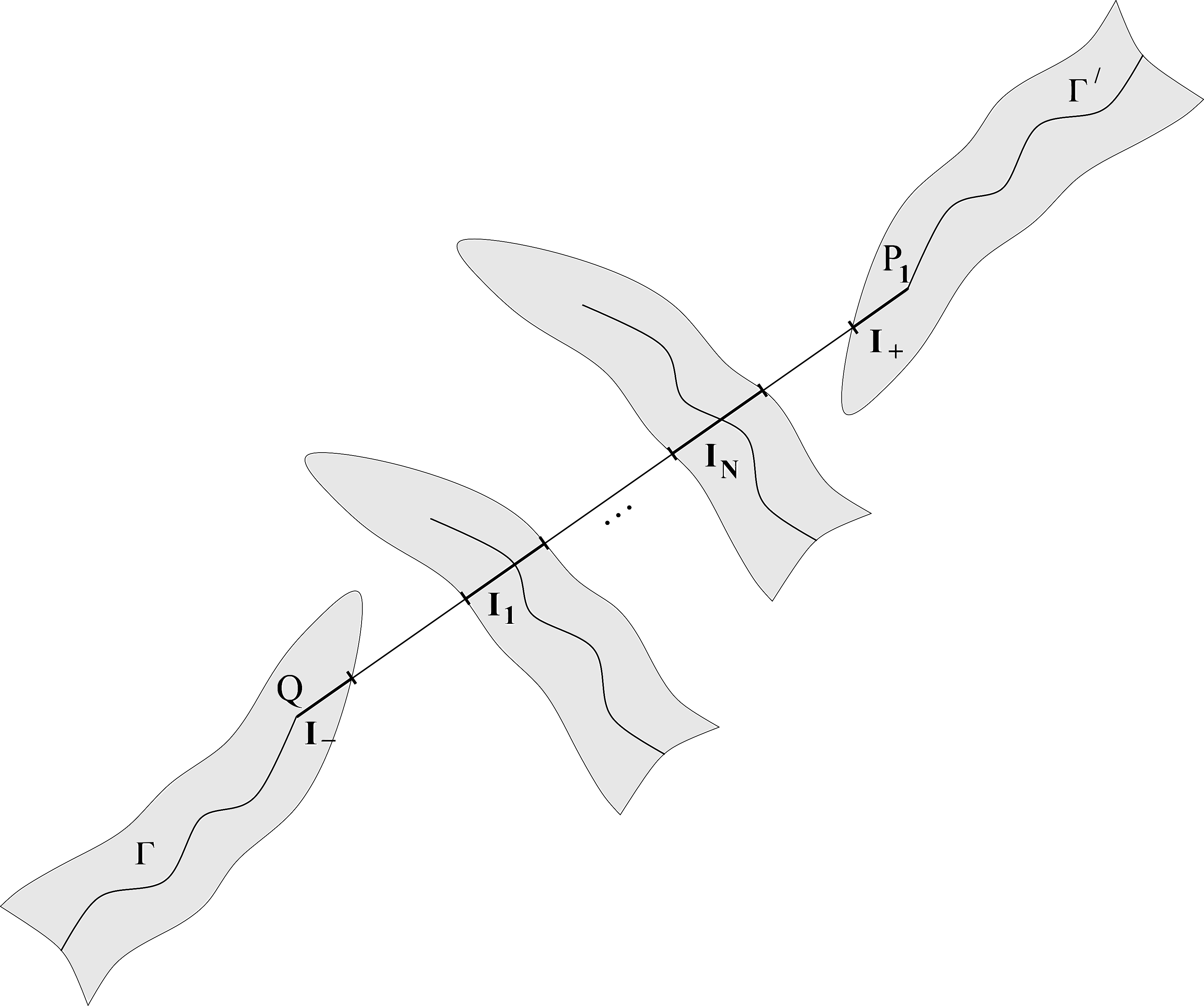}
\end{center}
\caption{The intersection of the segment $\, {\bf s} \, $ 
by the curves (\ref{piGammaSdvig}) and the corresponding 
segments $\, I_{p} \, $, $\, I_{-} \, $ and $\, I_{+} \, $.}
\label{Intersections}
\end{figure}

 The value of the potential $\, V (x, y) \, $ at the points 
$\, Q \, $ and $\, P_{1} \, $ also does not exceed 
$\, c_{0} - \Delta c \, $. These points can also be separated 
by segments $\, I_{-} \, $ and $\, I_{+} \, $ that do not 
intersect $\, I_{p} \, $, so that $\, V (x, y) \, = \, c_{0} \, $ 
at their upper and lower boundaries (Fig. \ref{Intersections}). 
Let us denote $\, I_{0} \, = \, I_{-} \cup I_{+} \, $.

 It is easy to see that the measure of at least one of the sets 
$\, I_{0} \, $, $\, I_{1} \, $, $\, \dots \, $, $\, I_{N} \, $ 
does not exceed
$$ {\sqrt{5} L \over 2} \Big/ {q^{2} \over 2} 
\,\,\, = \,\,\, {\sqrt{5} L \over q^{2}} $$ 

 According to Lagrange's theorem we then have
$$\Delta c \,\,\, \leq \,\,\, 
{\sqrt{5} C_{1} L \over 2 q^{2}} $$ 
and thus
$$q \,\,\, \leq \,\,\, 
\sqrt{ \sqrt{5} C_{1} L \over 2 \Delta c} $$
(similarly, for the set
$\, \Omega^{+}_{c_{0} + \Delta c} \, [V] $).

 The second of the possible cases (the asymmetric 
domain $\, \Omega$) is, in fact, almost no different 
from the first one. In this case, the path $\, \Gamma \, $ 
is determined by a non-self-intersecting curve connecting 
the maximally distant points in $\, \Omega \, $, and 
the first term in (\ref{FullIndex}) vanishes. It can also 
be seen that the segment $\, {\bf s} \, $ can intersect 
here with no more than two integer shifts of 
$\, \pi [ {\bf s} ] \, $. In other details, the 
consideration of this case completely repeats the 
above reasoning.

\vspace{1mm}

\begin{figure}[t]
\begin{center}
\includegraphics[width=0.7\linewidth]{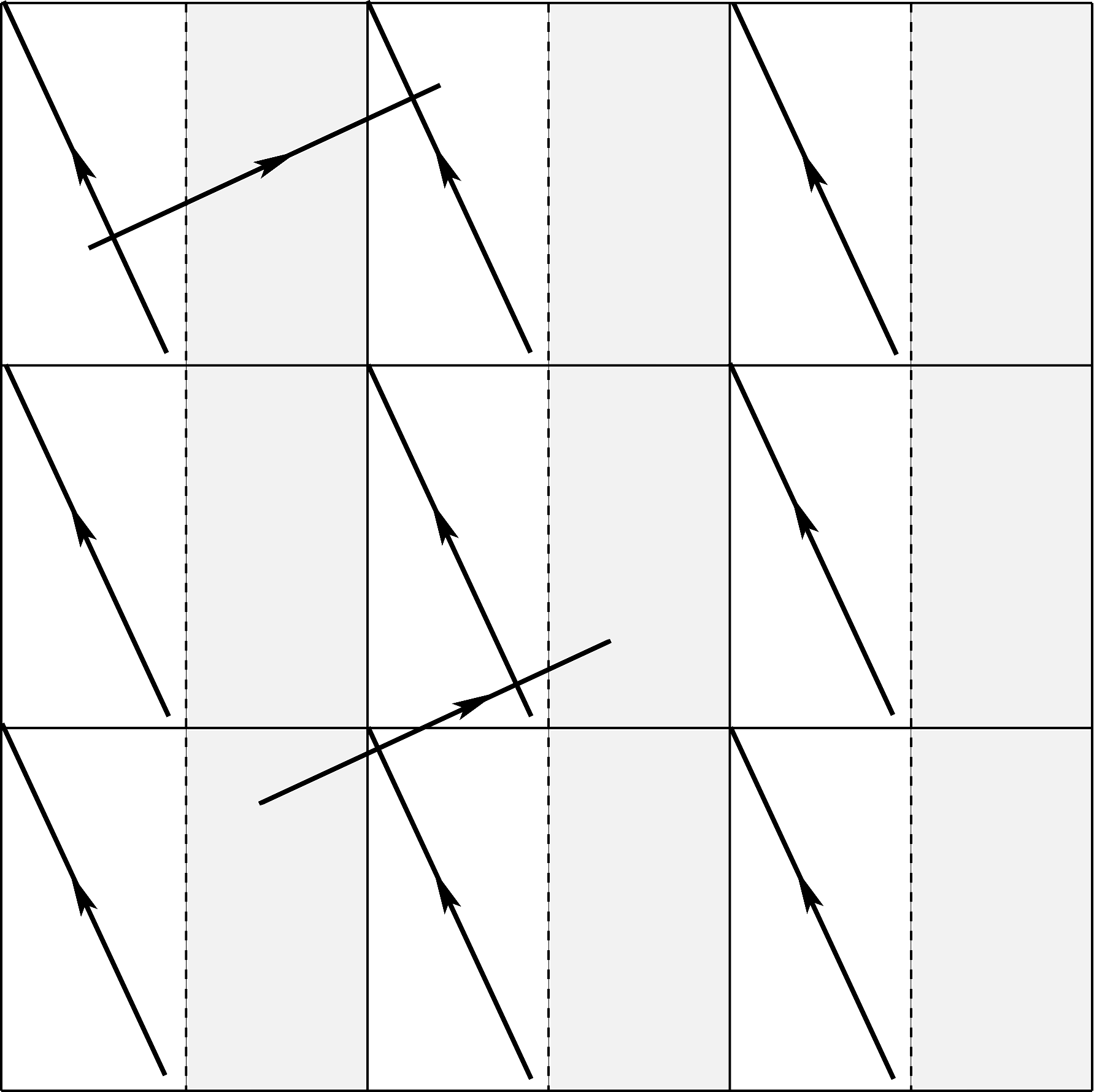}
\end{center}
\caption{Possible types of intersection of the 
segment $\, {\bf s} \, $ with integer shifts of
$\, \pi [ {\bf s} ] \, $ in the case of an asymmetric 
domain $\, \Omega \, $.} 
\label{SpisNonSym}
\end{figure}

 In the case of other symmetries we can generally assume 
the presence of the third-order symmetry, including the 
sixth-order symmetry as a special case. To define the 
torus $\, \mathbb{T}^{2} \, $ we use then vectors 
$\, {\bf e}_{1} \, $, $\, {\bf e}_{2} \, $, such that
$$\pi_{120^{\circ}} [ {\bf e}_{1} ] \,\,\, = \,\,\, {\bf e}_{2} $$

 When rotating by $120^{\circ}$, the cycle 
$\, {\bf a} \, = \, m \, {\bf e}_{1} \, + \, n \, {\bf e}_{2} \, $ 
acquires the following coordinates in the basis
$\, \{ {\bf e}_{1} , \, {\bf e}_{2} \} \, $
$$\left(
\begin{array}{cc}
0 &  - 1  \\
1 &  - 1
\end{array}  \right) 
\left(
\begin{array}{c}
m  \\
n
\end{array}  \right) 
\,\,\, = \,\,\, 
\left(
\begin{array}{c}
- n  \\
m - n
\end{array}  \right) $$

 The intersection index 
$\, {\bf a} \, \circ \, \pi_{120^{\circ}} [ {\bf a} ] \, $ 
is then equal to
$$m (m - n ) \, + \, n^{2} \,\,\, = \,\,\, m^{2} \, + \, n^{2} 
\, - \, m n  \,\,\, , $$
i.e. it also coincides with the square of the usual 
(Euclidean) length of the vector 
$\, m \, {\bf e}_{1} \, + \, n \, {\bf e}_{2} \, $, 
divided by $\, L^{2} \, $.

  The consideration of the case of an asymmetric domain 
$\, \Omega \, $ here obviously repeats the reasoning for 
the case of the fourth-order symmetry. For a symmetric domain 
$\, \Omega \, $ it suffices to choose a curve 
$\, \Gamma \subset \Omega \, $ connecting its most 
distant points such that the number of its intersections
with $\, \Gamma^{*} \, $ is equal to 0 or 1, after which 
the proof also repeats the above reasoning.

 A minor difference from the case of the 4th order symmetry 
is also present here in the estimate of the length of the 
segment $\, {\bf s} \, $, which now does not exceed $\, L \, $, 
which leads to a slightly different formulation of the Lemma 
in this case.

{\hfill The lemma is proven.}

\section{Conclusion}
\setcounter{equation}{0}

 In this paper, we consider the Novikov problem for a special 
class of quasiperiodic functions that play an important role in 
the physics of two-dimensional systems. More precisely, 
we consider the level lines of potentials that arise 
in ``two-layer'' structures as a result of a certain superposition 
of periodic functions. As is not difficult to show, such potentials 
represent functions with 4 quasiperiods on a plane.

 Here we study the case where the potentials have rotational 
symmetry, which occurs in many interesting systems. As is well known, 
in many such systems an important role is played by ``magic'' rotation 
angles, which correspond to the emergence of ``complex'' periodic 
potentials with extremely nontrivial properties. At the same time, 
the general case corresponding to the emergence of quasi-periodic 
potentials in dimension 2 is also of great interest.

 We mainly investigate here the emergence of open level 
lines of the described potentials, both for special cases 
of ``magic'' angles and for the generic case. In both cases, 
the emergence of open level lines has interesting features 
inherent to the specificity of each case. From our point of view, 
it is important, in particular, that the generic potentials 
have here ``chaotic'' open level lines at a single energy level, 
which unites them with ``chaotic'' potentials with 3 quasiperiods, 
as well as with random potentials on the plane.

\vspace{3mm}

 The author is grateful to Prof. I.A. Dynnikov for important
discussions during the work.

\end{document}